\documentclass[11pt]{article}

\usepackage{amsfonts,amssymb,chet,dsfont,graphicx,mathrsfs,pgfplots,tikz}
\allowdisplaybreaks

\newcommand{\1}{\mathds{1}}

\newcommand{\Op}[2]{\mathcal{O}_{#1}(\eta_{#2})}

\newcommand{\ee}[3]{(\eta_{#1}\cdot\eta_{#2})^{#3}}

\newcommand{\D}{\mathcal{D}}
\newcommand{\A}{\mathcal{A}}
\newcommand{\cOPE}[4]{{}_{#1}c_{#2#3}^{\phantom{#2#3}#4}}
\newcommand{\DOPE}[4]{{}_{#1}\D_{#2#3}^{\phantom{#2#3}#4}}
\newcommand{\tOPE}[6]{{}_{#1}t_{#2#3}^{#5#6#4}}

\newcommand{\aCF}[4]{{}_{#1}\alpha_{#2#3#4}}
\newcommand{\taCF}[4]{{}_{#1}\tilde{\alpha}_{#2#3#4}}
\newcommand{\FCF}[6]{{}_{#1}F_{#2#3#4}^{#5#6}}
\newcommand{\tFCF}[6]{{}_{#1}\tilde{F}_{#2#3#4}^{#5#6}}

\newcommand{\Vev}[1]{\left\langle{#1}\right\rangle}

\title{Conformal Conserved Currents\\in Embedding Space}

\author{Jean-Fran\c{c}ois Fortin$^{\ast,}$\email{jean-francois.fortin@phy.ulaval.ca}, Wen-Jie Ma$^{\ast,}$\email{wenjie.ma.1@ulaval.ca}, 
Valentina Prilepina$^{\ast,\$,}$\email{valentina.prilepina.1@ulaval.ca}, and Witold Skiba$^{\dagger,}$\email{witold.skiba@yale.edu}}

\affiliation{
$^\ast$D\'epartement de Physique, de G\'enie Physique et d'Optique\\Universit\'e Laval, Qu\'ebec, QC G1V 0A6, Canada\\
$^\$$Institute of Theoretical and Mathematical Physics (ITMP)\\Lomonosov Moscow State University\\Leninskie Gory, GSP-1, 119991 Moscow, Russian Federation\\
$^\dagger$Department of Physics, Yale University, New Haven, CT 06520, USA
}

\abstract{We study conformal conserved currents in arbitrary irreducible representations of the Lorentz group using the embedding space formalism.  With the help of the operator product expansion, we first show that conservation conditions can be fully investigated by considering only two- and three-point correlation functions.  We then find an explicitly conformally-covariant differential operator in embedding space that implements conservation based on the standard position space operator product expansion differential operator $\partial_\mu$, although the latter does not uplift to embedding space covariantly.  The differential operator in embedding space that imposes conservation is the same differential operator $\D_{ijA}$ used in the operator product expansion in embedding space.  We provide several examples including conserved currents in irreducible representations that are not symmetric and traceless.  With an eye on four-point conformal bootstrap equations for four conserved vector currents $\Vev{JJJJ}$ and four energy-momentum tensors $\Vev{TTTT}$, we mostly focus on conservation conditions for $\Vev{JJ\mathcal{O}}$ and $\Vev{TT\mathcal{O}}$.  Finally, we reproduce and extend the consequences of conformal Ward identities at coincident points by determining three-point coefficients in terms of charges.}

\date{December 2020} 

\begin{document}

\maketitle



\section{Introduction}\label{SecIntro}

Quantum field theories are often classified in terms of their symmetry group and their matter content, with the global symmetry group fixing the allowed charges of the matter fields.  Usually, for a continuous global symmetry group, the associated conserved currents lead to the charges, which describe the algebra of the continuous symmetry group.

In the context of conformal field theory (CFT), a conserved current is a quasi-primary operator for which a given descendant is also quasi-primary.  Since it appears in all CFTs, the most important quasi-primary operator of a CFT is its energy-momentum tensor, which is the conserved current responsible for the spacetime symmetry group of the CFT.  In order to carry out the conformal bootstrap program \cite{Ferrara:1973yt,Polyakov:1974gs} for four energy-momentum tensors, it is therefore necessary to investigate the conservation conditions associated to the energy-momentum tensor.

In this paper, we use the embedding space formalism of \cite{Fortin:2019fvx,Fortin:2019dnq} to analyse conservation conditions of quasi-primary operators in arbitrary irreducible representations of the Lorentz group.  Using the embedding space operator product expansion (OPE) differential operator, we determine conserved current differential operators in embedding space that are explicitly conformally covariant.  Our results are highly analogous to the ones in position space, where the differential operator responsible for conservation is the building block for the OPE differential operator.

We then show that conservation conditions are completely determined from two- and three-point correlation functions, with the former forcing the conformal dimensions of the conserved quasi-primary operators to saturate the unitarity bound \cite{Metsaev:1995re,Minwalla:1997ka,Grinstein:2008qk}.  Since our technique does not rely on the projection operators of the unconserved quasi-primary operators---only the projection operators of the conserved currents are necessary---we also study several examples, focusing primarily on the conservation conditions for $\Vev{JJ\mathcal{O}}$ and $\Vev{TT\mathcal{O}}$, where $J$ is a standard conserved vector current and $T$ is the energy-momentum tensor.

Conformal conserved currents have been studied extensively in the past.  One of the first analysis of anomalous conformal Ward identities can be found in \cite{Schreier:1971um}, while conserved symmetric-traceless quasi-primary operators in embedding space have been studied in \cite{Ferrara:1971zy,Ferrara:1971vh,Ferrara:1972xq,Ferrara:1973eg,Ferrara:1973yt}.  Conformal Ward identities at coincident points for the energy-momentum tensor and two scalars were analysed in \cite{Cardy:1987dg} where a relationship between three-point coefficients on one side and conformal dimensions and two-point coefficients on the other was uncovered.  Three-point correlation functions of conserved vector currents and energy-momentum tensors were discussed extensively and in all generality in \cite{Osborn:1993cr,Erdmenger:1996yc}, while conserved currents in the $O(N)$ model were investigated in \cite{Petkou:1994ad}.  The works of \cite{Shaynkman:2004vu,Alkalaev:2012rg} looked at the conditions under which differential equations involving conserved currents are conformally covariant.  The relevant result from the point of view of unitarity, where conformally-covariant equations contract the position space differential operator with a vector index of the smallest row of the Young tableau associated to a conserved current in a mixed irreducible representation, were used in \cite{Costa:2011mg,Costa:2014rya} to analyse conserved vector currents and the energy-momentum tensor.  A link between the number of independent tensor structures appearing in correlation functions of conserved currents and $S$-matrices was also uncovered and discussed in \cite{Costa:2011mg,Dymarsky:2013wla,Kravchuk:2016qvl}.  Conformal conserved currents appear as well in conformal collider physics and superconformal field theory, where they were studied for example in \cite{Li:2015itl,Meltzer:2018tnm} and \cite{Cordova:2016emh}, respectively.  There is also a very large body of work on higher-spin currents, \textit{i.e.}\ conserved currents that are not vector currents or the energy-momentum tensor, that cannot possibly be cited exhaustively (see \cite{Vasiliev:1999ba,Mikhailov:2002bp,Sezgin:2002rt,Eastwood:2002su,Giombi:2009wh,Giombi:2011rz,Maldacena:2011jn,Maldacena:2012sf,Alba:2015upa,Alba:2013yda} and references therein).  Theories with higher-spin currents are believed to be free \cite{Maldacena:2011jn,Alba:2015upa,Alba:2013yda}.

This paper is organized as follows: Section \ref{SecOPE} presents the OPE and demonstrate that the conservation conditions are analyzed more efficiently by examining the correlation functions.  A proof of the irrelevance of higher-point correlation functions in the study of conservation conditions, \textit{i.e.}\ correlation functions with more than three points, is also given.  Section \ref{SecUC} discusses the unitarity bound and finds the appropriate conformally-covariant conserved current differential operators that implement conservation of three-point correlation functions.  In Section \ref{SecCF}, the method is used to determine conservation conditions for three-point correlation functions of quasi-primary operators in arbitrary irreducible representations of the Lorentz group.\footnote{Since the method is general, we discuss arbitrary irreducible representations even though higher-spin theories are believed to be free.}  The results are given in terms of trivial substitutions.  Section \ref{SecEx} uses the formalism developed in the previous sections to investigate conservation conditions in several examples, focusing on conserved vector currents and the energy-momentum tensor.  Finally, we conclude in Section \ref{SecConc}, with Appendix \ref{SAppNot} containing a concise review of the embedding space formalism, Appendix \ref{SAppJEM} presenting the results for $\Vev{JJ\mathcal{O}}$ and $\Vev{TT\mathcal{O}}$, and Appendix \ref{SAppWard} describing Ward identities at coincident points for $\Vev{\mathcal{O}\mathcal{O}J}$ and $\Vev{\mathcal{O}\mathcal{O}T}$.

Apart from developing the techniques necessary to study conservation conditions in the embedding space formalism, this paper can be seen as a complement to \cite{Costa:2011mg}, where the conservation conditions for $\Vev{JJ\mathcal{O}}$ and $\Vev{TT\mathcal{O}}$ when $\mathcal{O}$ is in a symmetric-traceless irreducible representation were first computed.


\section{Conserved Currents and the OPE}\label{SecOPE}

In this section we discuss conserved currents from the point of view of the OPE.  We first explain why it is preferable to consider conserved currents in correlation functions rather than directly in the OPE.  Although we bypass the study of conserved currents from the point of view of the OPE, the latter is nevertheless used to show that it is only necessary to consider two- and three-point correlation functions to fully implement the conservation conditions \cite{Costa:2011mg,Costa:2011dw}.


\subsection{OPE and its Action}

The embedding space OPE, introduced in \cite{Fortin:2019fvx,Fortin:2019dnq}, transforms the product of two quasi-primary operators into an infinite sum of quasi-primary operators.  It can be expressed as
\eqn{
\begin{gathered}
\Op{i}{1}\Op{j}{2}=\sum_k\sum_{a=1}^{N_{ijk}}\cOPE{a}{i}{j}{k}\DOPE{a}{i}{j}{k}(\eta_1,\eta_2)\Op{k}{2},\\
\DOPE{a}{i}{j}{k}(\eta_1,\eta_2)=\frac{1}{\ee{1}{2}{p_{ijk}}}(\mathcal{T}_{12}^{\boldsymbol{N}_i}\Gamma)(\mathcal{T}_{21}^{\boldsymbol{N}_j}\Gamma)\cdot\tOPE{a}{i}{j}{k}{1}{2}\cdot\D_{12}^{(d,h_{ijk}-n_a/2,n_a)}(\mathcal{T}_{12\boldsymbol{N}_k}\Gamma)*,\\
p_{ijk}=\frac{1}{2}(\tau_i+\tau_j-\tau_k),\qquad h_{ijk}=-\frac{1}{2}(\chi_i-\chi_j+\chi_k),\\
\tau_{\mathcal{O}}=\Delta_{\mathcal{O}}-S_{\mathcal{O}},\qquad\chi_{\mathcal{O}}=\Delta_{\mathcal{O}}-\xi_{\mathcal{O}},\qquad\xi_{\mathcal{O}}=S_{\mathcal{O}}-\lfloor S_{\mathcal{O}}\rfloor,
\end{gathered}
}[EqOPE]
where the OPE differential operator is given by
\eqn{
\begin{gathered}
\D_{12}^{(d,h,n)A_1\cdots A_n}=\frac{1}{\ee{1}{2}{\frac{n}{2}}}\D_{12}^{2(h+n)}\eta_2^{A_1}\cdots\eta_2^{A_n},\\
\D_{12}^2=\D_{12}^A\D_{12A},\qquad\D_{12}^A=\ee{1}{2}{\frac{1}{2}}\A_{12}^{AB}\partial_{2B},\\
\A_{12}^{AB}=\frac{1}{\ee{1}{2}{}}[\ee{1}{2}{}g^{AB}-\eta_1^A\eta_2^B-\eta_1^B\eta_2^A],
\end{gathered}
}[EqDiff]
and the remaining quantities [including the half-projectors $(\mathcal{T}_{12}^{\boldsymbol{N}}\Gamma)$ that encode the irreducible representations of quasi-primary operators and the tensor structures $\tOPE{a}{i}{j}{k}{1}{2}$ that serve to properly contract embedding spinor and vector indices] are introduced in \cite{Fortin:2019dnq} (see Appendix \ref{SAppNot} for a brief review).

Relying on the knowledge of conserved currents in position space, we know that conservation conditions are obtained by applying and contracting a differential operator with the conserved current.  In position space, the differential operator is $\partial_\mu$, which is coincidentally the building block of the position space OPE differential operator.\footnote{Scalars and spinors must be considered separately.}  This procedure can be applied at the level of the conserved current, leading to an operator statement.  We note that in the embedding space formalism of \cite{Fortin:2019dnq}, the same statement cannot be easily made since quasi-primary operators in arbitrary irreducible representations are well defined through the OPE and its resulting correlation functions.

Moreover starting from the OPE, either in position space or in embedding space \eqref{EqOPE}, we observe that quasi-primary operators are not treated democratically.  Indeed, embedding space spinor indices for the quasi-primary operators on the LHS are free while they are contracted for the quasi-primary operators on the RHS.  Hence, it is trivial to impose conservation on the quasi-primary operators on the LHS of \eqref{EqOPE} but it is not clear how to proceed with the quasi-primary operators on the RHS of \eqref{EqOPE}.  Obviously, this issue does not occur when conservation conditions are analysed starting from correlation functions.

Another issue with the OPE \eqref{EqOPE} originates from the action of the OPE differential operator \eqref{EqDiff} on $(M-1)$-point correlation functions.  Indeed, a given $M$-point correlation function can be evaluated with the help of the OPE by replacing two quasi-primary operators with the RHS of \eqref{EqOPE}, effectively transforming the $M$-point correlation function of interest into an infinite sum of $(M-1)$-point correlation functions acted upon by the OPE differential operator \eqref{EqDiff}.  The action of the OPE differential operator \eqref{EqDiff} on the most general function of embedding space coordinates was found in \cite{Fortin:2019fvx,Fortin:2019dnq}.  From the explicit solution (see the $\bar{I}$-function in \cite{Fortin:2019dnq}), it is clear that the action of the OPE differential operator can lead to poles when the conformal dimensions of the exchanged quasi-primary operators are integers,\footnote{More precisely, when the twist $\tau_\mathcal{O}$ is an integer.} and the latter occurs precisely when quasi-primary operators are conserved.  This observation does not impede the use of the embedding space OPE since it is always possible to assume that conformal dimensions are generic until a change of basis of tensor structures (from the OPE basis to the three-point basis) has been performed, after which forcing conformal dimensions to their conserved values can be done without problem.  For example, for the first non-trivial case, corresponding to $M=3$, it was shown in \cite{Fortin:2019pep} that three-point correlation functions obtained from the action of the OPE are dressed with pre-factors that depend on the conformal dimension of the exchanged quasi-primary operator.  These pre-factors can blow up when the conformal dimension is an integer, but they are eliminated by changing basis of tensor structures, from the OPE basis to the three-point basis, with the help of rotation matrices as in \cite{Fortin:2020ncr}.  In the three-point basis, \textit{i.e.}\ from the three-point correlation functions without pre-factors, we are thus free to implement the conservation conditions.  As a final point, we want to stress that although one could change the proportionality constant appearing in the OPE differential operator \eqref{EqOPE} to remove the inconvenient pre-factors mentioned above, such a change would depend on the relevant quasi-primary operators in a non-trivial way, which is not very practical.  Moreover, it would depend on the correlation function under investigation, which is even more cumbersome since it precludes an optimal choice of proportionality constant.

From this discussion, we conclude that it is simpler to investigate the conservation conditions directly from correlation functions rather than from the OPE.


\subsection{Correlation Functions and OPE}

Although we just argued that conserved currents in embedding space are better studied starting from correlation functions, the OPE is still quite useful in proving that only two- and three-point correlation functions are necessary to implement the full set of conservation conditions (see \cite{Costa:2011mg,Costa:2011dw}).  At first glance, this result is expected as higher-point functions can be expressed, using the OPE repeatedly, in terms of lower-point functions.  However, the embedding space OPE involves both coordinates and the conservation condition might not automatically commute with the action of the OPE.

The two- and three-point correlation functions are given by \cite{Fortin:2019xyr}
\eqn{\Vev{\Op{i}{1}\Op{j}{2}}=(\mathcal{T}_{12}^{\boldsymbol{N}_i}\Gamma)\cdot(\mathcal{T}_{21}^{\boldsymbol{N}_j}\Gamma)\frac{\lambda_{\boldsymbol{N}_i}\cOPE{}{i}{j}{\1}}{\ee{1}{2}{\tau_i}},}[Eq2pt]
and \cite{Fortin:2019pep}
\eqn{
\begin{gathered}
\Vev{\Op{i}{1}\Op{j}{2}\Op{m}{3}}=\frac{(\mathcal{T}_{12}^{\boldsymbol{N}_i}\Gamma)(\mathcal{T}_{21}^{\boldsymbol{N}_j}\Gamma)(\mathcal{T}_{31}^{\boldsymbol{N}_m}\Gamma)\cdot\sum_{a=1}^{N_{ijm}}\aCF{a}{i}{j}{m}\mathscr{G}_{[a|}^{ij|m}}{\ee{1}{2}{\frac{1}{2}(\tau_i+\tau_j-\chi_m)}\ee{1}{3}{\frac{1}{2}(\chi_i-\chi_j+\tau_m)}\ee{2}{3}{\frac{1}{2}(-\chi_i+\chi_j+\chi_m)}},\\
\mathscr{G}_{[a|}^{ij|m}=(\bar{\eta}_3\cdot\Gamma)^{2\xi_m}\FCF{a}{i}{j}{m}{1}{2}(\A_{12},\Gamma_{12},\epsilon_{12};\A_{12}\cdot\bar{\eta}_3),
\end{gathered}
}[Eq3pt]
respectively.  In \eqref{Eq2pt} and \eqref{Eq3pt}, $\lambda_{\boldsymbol{N}_i}$ and $\mathscr{G}_{[a|}^{ij|m}$ are related to the tensor structures while $\cOPE{}{i}{j}{\1}$ and $\aCF{a}{i}{j}{m}$ are the three-point coefficients.\footnote{We note that $\lambda_{\boldsymbol{N}_i}$ can be set to $1$ without loss of generality.}  They arise from the OPE basis in \eqref{Eq2pt} and the three-point basis in \eqref{Eq3pt}.  Moreover, $\bar{\eta}_3^A=\eta_3^A\sqrt{\ee{1}{2}{}/[\ee{1}{3}{}\ee{2}{3}{}]}$, thus it is homogeneous of vanishing degree in all three embedding space coordinates.

We present here a schematic proof in embedding space using the fact that the differential operator implementing conservation in position space, given by $\partial_\mu$, is proportional to $\mathcal{L}_{A+}$ in embedding space with the embedding space coordinate $A=\mu$ (\textit{i.e.}\ non-covariant in embedding space).  The exact proof is easily derived from the discussion below and the results of the next sections.

Starting from an arbitrary $(M>3)$-point correlation function $\Vev{\Op{i_1}{1}\Op{i_2}{2}\cdots\Op{i_M}{M}}$ on which we want to impose conservation of $\Op{i_1}{1}$, it is easy to use the OPE \eqref{EqOPE} repetitively to express the conservation conditions schematically as
\eqn{\mathcal{L}_{1A+}\Vev{\Op{i_1}{1}\Op{i_2}{2}\cdots\Op{i_M}{M}}=\mathcal{L}_{1A+}\sum_jF_3(\eta_a\cdot\eta_b,\D_{ab})\Vev{\Op{i_1}{1}\Op{i_2}{2}\Op{j}{3}}=0,}
where $a,b\neq1,2$ for some complicated function $F_3$ built from the OPE \eqref{EqOPE}.  Here the differential operator $\mathcal{L}_{1A+}$ acts on $\eta_1$ as required, with its embedding space vector index contracted with the quasi-primary operator $\Op{i_1}{1}$ through some $\Gamma$-matrices as discussed below.  Since $\mathcal{L}_{1A+}$ commutes with $F_3$, the conservation conditions of three-point correlation functions imply the conservation conditions of the $(M>3)$-point correlation functions.

We could proceed a step further, using the OPE once more, to reach
\eqn{\mathcal{L}_{1A+}\Vev{\Op{i_1}{1}\Op{i_2}{2}\cdots\Op{i_M}{M}}=\mathcal{L}_{1A+}\sum_kF_2(\eta_a\cdot\eta_b,\D_{ab})\Vev{\Op{i_1}{1}\Op{k}{3}}=0,}
with $a\neq1$, $b\neq1,2$, and $F_2$ some function built from $F_3$ above and the OPE for $\Op{i_2}{2}\Op{j}{3}\sim\Op{k}{3}$.  Since $\mathcal{L}_{1A+}$ commutes with $F_2$, at first sight it seems that the conservation conditions of two-point correlation functions should imply the conservation conditions of three-point correlation functions, and hence all higher-point correlation functions.  But this conclusion is wrong due to the remaining two-point correlation function that forces the exchanged quasi-primary operator $\Op{k}{3}$ to be conserved with $\tau_k=\tau_{i_1}$ and $\chi_k=\chi_{i_1}$.  Therefore, contrary to the previous case, the OPE differential operator for $\Op{i_2}{2}\Op{j}{3}\sim\Op{k}{3}$ leads to poles originating from the conserved quasi-primary operator $\Op{k}{3}$ that is exchanged.  Thus, although the conservation condition on the two-point correlation function, implemented by $\mathcal{L}_{1A+}$, implies that the conformal dimension of the exchanged quasi-primary operator saturates the unitarity bound $\tau_k=\tau_{i_1}^*$ for some integer number $\tau_{i_1}^*$ and leads to a pre-factor $\tau_{i_1}-\tau_{i_1}^*$ times a modified two-point correlation function, it cannot be concluded that the conservation conditions for higher-point correlation functions are satisfied from the conservation conditions of two-point correlation functions.  Indeed, the OPE differential operator appearing in $\Op{i_2}{2}\Op{j}{3}\sim\Op{k}{3}$ generates poles in the conformal dimension $\tau_k=\tau_{i_1}$ that cancel the zero $\tau_{i_1}-\tau_{i_1}^*$ introduced by the conservation condition of the two-point correlation function.  These poles appear precisely because the exchanged quasi-primary operator $\Op{k}{3}$ is conserved.  We note that these poles are not removed by a change of basis---the OPE does not act on $\Vev{\Op{i_1}{1}\Op{k}{3}}$ but rather on $\mathcal{L}_{1A+}\Vev{\Op{i_1}{1}\Op{k}{3}}$.  Hence the conservation condition of the two-point correlation function does not imply the conservation conditions of higher-point correlation functions.

As a consequence, to fully investigate conserved currents, it is only necessary to analyse conservation conditions for two- and three-point correlation functions.


\section{Unitarity Bounds and Conformal Invariance}\label{SecUC}

This section discusses unitarity bounds and conformally-covariant conserved current differential operators.  Indeed, it is well known that conformal dimensions of conserved currents in general irreducible representations saturate the unitarity bound \cite{Metsaev:1995re,Minwalla:1997ka,Grinstein:2008qk}.  In fact, equations for conserved currents in non-trivial irreducible representations (\textit{i.e.} not scalars nor spinors) are conformally covariant as long as the unitarity bound is saturated and the contraction between the conserved current differential operator and the quasi-primary operator occurs with a Lorentz vector index from the smallest non-zero Dynkin index \cite{Shaynkman:2004vu,Alkalaev:2012rg,Costa:2011mg,Costa:2014rya}.

Throughout this section, we define $\eta_+=\eta_{d+1}+\eta_{d+2}=-\eta^{d+1}+\eta^{d+2}$ and $\mathcal{L}_{A+}=\mathcal{L}_{A,d+1}+\mathcal{L}_{A,d+2}$.  We also rely on the relations \cite{Fortin:2019dnq}
\eqn{
\begin{gathered}
-i\mathcal{L}_j^{AB}=\frac{1}{\ee{i}{j}{\frac{1}{2}}}(\eta_j^A\D_{ij}^B-\eta_j^B\D_{ij}^A)-\frac{1}{\ee{i}{j}{}}(\eta_i^A\eta_j^B-\eta_i^B\eta_j^A)\Theta_j,\\
\Gamma^{A_1\cdots A_nB}=\Gamma^{A_1\cdots A_n}\Gamma^B+\sum_{1\leq a\leq n}(-1)^{n+1-a}g^{A_aB}\Gamma^{A_1\cdots\widehat{A_a}\cdots A_n},
\end{gathered}
}[EqLD]
to find the conformally-covariant form of the conserved current differential operator in embedding space.  To gain intuition, we first discuss the special cases of scalars and spinors, before focusing on conserved quasi-primary operators in general irreducible representations of the Lorentz group.

The results of this section clearly show the separation of the conserved current differential operators into three parts: a part that can be made conformally covariant and that vanishes when acting on two-point correlation functions; a part that can always be discarded since it vanishes after acting on the conserved current thanks to the transversality condition and/or group theory; and a part that is not conformally covariant but is multiplied by a factor that vanishes when the unitarity bound is saturated.  Hence, conservation conditions for two-point correlation functions imply that conserved quasi-primary operators saturate their unitarity bound, while conservation conditions for three-point correlation functions can be obtained from conformally-covariant differential operators.  The latter are discussed in Section \ref{SecCF}.


\subsection{Unitarity Bounds}

It is known that conservation equations are conformally covariant only when the unitarity bound is saturated \cite{Shaynkman:2004vu,Alkalaev:2012rg,Costa:2011mg,Costa:2014rya}.\footnote{In \cite{Alkalaev:2012rg,Costa:2011mg}, another conformally-covariant equation was found that sometimes corresponds to the usual conservation equation.  This extra conformally-covariant equation will not be discussed here since in general it violates the unitarity bound.}  Using the conformal algebra, it is easy to see that the differential operator in the conservation equation must be contracted with a Lorentz vector index from the smallest non-vanishing Dynkin index of the irreducible representation of the quasi-primary operator \cite{Costa:2011mg,Costa:2014rya}.

The unitarity bounds have been studied in \cite{Metsaev:1995re,Minwalla:1997ka,Grinstein:2008qk}.  From their results, it is straightforward to see that the unitarity bound for non-trivial quasi-primary operators in general irreducible representations $\boldsymbol{N}$ [excluding scalars and spinors, for which $\tau\geq\tau^*=(d-2)/2$] can be expressed equivalently as
\eqn{
\begin{gathered}
\Delta\geq\Delta^*=d-1-\mathfrak{n}+S,\\
\tau\geq\tau^*=d-1-\mathfrak{n},\\
\chi\geq\chi^*=d-1-\mathfrak{n}+\lfloor S\rfloor.
\end{gathered}
}[EqU]
when written in terms of the different quantities introduced in \eqref{EqOPE}.  Here, $S$ is the ``spin'' of the irreducible representation $\boldsymbol{N}$ given by half the number of spinor indices\footnote{More precisely, the spin is given by $S=\sum_{1\leq i\leq r-1}N_i+N_r/2$ in odd dimensions and $S=\sum_{1\leq i\leq r-2}N_i+(N_{r-1}+N_r)/2$ in even dimensions.} while $\mathfrak{n}$ is the index of the smallest non-vanishing Dynkin index of the irreducible representation $\boldsymbol{N}=\sum_{1\leq i\leq r}N_i\boldsymbol{e}_i=\sum_{\mathfrak{n}\leq i\leq r}N_i\boldsymbol{e}_i$.\footnote{The proper definition for the parameter $\mathfrak{n}$ is the length of the smallest antisymmetric group of indices.  For odd dimensions, this definition matches the index of $\boldsymbol{e}_\mathfrak{n}$.  For even dimensions, this definition makes clear that $\mathfrak{n}=r-1$ for irreducible representations of the type $N_{r-1}\boldsymbol{e}_{r-1}+N_r\boldsymbol{e}_r$ with $N_{r-1}\geq1$ and $N_r\geq1$, while $\mathfrak{n}=r$ for irreducible representations of the types $N_{r-1}\boldsymbol{e}_{r-1}$ with $N_{r-1}\geq2$ and $N_r\boldsymbol{e}_r$ with $N_r\geq2$.}  Moreover $r$ is the rank of the Lorentz group and $(\boldsymbol{e}_i)_j=\delta_{ij}$ is the usual unit vector.  We note that \eqref{EqU} is correct for bosonic as well as fermionic irreducible representations.  Hence, when the unitarity bound is saturated, the twist $\tau=\tau^*$ (and for that matter $\chi=\chi^*$) is a non-negative integer.

From the relation between conserved currents and the unitarity bound, conserved currents have conformal dimensions saturating the unitarity bound \eqref{EqU}, thus their twist is a non-negative integer.  In general, in position space the action of the differential operator leading to the conservation conditions of a conserved current must separate into two non-trivial contributions: a contribution that is not conformally covariant multiplied by the saturated unitarity bound $\tau-\tau^*$, and a contribution that is conformally covariant.  Only the former contribution occurs for two-point correlation functions (the latter annihilates two-point correlation functions), while both contributions appear for three-point correlation functions.


\subsection{Conserved Scalars and Spinors}

Before discussing conservation conditions for quasi-primary operators in arbitrary irreducible representations, we investigate here scalars and spinors.  Conserved quasi-primary operators in scalar and spinor irreducible representations are special since their conservation conditions correspond to the free Klein-Gordon and Dirac equations, respectively.  Hence conserved scalars and spinors decouple from the theory.  From this point of view, they are not interesting.  They are however useful to argue for the correct conserved current differential operator in embedding space.

Starting with quasi-primary operators in the trivial irreducible representation, $\mathcal{O}^{(x)\boldsymbol{0}}(x)$, the conservation condition in position space is
\eqn{\partial^2\mathcal{O}^{(x)\boldsymbol{0}}(x)=0,}[Eq0PS]
which is satisfied for free scalars with $\tau=\tau^*=(d-2)/2$.\footnote{Here we exclude the identity operator $\1$ for which $\Delta_\1=0$.}  From the relation between quasi-primary operators in position space and in embedding space \cite{Fortin:2019dnq}, \eqref{Eq0PS} becomes
\eqna{
0&=\partial^2\mathcal{O}^{(x)\boldsymbol{0}}(x)=g^{\mu\nu}[iP_\mu,[iP_\nu,\mathcal{O}^{(x)\boldsymbol{0}}(x)]]\\
&=(\eta_+)^\tau g^{\mu\nu}[i(L_{\mu,d+1}+L_{\mu,d+2}),[i(L_{\nu,d+1}+L_{\nu,d+2}),\mathcal{O}^{\boldsymbol{0}}(\eta)]]\\
&=-(\eta_+)^\tau g^{\mu\nu}\mathcal{L}_{\mu+}\mathcal{L}_{\nu+}\mathcal{O}^{\boldsymbol{0}}(\eta)=-(\eta_+)^\tau g^{AB}\mathcal{L}_{A+}\mathcal{L}_{B+}\mathcal{O}^{\boldsymbol{0}}(\eta).
}
This translates into
\eqn{(-i\mathcal{L}_+)^2\mathcal{O}^{\boldsymbol{0}}(\eta)\equiv-g^{AB}\mathcal{L}_{A+}\mathcal{L}_{B+}\mathcal{O}^{\boldsymbol{0}}(\eta)=0,}[Eq0ES]
in embedding space.  It can be easily shown that \eqref{Eq0ES} gives the correct unitarity bound for scalars with the help of the explicit form of the two-point correlation functions \eqref{Eq2pt}, leading to
\eqn{(-i\mathcal{L}_{1+})^2\Vev{\mathcal{O}^{\boldsymbol{0}}(\eta_1)\mathcal{O}^{\boldsymbol{0}}(\eta_2)}=-g^{AB}\mathcal{L}_{1A+}\mathcal{L}_{1B+}\frac{\cOPE{}{\mathcal{O}}{\mathcal{O}}{\1}}{\ee{1}{2}{\tau}}=2\tau\left(\frac{d-2}{2}-\tau\right)\eta_{1+}\eta_{2+}\frac{\cOPE{}{\mathcal{O}}{\mathcal{O}}{\1}}{\ee{1}{2}{\tau+1}}=0,}
as expected.\footnote{Including the unitarity bound for the identity operator.}

To find the covariant form of the conserved current differential operator in embedding space, we use the first relation in \eqref{EqLD} to obtain
\eqn{(-i\mathcal{L}_{1+})^2=\frac{1}{\ee{1}{i}{}}(\eta_{1+})^2\D_{i1}^2-\frac{1}{\ee{1}{i}{}}\eta_{1+}[\ee{1}{i}{\frac{1}{2}}\D_{i1+}+\eta_{i+}\Theta_1](d-2+2\Theta_1).}[Eq0LD]
Equation \eqref{Eq0LD} clearly shows the conformally-covariant and non-covariant contributions to the conservation conditions.  It is trivial to see that the conformally-covariant differential operator $\D_{i1}^2$ (and for that matter $\D_{i1}^A$) annihilates the two-point correlation function for scalars, leaving only the non-covariant contribution times the saturated unitarity bound (after replacing $\Theta_1$ by $-\tau$).  Consequently, saturating the unitarity bound such that the two-point correlation function is conserved implies that the three-point correlation function conservation conditions for scalars can be derived from the conserved current differential operator $\D_{i1}^2$, which is covariant in embedding space.  As pointed out before, this is analogous to conservation in position space with $\partial^2\to\D_{i1}^2$, \textit{i.e.} with the building block of the position space OPE differential operator $\partial_\mu$ replaced by the building block of the embedding space OPE differential operator $\D_{i1A}$.

Turning to spinors $\mathcal{O}_\alpha^{(x)\boldsymbol{e}_r}(x)$, the conservation condition in position space is
\eqn{\gamma^\mu\partial_\mu\mathcal{O}^{(x)\boldsymbol{e}_r}(x)=0,}[EqerPS]
with equivalent equations for the two spinorial representations in even dimensions.  Equation \eqref{EqerPS} implies that conserved spinors are free fermions with ``twists'' $\tau=\tau^*=(d-2)/2$.  This can be rewritten in terms of quasi-primary operators in embedding space as
\eqna{
0&=C\gamma^\mu\partial_\mu\mathcal{O}^{(x)\boldsymbol{e}_r}(x)=C\gamma^\mu[iP_\mu,\mathcal{O}^{(x)\boldsymbol{e}_r}(x)]\\
&=-C\gamma^\mu[i(L_{\mu,d+1}+L_{\mu,d+2}),(\eta_+)^\tau\mathcal{O}_+^{\boldsymbol{e}_r}(\eta)]\\
&=i(\eta_+)^\tau C\gamma^\mu\mathcal{L}_{\mu+}\mathcal{O}_+^{\boldsymbol{e}_r}(\eta)=\frac{\alpha i}{2}(\eta_+)^\tau C_\Gamma\Gamma_+^{\phantom{+}A}\mathcal{L}_{A+}\mathcal{O}^{\boldsymbol{e}_r}(\eta),
}
where the parameter $\alpha$ and the matrices $C$ and $C_\Gamma$ are defined in \cite{Fortin:2019dnq}.  In other words, the conservation equation for spinors in embedding space can be expressed as
\eqn{\Gamma_+^{\phantom{+}A}(-i\mathcal{L}_{A+})\mathcal{O}^{\boldsymbol{e}_r}(\eta)=0,}[EqerES]
from which we obtain the appropriate unitarity bound
\eqn{\Gamma_+^{\phantom{+}A}(-i\mathcal{L}_{1A+})\Vev{\mathcal{O}^{\boldsymbol{e}_r}(\eta_1)\mathcal{O}^{\boldsymbol{e}_r^C}(\eta_2)}=-2\left(\frac{d-2}{2}-\tau\right)\eta_{1+}\Gamma_+\frac{\cOPE{}{\mathcal{O}}{\mathcal{O}}{\1}\eta_2\cdot\Gamma C_\Gamma^{-1}}{\ee{1}{2}{\tau+1}}=0,}
when acting on two-point correlation functions \eqref{Eq2pt} for spinors.

Again, the conformally-covariant conserved current differential operator in embedding space is computed with the help of \eqref{EqLD}, leading to
\eqna{
\Gamma_+^{\phantom{+}A}(-i\mathcal{L}_{1A+})&=-\frac{1}{2\ee{1}{i}{\frac{3}{2}}}\eta_{1+}\Gamma_+\eta_1\cdot\Gamma\eta_i\cdot\Gamma\Gamma^A\D_{i1A}\\
&\phantom{=}\qquad+\frac{1}{2\ee{1}{i}{\frac{3}{2}}}\Gamma_+\left[\eta_{1+}\eta_i\cdot\Gamma\Gamma^A\D_{i1A}+2\ee{1}{i}{}\D_{i1+}+2\ee{1}{i}{\frac{1}{2}}\eta_{i+}\Theta_1\right]\eta_1\cdot\Gamma\\
&\phantom{=}\qquad-\frac{1}{2\ee{1}{i}{}}\eta_{1+}\Gamma_+\eta_i\cdot\Gamma(d-2+2\Theta_1).
}[EqerLD]
Here, the first line of \eqref{EqerLD} is related to the conformally-covariant differential operator; the second line, which is not conformally covariant, vanishes by transversality when acting on the conserved spinor; and the last line implies the saturation of the unitarity bound (after replacing $\Theta_1$ by $-\tau$).  Indeed, the conformally-covariant differential operator $\eta_1\cdot\Gamma\,\Gamma^A\D_{i1A}\,\eta_i\cdot\Gamma$ annihilates the two-point correlation function for spinors, forcing the ``twist'' to be $\tau=\tau^*=(d-2)/2$ for the spinor to be conserved.  As a consequence, the spinor conservation conditions for three-point correlation functions can be obtained from $\eta_1\cdot\Gamma\,\Gamma^A\D_{i1A}\,\eta_i\cdot\Gamma$, which is the natural choice in embedding space, analogous to the position space conserved current differential operator $\gamma^\mu\partial_\mu$.\footnote{The extra $\eta_{1+}\Gamma_+$ factor in $\eta_{1+}\Gamma_+\eta_1\cdot\Gamma\eta_i\cdot\Gamma\Gamma^A\D_{i1A}$ (which is not conformally covariant) appearing in the spinor conservation conditions \eqref{EqerLD} can be easily removed with the help of left multiplication by $(\eta_{1+})^{-2}\eta_1\cdot\Gamma$.  This procedure demonstrates that $\eta_1\cdot\Gamma\,\Gamma^A\D_{i1A}\,\eta_i\cdot\Gamma$ leads to the same conservation conditions as the first line of \eqref{EqerLD}.}

We therefore conclude that for conserved scalars and spinors in embedding space, conservation conditions for three-point correlation functions are generated from the following conformally-covariant equations,
\eqn{\frac{1}{\ee{1}{i}{}}\D_{i1}^2\Vev{\mathcal{O}^{\boldsymbol{0}}(\eta_1)\Op{j}{2}\Op{m}{3}}=0,\qquad\frac{\eta_1\cdot\Gamma\,\Gamma^A\D_{i1A}\,\eta_i\cdot\Gamma}{\ee{1}{i}{\frac{3}{2}}}\Vev{\mathcal{O}^{\boldsymbol{e}_r}(\eta_1)\Op{j}{2}\Op{m}{3}}=0,}[EqU0er]
once the unitarity bound has been saturated.  Clearly, the free embedding coordinate $\eta_i$ can be chosen to be $\eta_2$ or $\eta_3$ in \eqref{EqU0er}.  Finally, it is important to stress again that conserved scalars and spinors are special since they correspond to free quasi-primary operators.  Therefore they decouple from the theory and all three-point correlation correlations involving conserved scalars and/or spinors are somehow trivial.


\subsection{Conserved Quasi-Primary Operators in General Irreducible Representations}

We are now ready to investigate the conservation conditions for quasi-primary operators in general irreducible representations $\boldsymbol{N}=\sum_{\mathfrak{n}\leq i\leq r}N_i\boldsymbol{e}_i$, excluding scalars and spinors in the defining representations.  As pointed out before, $\mathfrak{n}$ corresponds more precisely to the number of indices in the smallest antisymmetric group of indices in $\boldsymbol{N}$, with $\mathfrak{n}\geq1$ since scalars and spinors are not considered.

In position space, where the conservation conditions are
\eqn{C\gamma^{\mu_1\cdots\mu_\mathfrak{n}}\partial_{\mu_1}*\mathcal{O}^{(x)\boldsymbol{N}}(x)=0,}[EqNPS]
with the $*$-product representing full spinor contraction,\footnote{Obviously, the spinor contraction must be performed with two spinor indices on the quasi-primary operator associated with vector indices in one $\boldsymbol{e}_\mathfrak{n}$ (and its generalization to even dimensions).} conserved currents must saturate the unitarity bound \eqref{EqU}, \textit{i.e.}\ $\tau=\tau^*=d-1-\mathfrak{n}$.  In embedding space, the conservation conditions become
\eqna{
0&=C\gamma^{\mu_1\cdots\mu_\mathfrak{n}}\partial_{\mu_1}*\mathcal{O}^{(x)\boldsymbol{N}}(x)=C\gamma^{\mu_1\cdots\mu_\mathfrak{n}}*[iP_{\mu_1},\mathcal{O}^{(x)\boldsymbol{N}}(x)]\\
&=-C\gamma^{\mu_1\cdots\mu_\mathfrak{n}}*[i(L_{\mu_1,d+1}+L_{\mu_1,d+2}),(\eta_+)^\tau\mathcal{O}_+^{\boldsymbol{N}}(\eta)]\\
&=i(\eta_+)^\tau C\gamma^{\mu_1\cdots\mu_\mathfrak{n}}\mathcal{L}_{\mu_1+}*\mathcal{O}_+^{\boldsymbol{N}}(\eta)=\frac{\alpha i}{2}(\eta_+)^\tau\mathcal{L}_{A_1+}\left[C_\Gamma\Gamma_+^{\phantom{+}A_1\mu_2\cdots\mu_\mathfrak{n}}*\mathcal{O}^{\boldsymbol{N}}(\eta)\right]_+,
}
or equivalently
\eqn{C_\Gamma\Gamma_+^{\phantom{+}A_1\cdots A_\mathfrak{n}}(-i\mathcal{L}_{A_1+})*\mathcal{O}^{\boldsymbol{N}}(\eta)=0.}[EqNES]
We note here that \eqref{EqNES} is closer to the conformally-covariant version of the conservation conditions since the indices $\mu_{i\geq2}$ were replaced by $A_{i\geq2}$ and the uncontracted position space spinor indices were extended to uncontracted embedding space spinor indices.  These changes can be implemented by performing suitable conformal transformations, contrary to the remaining $+$ indices in \eqref{EqNES}.

As before, we obtain the conformally-covariant conserved current differential operator for quasi-primary operators in general irreducible representations from \eqref{EqNES} with the help of \eqref{EqLD}, leading to
\eqna{
C_\Gamma\Gamma_+^{\phantom{+}A_1\cdots A_\mathfrak{n}}(-i\mathcal{L}_{1A_1+})&=-\frac{\mathfrak{n}g_{+A_0}}{\ee{1}{i}{\frac{3}{2}}}\eta_{1+}\eta_1^{[A_0}\D_{i1A_1}\eta_{iB_0}\A_{1iB_1}^{\phantom{i1B_1}|A_1|}\cdots\A_{1iB_\mathfrak{n}}^{\phantom{i1B_\mathfrak{n}}A_\mathfrak{n}]}C_\Gamma\Gamma^{B_0\cdots B_\mathfrak{n}}\\
&\phantom{=}\qquad+\mathcal{Z}^{A_2\cdots A_\mathfrak{n}}-\frac{1}{\ee{1}{i}{}}\eta_{1+}\eta_{iA_1}C_\Gamma\Gamma_+^{\phantom{+}A_1\cdots A_\mathfrak{n}}(d-1-\mathfrak{n}+\Theta_1),
}[EqNLD]
where
\eqna{
\mathcal{Z}^{A_2\cdots A_\mathfrak{n}}&=\frac{1}{\ee{1}{i}{}}\left[\ee{1}{i}{\frac{1}{2}}\D_{i1+}+\eta_{i+}(\Theta_1-1)\right]\eta_{1A_1}C_\Gamma\Gamma_+^{\phantom{+}A_1\cdots A_\mathfrak{n}}+\frac{2\eta_{1+}\eta_{i+}}{\ee{1}{i}{2}}\eta_{iA_0}\eta_{1A_1}C_\Gamma\Gamma^{A_0\cdots A_\mathfrak{n}}\\
&\phantom{=}\qquad-\frac{1}{\ee{1}{i}{2}}\sum_{2\leq a\leq\mathfrak{n}}(-1)^a(\eta_{i+}\eta_1^{A_a}+\eta_{1+}\eta_i^{A_a})\eta_{iA_0}\eta_{1A_1}C_\Gamma\Gamma_+^{\phantom{+}A_0\cdots\widehat{A_a}\cdots A_\mathfrak{n}}\\
&\phantom{=}\qquad-\frac{1}{\ee{1}{i}{\frac{3}{2}}}\D_{i1A_1}\eta_{1+}\eta_{iB_0}\left[C_\Gamma\Gamma_+^{\phantom{+}A_1\cdots A_\mathfrak{n}B_0}\eta_1\cdot\Gamma-\eta_{1B_1}C_\Gamma\Gamma_+^{\phantom{+}A_1\cdots A_\mathfrak{n}B_0B_1}\right]\\
&\phantom{=}\qquad-\frac{\eta_{1+}\eta_{i+}}{\ee{1}{i}{\frac{5}{2}}}\sum_{2\leq a\leq\mathfrak{n}}(-1)^a\eta_1^{A_a}\D_{i1A_1}\eta_{iB_0}\eta_{1B_1}C_\Gamma\Gamma^{B_0B_1A_1\cdots\widehat{A_a}\cdots A_\mathfrak{n}}\\
&\phantom{=}\qquad+\frac{(\mathfrak{n}-1)g_{+A_0}}{\ee{1}{i}{\frac{5}{2}}}\eta_{1+}\sum_{2\leq a\leq\mathfrak{n}}(-1)^a\eta_1^{[A_0}\D_{i1A_1}\eta_{iB_0}\eta_{1B_1}\eta_i^{|A_a}C_\Gamma\Gamma^{B_0B_1A_1|A_2\cdots\widehat{A_a}\cdots A_\mathfrak{n}]}.
}
From \eqref{EqNLD}, we observe that a pattern similar to the previous cases with scalars and spinors emerges.  Indeed, the first term, which can be made conformally covariant in embedding space (see below), annihilates two-point correlation functions (see Section \ref{SecCF}).  The second term, denoted by $\mathcal{Z}^{A_2\cdots A_\mathfrak{n}}$, can be discarded altogether since it vanishes when acting on the conserved current through a combination of the transversality condition and group theory arguments (spinor contractions of products of two antisymmetric matrices vanish when the numbers of antisymmetric indices do not match).  Finally, the third term does not annihilate two-point correlation functions and thus forces the saturation of the unitarity bound \eqref{EqU} (after replacing $\Theta_1$ by $-\tau$).

As a result, after imposing the unitarity bound, only the first contribution to the differential operator \eqref{EqNLD} is non-trivial, thus conservation conditions on correlation functions can be derived from
\eqn{g_+^{\phantom{+}A_0}\eta_{1[A_0}\D_{i1}^{A_1}(\mathcal{T}_{i1\boldsymbol{e}_\mathfrak{n}}\Gamma)_{|A_1|A_2\cdots A_\mathfrak{n}]}*\Vev{\mathcal{O}^{\boldsymbol{N}}(\eta_1)\Op{j}{2}\Op{m}{3}}=0,}[EqUNp1]
when concentrating on three-point correlation functions, which gives
\eqn{g_+^{\phantom{+}A_0}\D_{i1}^{A_1}(\mathcal{T}_{i1\boldsymbol{e}_\mathfrak{n}}\Gamma)_{A_1A_0A_3\cdots A_\mathfrak{n}}*\Vev{\mathcal{O}^{\boldsymbol{N}}(\eta_1)\Op{j}{2}\Op{m}{3}}=0,}[EqUNp2]
after contracting with $\eta_i^{A_2}$.  Combining \eqref{EqUNp1} and \eqref{EqUNp2} finally implies that the three-point conservation conditions can be obtained from
\eqn{\D_{i1}^{A_1}(\mathcal{T}_{i1\boldsymbol{e}_\mathfrak{n}}\Gamma)_{A_1\cdots A_\mathfrak{n}}*\Vev{\mathcal{O}^{\boldsymbol{N}}(\eta_1)\Op{j}{2}\Op{m}{3}}=0,}[EqUN]
which is conformally covariant, as desired.  We note that $\eta_i$ is arbitrary in \eqref{EqUN}, therefore it can be chosen to be $\eta_2$ or $\eta_3$ since the conservation conditions do not depend on that particular choice.  Moreover, the form $\D_{i1}^{A_1}(\mathcal{T}_{i1\boldsymbol{e}_\mathfrak{n}}\Gamma)_{A_1\cdots A_\mathfrak{n}}*$ is the expected embedding space analog of the conserved current differential operator in position space $\partial^{\mu_1}(\mathcal{T}_{\boldsymbol{e}_\mathfrak{n}})_{\mu_1\cdots\mu_\mathfrak{n}}*$.


\subsection{Summary}

Before proceeding, we summarize the findings of this section.  The conserved current differential operators in position space can be uplifted to the embedding space.  The final results can be separated in three distinct parts: one contribution that can be made conformally covariant and that annihilates two-point correlation functions; one contribution that can be discarded since it annihilates all correlation functions; and one contribution that is not conformally covariant and that vanishes only when the unitarity bound is saturated.

Hence, once the unitarity bound is saturated to implement conservation of two-point correlation functions---with ``twists'' $\tau^*=(d-2)/2$ for scalars and spinors and $\tau=\tau^*=d-1-\mathfrak{n}$ for quasi-primary operators in general irreducible representations $\boldsymbol{N}$---the conformally-covariant conserved current differential operators in embedding space are directly derived from their position space counterparts as
\eqna{
\boldsymbol{0}&:\quad\tau^*=\frac{d-2}{2}\text{ with }\partial_1^2\to\D_{i1}^{\boldsymbol{0}}=\frac{1}{\ee{1}{i}{}}\D_{i1}^2,\\
\boldsymbol{e}_r&:\quad\tau^*=\frac{d-2}{2}\text{ with }\gamma^\mu\partial_{1\mu}\to\D_{i1}^{\boldsymbol{e}_r}=\frac{\eta_1\cdot\Gamma\,\Gamma^A\D_{i1A}\,\eta_i\cdot\Gamma}{\ee{1}{i}{\frac{3}{2}}},\\
\boldsymbol{N}&:\quad\tau^*=d-1-\mathfrak{n}\text{ with }(\mathcal{T}^{\boldsymbol{e}_{\mathfrak{n}-1}})^{\mu_2\cdots\mu_\mathfrak{n}}\partial_1^{\mu_1}(\mathcal{T}_{\boldsymbol{e}_\mathfrak{n}})_{\mu_\mathfrak{n}\cdots\mu_1}*\\
&\qquad\qquad\qquad\qquad\qquad\qquad\to\D_{i1}^{\boldsymbol{e}_\mathfrak{n}}=\frac{1}{\ee{1}{i}{\frac{1}{2}}}(\mathcal{T}_{1i}^{\boldsymbol{e}_{\mathfrak{n}-1}}\Gamma)^{A_2\cdots A_\mathfrak{n}}\D_{i1}^{A_1}(\mathcal{T}_{i1\boldsymbol{e}_\mathfrak{n}}\Gamma)_{A_\mathfrak{n}\cdots A_1}*.
}[EqUAll]
Their action on three-point correlation functions then leads to the complete set of conservation conditions.  We note that the conformally-covariant conserved current differential operators in \eqref{EqUAll} are scaled to have vanishing degree of homogeneity in $\eta_i$ and $-1$ in $\eta_1$, in analogy with their position space counterparts.  This is in agreement with the relation between the conformally-covariant conserved current differential operators in embedding space \eqref{EqUAll} and the equivalent objects in position space: in both cases, the conserved current differential operators are constructed from the appropriate OPE differential operator building block.

Moreover, we observe that for the two special cases corresponding to conserved scalars and spinors, the conserved current differential operators do not change the irreducible representations of the conserved quasi-primary operators.  In other words, conservation conditions for scalars and spinors involve scalars and spinors, respectively.  This is not the case for conserved quasi-primary operators in general irreducible representations.  Indeed, following \eqref{EqUAll} conservation conditions for a quasi-primary operator in the irreducible representation $\boldsymbol{N}$ involve a quasi-primary operator in the irreducible representation $\overline{\boldsymbol{N}}=\boldsymbol{N}-\boldsymbol{e}_\mathfrak{n}+\boldsymbol{e}_{\mathfrak{n}-1}$ with $\boldsymbol{e}_0\equiv\boldsymbol{0}$.  We conclude that there are as many conservation conditions as there are tensor structures for three-point correlation functions involving one conserved scalar or spinor, while the number of conservation conditions is not related to the number of tensor structures for three-point correlation functions with one conserved current in a non-trivial irreducible representation.


\section{Conserved Currents and Correlation Functions}\label{SecCF}

In this section we investigate conservation conditions for two- and three-point correlation functions from the conformally-covariant conserved current differential operators found in Section \ref{SecUC}.  Before proceeding, we first discuss one important ingredient, the Fock conditions.  Fock conditions are useful in rewriting seemingly different quantities in the same manner, allowing for an easy treatment of the conservation conditions.  Then we derive some identities from which it is straightforward to verify that the conformally-covariant differential operators \eqref{EqUAll} annihilate two-point correlation functions.  From the previous identities, we also compute the full action of the conserved current differential operators on three-point correlation functions with conserved quasi-primary operators to determine the conformal Ward identities.


\subsection{Fock Conditions}

Fock conditions are usually discussed for $SU(d)$ groups, but they generalize quite simply to $SO(d)$ groups \cite{Fortin}.  They state that quasi-primary operators in irreducible representation $\boldsymbol{N}=\sum_{1\leq i\leq r}N_i\boldsymbol{e}_i$ can be written as sums with permuted indices.

More precisely, the sum permutes a fixed index associated to one antisymmetric group of indices in $\boldsymbol{e}_a$ with all possible indices associated to one antisymmetric group of indices in $\boldsymbol{e}_{b\geq a}$.  For example, for $2\boldsymbol{e}_1+2\boldsymbol{e}_2$ with the two groups of $\boldsymbol{e}_1$ indices given by $\{A_1\}$ and $\{A_2\}$ and the two groups of $\boldsymbol{e}_2$ indices given by $\{B_1,B_2\}$ and $\{B_3,B_4\}$, the Fock conditions in terms of the half-projectors are
\eqna{
(\mathcal{T}_{ij}^{\boldsymbol{N}}\Gamma)^{A_1A_2B_1B_2B_3B_4}&=(\mathcal{T}_{ij}^{\boldsymbol{N}}\Gamma)^{A_2A_1B_1B_2B_3B_4}\\
&=(\mathcal{T}_{ij}^{\boldsymbol{N}}\Gamma)^{B_1A_2A_1B_2B_3B_4}+(\mathcal{T}_{ij}^{\boldsymbol{N}}\Gamma)^{B_2A_2B_1A_1B_3B_4}\\
&=(\mathcal{T}_{ij}^{\boldsymbol{N}}\Gamma)^{B_3A_2B_1B_2A_1B_4}+(\mathcal{T}_{ij}^{\boldsymbol{N}}\Gamma)^{B_4A_2B_1B_2B_3A_1},
}
when choosing $A_1$ as the fixed index and
\eqn{(\mathcal{T}_{ij}^{\boldsymbol{N}}\Gamma)^{A_1A_2B_1B_2B_3B_4}=(\mathcal{T}_{ij}^{\boldsymbol{N}}\Gamma)^{A_1A_2B_3B_2B_1B_4}+(\mathcal{T}_{ij}^{\boldsymbol{N}}\Gamma)^{A_1A_2B_4B_2B_3B_1},}
when choosing $B_1$ as the fixed index.  The Fock conditions are also expressed in a similar way for projection operators $\hat{\mathcal{P}}^{\boldsymbol{N}}$.

Clearly, the Fock conditions are very important when conservation conditions are analysed since they help in writing the latter in terms of linearly-independent contributions.  We point out here that the knowledge of the actual projection operator for $\boldsymbol{N}$ is not always necessary to determine the conservation conditions, the Fock conditions may suffice.  In general, the explicit projection operator for the conserved quasi-primary operator in the irreducible representation $\boldsymbol{N}$ is necessary while the Fock conditions for its descendant in the irreducible representation $\overline{\boldsymbol{N}}$ and the two unconserved quasi-primary operators are sufficient.


\subsection{Identities}

To compute the conservation conditions in all generality, it is convenient to introduce a set of identities.  Most of the identities included here can be found in \cite{Fortin:2019dnq}, the remaining ones are easily derived from the original ones.  For completeness, we present the most important identities in the context of conserved currents.  We first focus on the basic quantities: the OPE differential operator, the half-projectors, and the projection operators.

The embedding space OPE differential operator \eqref{EqDiff} satisfies the following identities
\eqn{
\begin{gathered}
\D_{ij}^A\eta_j^B=\ee{i}{j}{\frac{1}{2}}\A_{ij}^{AB},\qquad\D_{ij}^A\A_{ij}^{BC}=-\frac{2}{\ee{i}{j}{\frac{1}{2}}}\A_{ij}^{A(B}\eta_i^{C)},\\
\D_{ij}^A\frac{1}{\ee{j}{k}{p}}=-p\frac{\ee{i}{j}{\frac{1}{2}}}{\ee{j}{k}{p+1}}(\A_{ij}\cdot\eta_k)^A,\qquad\D_{ijA}\A_{jk}^{BC}=-\frac{2\ee{i}{j}{\frac{1}{2}}}{\ee{j}{k}{}}\A_{ijAD}\A_{jk}^{D(B}\eta_k^{C)},
\end{gathered}
}[EqIdD]
while the half-projectors and the projection operators verify
\eqn{
\begin{gathered}
(\mathcal{T}_{ij}^{\boldsymbol{N}}\Gamma)=\frac{\ee{i}{k}{\frac{1}{2}(S-\xi)}}{\ee{i}{j}{\frac{1}{2}(S-\xi)}}(\mathcal{T}_{ik}^{\boldsymbol{N}}\Gamma)\cdot\left(\frac{(\eta_i\cdot\Gamma)^{2\xi}\hat{\mathcal{P}}_{ij}^{\boldsymbol{N}}(\eta_j\cdot\Gamma)^{2\xi}}{(2\eta_i\cdot\eta_j)^{2\xi}}\right),\\
(\eta_j\cdot\Gamma)^{2\xi}\hat{\mathcal{P}}_{ij}^{\boldsymbol{N}}=(\eta_j\cdot\Gamma)^{2\xi}(\A_{ij})^{n_v}\hat{\mathcal{P}}_{kj}^{\boldsymbol{N}}(\A_{ij})^{n_v}.
\end{gathered}
}[EqIdTP]
Identities \eqref{EqIdD} and \eqref{EqIdTP} can be found in \cite{Fortin:2019dnq}.

Important identities useful when applying the conformally-covariant conserved current differential operators \eqref{EqUAll} on correlation functions are
\eqn{(\mathcal{T}_{ji\boldsymbol{e}_\mathfrak{n}}\Gamma)_{\{A'\}}*(\mathcal{T}_{ij}^{\boldsymbol{N}}\Gamma)^{\{Aa\}}=(\mathcal{T}_{ij}^{\boldsymbol{N}-\boldsymbol{e}_\mathfrak{n}}\Gamma)^{\{A'a'\}}(\hat{\mathcal{P}}_{ij}^{\boldsymbol{N}})_{\{a'A'\}}^{\phantom{\{a'A'\}}\{Aa\}},}[EqIdTT]
and
\eqna{
&\D_{ji}^{A'_1}(\mathcal{T}_{ji\boldsymbol{e}_\mathfrak{n}}\Gamma)_{A'_\mathfrak{n}\cdots A'_1}*(\mathcal{T}_{ij}^{\boldsymbol{N}}\Gamma)^{\{Aa\}}\\
&\qquad=\D_{ji}^{A'_1}(\mathcal{T}_{ij}^{\boldsymbol{N}-\boldsymbol{e}_\mathfrak{n}}\Gamma)^{\{A'a'\}}(\hat{\mathcal{P}}_{ij}^{\boldsymbol{N}})_{\{a'A'\}}^{\phantom{\{a'A'\}}\{Aa\}}\\
&\qquad=(\mathcal{T}_{ij}^{\boldsymbol{N}-\boldsymbol{e}_\mathfrak{n}}\Gamma)^{\{A'a'\}}\D_{ji}^{A'_1}(\hat{\mathcal{P}}_{ij}^{\boldsymbol{N}})_{\{a'A'\}}^{\phantom{\{a'A'\}}\{Aa\}}\\
&\qquad=-\frac{1}{\ee{i}{j}{\frac{1}{2}}}(\mathcal{T}_{ij}^{\boldsymbol{N}-\boldsymbol{e}_\mathfrak{n}}\Gamma)^{\{A'a'\}}\sum_{1\leq b\leq n_v}(\hat{\mathcal{P}}_{ij}^{\boldsymbol{N}})_{\{a'A'\}}^{\phantom{\{a'A'\}}A_1\cdots A_{b-1}A'_1A_{b+1}\cdots A_{n_v}a}\eta_j^{A_b}.
}[EqIdDTT]
Equation \eqref{EqIdTT} can be derived from the identities found in \cite{Fortin:2019dnq} and some group theory arguments while \eqref{EqIdDTT} has been simplified with the help of the identities \eqref{EqIdD}, \eqref{EqIdTP} and \eqref{EqIdTT}, the Fock conditions and the tracelessness condition.  On the RHS of \eqref{EqIdTT}, index contraction is performed between all the indices of the half-projector and the associated indices of the projection operator---the projection operator thus has some of its primed indices, the $e_\mathfrak{n}$-indices, uncontracted.

With \eqref{EqUAll} in mind and the identity \eqref{EqIdDTT}, we also have
\eqna{
&(\mathcal{T}_{ij}^{\boldsymbol{e}_{\mathfrak{n}-1}}\Gamma)^{A'_2\cdots A'_\mathfrak{n}}(\mathcal{T}_{ij}^{\boldsymbol{N}-\boldsymbol{e}_\mathfrak{n}}\Gamma)^{\{A'a'\}}(\hat{\mathcal{P}}_{ij}^{\boldsymbol{N}})_{\{a'A'\}}^{\phantom{\{a'A'\}}\{Aa\}}\\
&\qquad=(\mathcal{T}_{ij}^{\boldsymbol{N}-\boldsymbol{e}_\mathfrak{n}+\boldsymbol{e}_{\mathfrak{n}-1}}\Gamma)^{\{A'a'\}}(\hat{\mathcal{P}}_{ij}^{\boldsymbol{N}})_{\{a'A'\}}^{\phantom{\{a'A'\}}\{Aa\}}=(\mathcal{T}_{ij}^{\overline{\boldsymbol{N}}}\Gamma)^{\{A'a'\}}(\hat{\mathcal{P}}_{ij}^{\boldsymbol{N}})_{\{a'A'\}}^{\phantom{\{a'A'\}}\{Aa\}},
}[EqIdTTP]
with missing $A'_1$ embedding space index on the half-projector and free $A'_1$ embedding space index on the projection operator.  Relying on group theory, \eqref{EqIdTTP} is true in part due to the definition of $\mathfrak{n}$.  It would not necessarily be correct when another Dynkin index is extracted.  From \eqref{EqIdTTP}, it is clear that the initial half-projector for the conserved quasi-primary operator $\mathcal{O}^{(x)\boldsymbol{N}}(x)$ in the irreducible representation $\boldsymbol{N}$ is modified to the half-projector for the new quasi-primary operator $\partial\cdot\mathcal{O}^{(x)\boldsymbol{N}}(x)$ in the irreducible representation $\overline{\boldsymbol{N}}$.

The last identity of interest for half-projectors is given by
\eqna{
\D_{ji}^{A'}(\mathcal{T}_{ki}^{\boldsymbol{N}}\Gamma)^{\{Bb\}}&=-\frac{S+3\xi}{2}\frac{\ee{i}{j}{\frac{1}{2}}}{\ee{i}{k}{}}(\A_{ij}\cdot\eta_k)^{A'}(\mathcal{T}_{ki}^{\boldsymbol{N}}\Gamma)^{\{Bb\}}\\
&\phantom{=}\qquad-\frac{\ee{i}{j}{\frac{1}{2}}}{\ee{i}{k}{}}(\mathcal{T}_{ki}^{\boldsymbol{N}}\Gamma)^{\{B'b'\}}\sum_{1\leq c\leq n_v}(\hat{\mathcal{P}}_{ki}^{\boldsymbol{N}})_{\{b'B'\}}^{\phantom{\{b'B'\}}B_1\cdots B_{c-1}CB_{c+1}\cdots B_{n_v}b}\A_{ijC}^{\phantom{ijB}A'}\eta_k^{B_c}\\
&\phantom{=}\qquad+\xi\frac{\ee{i}{j}{\frac{1}{2}}}{\ee{i}{k}{}}(\mathcal{T}_{ki}^{\boldsymbol{N}}\Gamma)^{\{Bb'\}}(\eta_k\cdot\Gamma\Gamma_{ij}^{A'})_{b'}^{\phantom{b'}b},
}[EqIdDT]
which can be proven by using \eqref{EqIdTP} and \eqref{EqIdD}.


\subsection{Two-Point Correlation Functions Revisited}

We are now ready to prove that the conformally-covariant conserved current differential operators \eqref{EqUAll} annihilate two-point correlation functions.  We already completed the proof for the special cases of conserved scalars and fermions in Section \ref{SecUC}, we therefore proceed with conserved quasi-primary operators in general irreducible representations $\boldsymbol{N}$.

From the two-point correlation functions \eqref{Eq2pt}, the action of the conformally-covariant conserved current differential operators is
\eqna{
&\D_{21}^{\boldsymbol{e}_\mathfrak{n}}*\Vev{\mathcal{O}^{\boldsymbol{N}}(\eta_1)\mathcal{O}^{\boldsymbol{N}^C}(\eta_2)}\\
&\qquad=(\mathcal{T}_{12}^{\boldsymbol{e}_{\mathfrak{n}-1}}\Gamma)^{A'_2\cdots A'_\mathfrak{n}}\D_{21}^{A'_1}(\mathcal{T}_{21\boldsymbol{e}_\mathfrak{n}}\Gamma)_{A'_\mathfrak{n}\cdots A'_1}*(\mathcal{T}_{12}^{\boldsymbol{N}}\Gamma)\cdot(\mathcal{T}_{21}^{\boldsymbol{N}^C}\Gamma)\frac{\cOPE{}{\mathcal{O}}{\mathcal{O}}{\1}}{\ee{1}{2}{\tau+1/2}}\\
&\qquad=-(\mathcal{T}_{12}^{\overline{\boldsymbol{N}}}\Gamma)^{\{A'a'\}}\left[\sum_{1\leq b\leq n_v}(\hat{\mathcal{P}}_{12}^{\boldsymbol{N}})_{\{a'A'\}}^{\phantom{\{a'A'\}}A_1\cdots A_{b-1}A'_1A_{b+1}\cdots A_{n_v}a}\eta_2^{A_b}\right.\\
&\qquad\phantom{=}+(\hat{\mathcal{P}}_{12}^{\boldsymbol{N}})_{\{a'A'\}}^{\phantom{\{a'A'\}}\{B'b'\}}\sum_{1\leq c\leq n_v}\eta_{2B'_c}g^{A'_1C}(\hat{\mathcal{P}}_{12}^{\boldsymbol{N}})_{b'B'_{n_v}\cdots B'_{c+1}CB'_{c+1}\cdots B'_1}^{\phantom{b'B'_{n_v}\cdots B'_{c+1}CB'_{c+1}\cdots B'_1}\{Aa\}}\\
&\qquad\phantom{=}\left.-\xi(\hat{\mathcal{P}}_{12}^{\boldsymbol{N}})_{\{a'A'\}}^{\phantom{\{a'A'\}}\{Ab'\}}(\eta_2\cdot\Gamma\Gamma_{12}^{A'_1})_{b'}^{\phantom{b'}a}\right](\mathcal{T}_{21}^{\boldsymbol{N}^C}\Gamma)_{\{aA\}}\frac{\cOPE{}{\mathcal{O}}{\mathcal{O}}{\1}}{\ee{1}{2}{\tau+1}}=0,
}[EqD2pt]
following \eqref{EqUAll} and using \eqref{EqIdDTT}, \eqref{EqIdTTP}, \eqref{EqIdTT}, and \eqref{EqIdDT}.  The last equality comes from the transversality condition, exemplified here by the vanishing contractions of $\eta_2^{A_b}$ in the first sum, of $\eta_{2B'_c}$ in the second sum, and $\eta_2\cdot\Gamma$ in the last term.  We note also that \eqref{EqD2pt} is a two-point correlation function of two different quasi-primary operators with conformal dimensions and irreducible representations $(\Delta+1,\overline{\boldsymbol{N}})$ and $(\Delta,\boldsymbol{N})$, respectively, therefore it had to vanish.

Hence, the conformally-covariant conserved current differential operators \eqref{EqUAll} annihilate two-point correlation functions, forcing the saturation of the unitarity bound for conserved currents following \eqref{EqNLD}.  In turn, the complete set of conservation conditions can be derived from the action of the conformally-covariant conserved current differential operators \eqref{EqUAll} on three-point correlation functions.


\subsection{Three-Point Correlation Functions}

We are finally in position to derive the complete set of conservation conditions by investigating the action of the conserved current differential operators \eqref{EqUAll} on three-point correlation functions.  With the results of the previous sections, it is actually straightforward to proceed in all generality.  We focus on conserved quasi-primary operators in general irreducible representations since the special cases of conserved scalars and spinors are of no interest.

For conserved quasi-primary operators $\Op{i}{1}$ in general irreducible representations $\boldsymbol{N}_i$, we have the conservation conditions
\eqna{
0&=\D_{21}^{\boldsymbol{e}_{\mathfrak{n}_i}}*\Vev{\Op{i}{1}\Op{j}{2}\Op{m}{3}}\\
&=\frac{(\mathcal{T}_{12}^{\overline{\boldsymbol{N}}_i}\Gamma)^{\{A'a'\}}(\mathcal{T}_{21}^{\boldsymbol{N}_j}\Gamma)^{\{Bb'\}}\left(\mathcal{T}_{31}^{\boldsymbol{N}_m}\Gamma(\bar{\eta}_3\cdot\Gamma)^{2\xi_m}\right)^{\{E'e'\}}}{\ee{1}{2}{\frac{1}{2}(\tau_i+1+\tau_j-\chi_m)}\ee{1}{3}{\frac{1}{2}(\chi_i+1-\chi_j+\tau_m)}\ee{2}{3}{\frac{1}{2}(-\chi_i-1+\chi_j+\chi_m)}}(\hat{\mathcal{P}}_{12}^{\boldsymbol{N}_i})_{\{a'A'\}}^{\phantom{\{a'A'\}}\{Aa\}}\\
&\phantom{=}\qquad\times\sum_{r=1}^{N_{ijm}}\aCF{r}{i}{j}{m}\left[-\frac{\chi_i-\chi_j+\chi_m+6\xi_m}{2}\delta_{b'}^{\phantom{b'}b}\bar{\eta}_3^{A'_1}(\hat{\mathcal{P}}_{31}^{\boldsymbol{N}_m})_{\{e'E'\}}^{\phantom{\{e'E'\}}\{Ee\}}\right.\\
&\phantom{=}\qquad-\sum_{1\leq s\leq n_v^m}\delta_{b'}^{\phantom{b'}b}(\hat{\mathcal{P}}_{31}^{\boldsymbol{N}_m})_{\{e'E'\}}^{\phantom{\{e'E'\}}E_1\cdots E_{s-1}A'_1E_{s+1}\cdots E_{n_v^m}e}\bar{\eta}_3^{E_s}+\xi_j(\bar{\eta}_2\cdot\Gamma\Gamma_{12}^{A'_1})_{b'}^{\phantom{b'}b}(\hat{\mathcal{P}}_{31}^{\boldsymbol{N}_m})_{\{e'E'\}}^{\phantom{\{e'E'\}}\{Ee\}}\\
&\phantom{=}\qquad\left.+\xi_m\delta_{b'}^{\phantom{b'}b}(\hat{\mathcal{P}}_{31}^{\boldsymbol{N}_m})_{\{e'E'\}}^{\phantom{\{e'E'\}}\{Ee''\}}(\Gamma_{12}^{A'_1}\bar{\eta}_3\cdot\Gamma)_{e''}^{\phantom{e''}e}+\delta_{b'}^{\phantom{b'}b}(\hat{\mathcal{P}}_{31}^{\boldsymbol{N}_m})_{\{e'E'\}}^{\phantom{\{e'E'\}}\{Ee\}}\frac{\ee{1}{3}{\frac{1}{2}}}{\ee{2}{3}{\frac{1}{2}}}\D_{21}^{A'_1}\right]\\
&\phantom{=}\qquad\times(\FCF{r}{i}{j}{m}{1}{2})_{\{aA\}\{bB\}\{eE\}},
}[Eq3pt1]
where the derivative acts on the tensor structure building blocks as
\eqn{
\begin{gathered}
\frac{\ee{1}{3}{\frac{1}{2}}}{\ee{2}{3}{\frac{1}{2}}}\D_{21}^{A'_1}\A_{12}^{F_1F_2}=-2\A_{12}^{A'_1(F_1}\bar{\eta}_2^{F_2)},\\
\frac{\ee{1}{3}{\frac{1}{2}}}{\ee{2}{3}{\frac{1}{2}}}\D_{21}^{A'_1}\Gamma_{12}^{F_1\cdots F_n}=-\sum_{1\leq s\leq n}\left[\Gamma_{12}^{F_1\cdots F_{s-1}A'_1F_{s+1}\cdots F_n}\bar{\eta}_2^{F_s}+(-1)^{s-1}\bar{\eta}_2\cdot\Gamma\Gamma_{12}^{F_1\cdots\widehat{F_s}\cdots F_n}\A_{12}^{A'_1F_s}\right],\\
\frac{\ee{1}{3}{\frac{1}{2}}}{\ee{2}{3}{\frac{1}{2}}}\D_{21}^{A'_1}\epsilon_{12}^{F_1\cdots F_d}=-\sum_{1\leq s\leq d}\epsilon_{12}^{F_1\cdots F_{s-1}A'_1F_{s+1}\cdots F_d}\bar{\eta}_2^{F_s},\\
\frac{\ee{1}{3}{\frac{1}{2}}}{\ee{2}{3}{\frac{1}{2}}}\D_{21}^{A'_1}(\A_{12}\cdot\bar{\eta}_3)^F=-\A_{12}^{A'_1F}-(\A_{12}\cdot\bar{\eta}_3)^{A'_1}\bar{\eta}_2^F-\frac{1}{2}(\A_{12}\cdot\bar{\eta}_3)^{A'_1}(\A_{12}\cdot\bar{\eta}_3)^F.
\end{gathered}
}[Eq3pt1subs]

Similarly, the conservation conditions for conserved quasi-primary operators $\Op{j}{2}$ in general irreducible representations $\boldsymbol{N}_j$ are
\eqna{
0&=\D_{12}^{\boldsymbol{e}_{\mathfrak{n}_j}}*\Vev{\Op{i}{1}\Op{j}{2}\Op{m}{3}}\\
&=\frac{(\mathcal{T}_{12}^{\boldsymbol{N}_i}\Gamma)^{\{Aa'\}}(\mathcal{T}_{21}^{\overline{\boldsymbol{N}}_j}\Gamma)^{\{B'b'\}}\left(\mathcal{T}_{31}^{\boldsymbol{N}_m}\Gamma(\bar{\eta}_3\cdot\Gamma)^{2\xi_m}\right)^{\{Ee\}}}{\ee{1}{2}{\frac{1}{2}(\tau_i+\tau_j+1-\chi_m)}\ee{1}{3}{\frac{1}{2}(\chi_i-\chi_j-1+\tau_m)}\ee{2}{3}{\frac{1}{2}(-\chi_i+\chi_j+1+\chi_m)}}(\hat{\mathcal{P}}_{21}^{\boldsymbol{N}_j})_{\{b'B'\}}^{\phantom{\{b'B'\}}\{Bb\}}\\
&\phantom{=}\qquad\times\sum_{r=1}^{N_{ijm}}\aCF{r}{i}{j}{m}\left[-\frac{-\chi_i+\chi_j+\chi_m+2\xi_m}{2}\delta_{a'}^{\phantom{a'}a}\bar{\eta}_3^{B'_1}+\xi_i(\bar{\eta}_1\cdot\Gamma\Gamma_{12}^{B'_1})_{a'}^{\phantom{a'}a}+\delta_{a'}^{\phantom{a'}a}\frac{\ee{2}{3}{\frac{1}{2}}}{\ee{1}{3}{\frac{1}{2}}}\D_{12}^{B'_1}\right]\\
&\phantom{=}\qquad\times(\FCF{r}{i}{j}{m}{1}{2})_{\{aA\}\{bB\}\{eE\}},
}[Eq3pt2]
with the substitutions
\eqn{
\begin{gathered}
\frac{\ee{2}{3}{\frac{1}{2}}}{\ee{1}{3}{\frac{1}{2}}}\D_{12}^{B'_1}\A_{12}^{F_1F_2}\to0,\\
\frac{\ee{2}{3}{\frac{1}{2}}}{\ee{1}{3}{\frac{1}{2}}}\D_{12}^{B'_1}\Gamma_{12}^{F_1\cdots F_n}\to\sum_{1\leq s\leq n}(-1)^s\bar{\eta}_1\cdot\Gamma\Gamma_{12}^{F_1\cdots\widehat{F_s}\cdots F_n}\A_{12}^{B'_1F_s},\\
\frac{\ee{2}{3}{\frac{1}{2}}}{\ee{1}{3}{\frac{1}{2}}}\D_{12}^{B'_1}\epsilon_{12}^{F_1\cdots F_d}\to0,\\
\frac{\ee{2}{3}{\frac{1}{2}}}{\ee{1}{3}{\frac{1}{2}}}\D_{12}^{B'_1}(\A_{12}\cdot\bar{\eta}_3)^F\to-\A_{12}^{B'_1F}-\frac{1}{2}\bar{\eta}_3^{B'_1}(\A_{12}\cdot\bar{\eta}_3)^F.
\end{gathered}
}[Eq3pt2subs]
for the derivative on the tensor structures.

Finally, for conserved quasi-primary operators $\Op{m}{3}$ in general irreducible representations $\boldsymbol{N}_m$, we reach the conservation conditions
\eqna{
0&=\D_{13}^{\boldsymbol{e}_{\mathfrak{n}_m}}*\Vev{\Op{i}{1}\Op{j}{2}\Op{m}{3}}\\
&=\frac{(\mathcal{T}_{12}^{\boldsymbol{N}_i}\Gamma)^{\{Aa\}}(\mathcal{T}_{21}^{\boldsymbol{N}_j}\Gamma)^{\{Bb\}}\left(\mathcal{T}_{31}^{\overline{\boldsymbol{N}}_m}\Gamma(\bar{\eta}_3\cdot\Gamma)^{2\xi_m}\right)^{\{E'e'\}}}{\ee{1}{2}{\frac{1}{2}(\tau_i+\tau_j-\chi_m-1)}\ee{1}{3}{\frac{1}{2}(\chi_i-\chi_j+\tau_m+1)}\ee{2}{3}{\frac{1}{2}(-\chi_i+\chi_j+\chi_m+1)}}(\hat{\mathcal{P}}_{31}^{\boldsymbol{N}_m})_{\{e'E'\}}^{\phantom{\{e'E'\}}\{Ee\}}\\
&\phantom{=}\qquad\times\sum_{r=1}^{N_{ijm}}\aCF{r}{i}{j}{m}\left[\frac{-\chi_i+\chi_j+\chi_m+2\xi_m}{2}(\A_{12}\cdot\bar{\eta}_3)^{E'_1}+\frac{\ee{2}{3}{\frac{1}{2}}}{\ee{1}{2}{\frac{1}{2}}}\D_{13}^{E'_1}\right](\FCF{r}{i}{j}{m}{1}{2})_{\{aA\}\{bB\}\{eE\}},
}[Eq3pt3]
where the last derivative on the tensor structures can be implemented by substituting
\eqn{\frac{\ee{2}{3}{\frac{1}{2}}}{\ee{1}{2}{\frac{1}{2}}}\D_{13}^{E'_1}(\A_{12}\cdot\bar{\eta}_3)^F\to\A_{12}^{E'_1F}+\frac{1}{2}(\A_{12}\cdot\bar{\eta}_3)^{E'_1}(\A_{12}\cdot\bar{\eta}_3)^F,}[Eq3pt3subs]
with the other derivatives being zero.

Here, to obtain the conservation conditions \eqref{Eq3pt1}, \eqref{Eq3pt2}, and \eqref{Eq3pt3} as well as their respective substitutions \eqref{Eq3pt1subs}, \eqref{Eq3pt2subs}, and \eqref{Eq3pt3subs}, we used the three-point correlation functions \eqref{Eq3pt}, the identities of the previous sections, the tracelessness condition, and the transversality condition.

The conservation conditions above constrain three-point correlation functions by setting some linear combinations of three-point coefficients (one condition for each tensor structure in three-point correlation function in which the conserved operator is contracted with the derivative operator) to zero.  Since the linear combinations depend on the conformal dimensions the conservation conditions may alternatively impose relations among the conformal dimensions of the remaining quasi-primary operators.  In the first case, three-point correlation functions with conserved currents can be obtained from the associated three-point correlation functions without conserved currents by fixing some of the three-point coefficients such that the linear combinations vanish, effectively implementing the conservation conditions.  The second case occurs when a particular constraint depends on only one three-point coefficient, forcing either the coefficient or its pre-factor to zero.  In both cases, the conformal dimension of the conserved quasi-primary operator must be set to the unitarity bound \eqref{EqU}.

We observe also that the conservation conditions \eqref{Eq3pt1} correspond to three-point correlation functions of quasi-primary operators with conformal dimensions and irreducible representations $(\Delta_i+1,\overline{\boldsymbol{N}}_i)$, $(\Delta_j,\boldsymbol{N}_j)$, and $(\Delta_m,\boldsymbol{N}_m)$, respectively, for which there are $N_{\bar{\imath}jm}$ tensor structures.  Similarly, \eqref{Eq3pt2} includes $(\Delta_i,\boldsymbol{N}_i)$, $(\Delta_j+1,\overline{\boldsymbol{N}}_j)$, and $(\Delta_m,\boldsymbol{N}_m)$ with $N_{i\bar{\jmath}m}$ tensor structures while \eqref{Eq3pt3} contains $(\Delta_i,\boldsymbol{N}_i)$, $(\Delta_j,\boldsymbol{N}_j)$, and $(\Delta_m+1,\overline{\boldsymbol{N}}_m)$ with $N_{ij\bar{m}}$ tensor structures.  As a consequence, due to the possibility that the conservation conditions are linearly dependent, the maximum number of constraints coming from conservation conditions is $N_{\bar{\imath}jm}$ ($N_{i\bar{\jmath}m}$) [$N_{ij\bar{m}}$] when the first (second) [third] quasi-primary operator is conserved.  Therefore, there are no constraints from conservation for a three-point correlation function where the first (second) [third] quasi-primary operator is conserved if $N_{\bar{\imath}jm}=0$ ($N_{i\bar{\jmath}m}=0$) [$N_{ij\bar{m}}=0$].

Obviously, when two or more quasi-primary operators are conserved, the conservation conditions of each conserved quasi-primary operators must be considered simultaneously.  We note that the numbers of constraints do not necessarily add since some constraints may be linearly dependent.

Finally, tensor structures in three-point correlation functions with identical quasi-primary operators must be invariant under the associated exchange symmetry.  These cases can be dealt with in the same manner as before.  When more than one quasi-primary operator is in a given irreducible representation it is always possible to choose a basis of tensor structures that transform manifestly under the exchange symmetry.  In other words, in this new basis tensor structures lie in irreducible representations of the exchange symmetry\footnote{The exchange symmetry group is $\mathbb{Z}_2$ ($S_3$) when there are exactly two (three) quasi-primary operators in the same irreducible representation.  For $\mathbb{Z}_2$, there are two one-dimensional irreducible representations (called trivial and sign) while for $S_3$ there are two one-dimensional irreducible representations (also called trivial and sign) and one two-dimensional irreducible representation (called standard).} and the associated three-point coefficients must transform accordingly under the exchange symmetry to generate singlets.\footnote{We point out here that the exchange symmetry implies that embedding space coordinates are also exchanged.  Therefore the half-projectors in the three-point correlation functions \eqref{Eq3pt} may be modified under the exchange symmetry and they must be re-expressed in terms of the original half-projectors, using the identities in \cite{Fortin:2019dnq}, to determine the correct linear combinations of tensor structures with manifest behaviors under the exchange symmetry.}  Since the associated three-point coefficients are invariant under the appropriate exchange symmetry when the quasi-primary operators in a given irreducible representation are identical, only the tensor structures in the trivial irreducible representation are allowed for identical quasi-primary operators.


\section{Examples}\label{SecEx}

In this section, the formalism developed so far is used in the context of simple examples.  The examples involve conserved currents in different irreducible representations including symmetric-traceless, fermionic, and defining irreducible representations.  We also investigate all the relevant cases necessary for the conformal bootstrap of four conserved vector currents and four energy-momentum tensors.  For the latter examples, most of the results are left for Appendix \ref{SAppJEM}.  For notational simplicity, three-point correlation functions are denoted by the irreducible representations of the quasi-primary operators which are delimited by square brackets, \textit{i.e.}\ $[\boldsymbol{N}_i,\boldsymbol{N}_j,\boldsymbol{N}_m]$, with the first (second) [third] quasi-primary operator always indexed by $i$ ($j$) [$m$], respectively.


\subsection{Conserved \texorpdfstring{$\ell\boldsymbol{e}_1$}{le1}}

We first investigate conserved quasi-primary operators in the $\ell\boldsymbol{e}_1$ irreducible representation.

We start with three-point correlation functions including two scalar quasi-primary operators.  We note that there is only one tensor structure for both the initial (with the conserved quasi-primary operator) and final (with the conserved quasi-primary operator acted upon by the conserved current differential operator) three-point correlation functions.  Hence, unless the conformal dimensions of the two scalars are related, the original three-point correlation functions should vanish.  Moreover, to fully illustrate the method, we proceed with the three different starting points, \textit{i.e.}\ the conservation conditions \eqref{Eq3pt1}, \eqref{Eq3pt2}, and \eqref{Eq3pt3}.

In the case $[\ell\boldsymbol{e}_1,\boldsymbol{0},\boldsymbol{0}]$, the tensor structure is
\eqn{(\FCF{1}{i}{j}{m}{1}{2})_{A_\ell\cdots A_1}=(\A_{12}\cdot\bar{\eta}_3)_{A_1}\cdots(\A_{12}\cdot\bar{\eta}_3)_{A_\ell},}
and the conservation conditions \eqref{Eq3pt1} and \eqref{Eq3pt1subs} give
\eqna{
0&=(\mathcal{T}_{12}^{(\ell-1)\boldsymbol{e}_1}\Gamma)^{A'_2\cdots A'_\ell}(\hat{\mathcal{P}}_{12}^{\ell\boldsymbol{e}_1})_{A'_\ell\cdots A'_1}^{\phantom{A'_\ell\cdots A'_1}A_1\cdots A_\ell}\aCF{1}{i}{j}{m}\left\{\frac{d+\ell-2-\Delta_j+\Delta_m}{2}\bar{\eta}_3^{A'_1}(\A_{12}\cdot\bar{\eta}_3)_{A_1}\right.\\
&\phantom{=}\qquad\left.+\ell\left[\A_{12A_1}^{\phantom{12A_1}A'_1}+\frac{1}{2}\bar{\eta}_3^{A'_1}(\A_{12}\cdot\bar{\eta}_3)_{A_1}\right]\right\}(\A_{12}\cdot\bar{\eta}_3)_{A_2}\cdots(\A_{12}\cdot\bar{\eta}_3)_{A_\ell},
}[Eqle100p]
after straightforward simplifications using the unitarity bound \eqref{EqU} with $S=\ell$ and $\mathfrak{n}=1$ as well as the Fock conditions for $\ell\boldsymbol{e}_1$.  Although there seems to be two contributions to \eqref{Eqle100p}, both contributions must merge into the sole tensor structure.  Extracting indices from the projection operator $\hat{\mathcal{P}}_{12}^{\ell\boldsymbol{e}_1}$ following \cite{Fortin:2020ncr}, \eqref{Eqle100p} becomes
\eqn{0=(\mathcal{T}_{12}^{(\ell-1)\boldsymbol{e}_1}\Gamma)^{A_2\cdots A_\ell}\aCF{1}{i}{j}{m}\frac{d+\ell-3}{d+2\ell-4}(\Delta_j-\Delta_m)(\A_{12}\cdot\bar{\eta}_3)_{A_2}\cdots(\A_{12}\cdot\bar{\eta}_3)_{A_\ell}.}[Eqle100]
Hence, in general dimensions, three-point correlation functions $[\ell\boldsymbol{e}_1,\boldsymbol{0},\boldsymbol{0}]$ with conserved $\ell\boldsymbol{e}_1$ vanish unless the two scalar quasi-primary operators have the same conformal dimension \cite{Ferrara:1973eg}.

For $[\boldsymbol{0},\ell\boldsymbol{e}_1,\boldsymbol{0}]$, we have instead
\eqna{
0&=(\mathcal{T}_{21}^{(\ell-1)\boldsymbol{e}_1}\Gamma)^{B'_2\cdots B'_\ell}(\hat{\mathcal{P}}_{21}^{\ell\boldsymbol{e}_1})_{B'_\ell\cdots B'_1}^{\phantom{B'_\ell\cdots B'_1}B_1\cdots B_\ell}\aCF{1}{i}{j}{m}\left\{\frac{-\Delta_i+d+\ell-2+\Delta_m}{2}\bar{\eta}_3^{B'_1}(\A_{12}\cdot\bar{\eta}_3)_{B_1}\right.\\
&\phantom{=}\qquad\left.+\ell\left[\A_{12B_1}^{\phantom{12B_1}B'_1}+\frac{1}{2}\bar{\eta}_3^{B'_1}(\A_{12}\cdot\bar{\eta}_3)_{B_1}\right]\right\}(\A_{12}\cdot\bar{\eta}_3)_{B_2}\cdots(\A_{12}\cdot\bar{\eta}_3)_{B_\ell},
}[Eq0le10p]
from \eqref{Eq3pt2} and \eqref{Eq3pt2subs}, which is equivalent to \eqref{Eqle100p} above with the appropriate substitutions.  Hence, extracting indices and contracting leads to
\eqn{0=(\mathcal{T}_{21}^{(\ell-1)\boldsymbol{e}_1}\Gamma)^{B_2\cdots B_\ell}\aCF{1}{i}{j}{m}\frac{d+\ell-3}{d+2\ell-4}(\Delta_i-\Delta_m)(\A_{12}\cdot\bar{\eta}_3)_{B_2}\cdots(\A_{12}\cdot\bar{\eta}_3)_{B_\ell},}[Eq0le10]
in agreement with \eqref{Eqle100}.  Once again, we conclude that three-point correlation functions of two scalars and one conserved $\ell\boldsymbol{e}_1$ vanish unless both scalars have the same conformal dimension.

Finally, considering  $[\boldsymbol{0},\boldsymbol{0},\ell\boldsymbol{e}_1]$, the conservation conditions \eqref{Eq3pt3} with \eqref{Eq3pt3subs} lead to
\eqna{
0&=(\mathcal{T}_{31}^{(\ell-1)\boldsymbol{e}_1}\Gamma)^{E'_2\cdots E'_\ell}(\hat{\mathcal{P}}_{31}^{\ell\boldsymbol{e}_1})_{E'_\ell\cdots E'_1}^{\phantom{E'_\ell\cdots E'_1}E_1\cdots E_\ell}\aCF{1}{i}{j}{m}\left\{\frac{-\Delta_i+\Delta_j+d+\ell-2}{2}(\A_{12}\cdot\bar{\eta}_3)^{E'_1}(\A_{12}\cdot\bar{\eta}_3)_{E_1}\right.\\
&\phantom{=}\qquad\left.+\ell\left[\A_{12E_1}^{\phantom{12E_1}E'_1}+\frac{1}{2}\bar{\eta}_3^{E'_1}(\A_{12}\cdot\bar{\eta}_3)_{E_1}\right]\right\}(\A_{12}\cdot\bar{\eta}_3)_{E_2}\cdots(\A_{12}\cdot\bar{\eta}_3)_{e_\ell},
}[Eq00le1p]
as expected from \eqref{Eqle100p} and \eqref{Eq0le10p}.  Therefore, proceeding as before, we reach
\eqn{0=(\mathcal{T}_{31}^{(\ell-1)\boldsymbol{e}_1}\Gamma)^{E_2\cdots E_\ell}\aCF{1}{i}{j}{m}\frac{d+\ell-3}{d+2\ell-4}(\Delta_i-\Delta_j)(\A_{12}\cdot\bar{\eta}_3)_{E_2}\cdots(\A_{12}\cdot\bar{\eta}_3)_{E_\ell}.}[Eq00le1]
Since \eqref{Eqle100p} = \eqref{Eq0le10p} = \eqref{Eq00le1p} and \eqref{Eqle100} = \eqref{Eq0le10} = \eqref{Eq00le1}, this simple exercise shows that the conservation conditions are the same irrespective of the location of the conserved quasi-primary operator.\footnote{Although with generic irreducible representations, one must be careful relating tensor structures (see for example \cite{Fortin:2019pep}).}

Since the OPE has already been used to compute three-point correlation functions with $[\boldsymbol{0},\boldsymbol{0},\ell\boldsymbol{e}_1]$ in \cite{Fortin:2019pep}, this last case can also highlight why conservation conditions for two-point correlation functions are not sufficient.  Concentrating on $\ell=1$, the OPE \eqref{EqOPE} leads to
\eqn{\Vev{\Op{i}{1}\Op{j}{2}\Op{m}{3}}=\frac{R_{ijm}\aCF{1}{i}{j}{m}}{\ee{1}{2}{p_{ijm}}}\A_{12D}^{\phantom{12D}C}\D_{12}^{(d,h_{ijm}-1/2,1)D}(\mathcal{T}_{12\boldsymbol{e}_1}\Gamma)_C*\Vev{\Op{m}{2}\Op{m}{3}},}
where $R_{ijm}$ is the rotation matrix
\eqn{\frac{1}{R_{ijm}}=\lambda_{\boldsymbol{e}_1}(-2)^{h_{ijm}+1/2}(-h_{ijm}-1/2)(\Delta_m+1-d)(\Delta_m+1)_{h_{ijm}-1/2}(\Delta_m+1-d/2)_{h_{ijm}-1/2},}
such that the three-point correlation functions are
\eqn{\Vev{\Op{i}{1}\Op{j}{2}\Op{m}{3}}=\frac{(\mathcal{T}_{31}^{\ell\boldsymbol{e}_1}\Gamma)^E\aCF{1}{i}{j}{m}(\A_{12}\cdot\bar{\eta}_3)_E}{\ee{1}{2}{\frac{1}{2}(\tau_i+\tau_j-\chi_m)}\ee{1}{3}{\frac{1}{2}(\chi_i-\chi_j+\tau_m)}\ee{2}{3}{\frac{1}{2}(-\chi_i+\chi_j+\chi_m)}}.}
We can already see the pole at the unitarity bound $\Delta_m=d-1$ in the rotation matrix that ultimately cancels the zero originating from the conservation conditions for two-point correlation functions.  Indeed, the conservation conditions \eqref{EqNES} lead to
\eqna{
0&=C_\Gamma\Gamma_+^{\phantom{+}E}(-i\mathcal{L}_{3E+})*\Vev{\Op{i}{1}\Op{j}{2}\Op{m}{3}}\\
&=-\frac{R_{ijm}\aCF{1}{i}{j}{m}}{\ee{1}{2}{p_{ijm}}}\A_{12D}^{\phantom{12D}C}\D_{12}^{(d,h_{ijm}-1/2,1)D}(\mathcal{T}_{12\boldsymbol{e}_1}\Gamma)_C^{ba}(C_\Gamma\Gamma_+^{\phantom{+}E})^{dc}\\
&\phantom{=}\qquad\times\frac{\eta_{3+}\eta_{2E}}{\ee{2}{3}{}}(d-2+\Theta_3)\Vev{\Op{ab}{2}\Op{cd}{3}},
}
using \eqref{EqNES} and then \eqref{EqNLD} when $\mathcal{L}_{3E+}$ acts on the two-point correlation functions.  Replacing $\Theta_3$ by $-\Delta_m+1$ shows explicitly the aforementioned cancellation, proving that two-point conservation conditions do not imply three-point conservation conditions.  Obviously, completing the computation above leads to the same conservation conditions as \eqref{Eq00le1} for $\ell=1$, \textit{i.e.}\ $\aCF{1}{i}{j}{m}(\Delta_i-\Delta_j)=0$.

Another example of interest is given by $[\boldsymbol{e}_1,\boldsymbol{e}_1,\boldsymbol{0}]$ in $3d$ where the $\epsilon$-tensor appears in one of the tensor structures.  In this case, there are three tensor structures given respectively by
\eqn{(\FCF{1}{i}{j}{m}{1}{2})_{AB}=\A_{12AB},\qquad(\FCF{2}{i}{j}{m}{1}{2})_{AB}=(\A_{12}\cdot\bar{\eta}_3)_A(\A_{12}\cdot\bar{\eta}_3)_B,\qquad(\FCF{1}{i}{j}{m}{1}{2})_{AB}=\epsilon_{12ABF}(\A_{12}\cdot\bar{\eta}_3)^F.}
Demanding that the first quasi-primary operator is conserved, we expect one constraint from conservation since there is only one tensor structure for $[\boldsymbol{0},\boldsymbol{e}_1,\boldsymbol{0}]$, and the constraint should not involve the third tensor structure with the $\epsilon$-tensor.  The conservation conditions \eqref{Eq3pt1} and \eqref{Eq3pt1subs} indeed give
\eqna{
0&=(\mathcal{T}_{21}^{\boldsymbol{N}_j}\Gamma)^B(\hat{\mathcal{P}}_{12}^{\boldsymbol{N}_i})_{A'}^{\phantom{A'}A}\sum_{r=1}^{N_{ijm}}\aCF{r}{i}{j}{m}\left[-\frac{2-\Delta_j+\Delta_m}{2}\bar{\eta}_3^{A'}+\frac{\ee{1}{3}{\frac{1}{2}}}{\ee{2}{3}{\frac{1}{2}}}\D_{21}^{A'}\right](\FCF{r}{i}{j}{m}{1}{2})_{AB}\\
&=(\mathcal{T}_{21}^{\boldsymbol{N}_j}\Gamma)^B\left[-\frac{2-\Delta_j+\Delta_m}{2}\aCF{1}{i}{j}{m}+(-\Delta_j+\Delta_m)\aCF{2}{i}{j}{m}\right](\A_{12}\cdot\bar{\eta}_3)_B,
}
corresponding to one linear combination not involving $\aCF{3}{i}{j}{m}$ being set to zero.  From the example above, the case $[\boldsymbol{e}_1,\boldsymbol{e}_1,\boldsymbol{e}_1]$ in $3d$ with the first quasi-primary operator being conserved would most likely results in constraints on the three-point coefficients associated to tensor structures with $\epsilon$-tensors.

It is also interesting to study two other cases with conserved symmetric-traceless, mainly $[\boldsymbol{e}_n,\boldsymbol{e}_{n+2},\ell\boldsymbol{e}_1]$ for any $\ell$ and $n$ as well as $[\boldsymbol{0},n\boldsymbol{e}_2,\ell\boldsymbol{e}_1]$ for $\ell\geq n$.  In both cases, there is only one tensor structure for the initial three-point correlation functions, given by
\eqn{
\begin{gathered}
(\FCF{1}{i}{j}{m}{1}{2})_{A_n\cdots A_1B_{n+2}\cdots B_1E_\ell\cdots E_1}=\left[\prod_{1\leq k\leq n}\A_{12A_kB_k}\right]\A_{12B_{n+1}E_1}(\A_{12}\cdot\bar{\eta}_3)_{B_{n+2}}\left[\prod_{2\leq k\leq\ell}(\A_{12}\cdot\bar{\eta}_3)_{E_k}\right],\\
(\FCF{1}{i}{j}{m}{1}{2})_{B_{2n}\cdots B_1E_\ell\cdots E_1}=\left[\prod_{1\leq k\leq n}\A_{12B_{2k-1}E_k}\right]\left[\prod_{1\leq k\leq n}(\A_{12}\cdot\bar{\eta}_3)_{B_{2k}}\right]\left[\prod_{1\leq k\leq\ell-n}(\A_{12}\cdot\bar{\eta}_3)_{E_{k+n}}\right],
\end{gathered}
}
respectively, and one tensor structure for the final three-point correlation functions.  As for the conserved symmetric-traceless with two scalars, we thus expect that these three-point correlation functions vanish unless the conformal dimensions of the two unconserved quasi-primary operators are related, even though they are in different irreducible representations.  Using \eqref{Eq3pt3}, \eqref{Eq3pt3subs} and the identity for index extraction of \cite{Fortin:2020ncr}, it is easy to see that the conservation conditions are
\eqn{
\begin{gathered}
0=\aCF{1}{i}{j}{m}\frac{(\ell-1)(d+\ell-2)}{\ell(d+2l-4)}(\Delta_i-\Delta_j),\\
0=\aCF{1}{i}{j}{m}\frac{(\ell-n)(d+\ell+n-3)}{\ell(d+2l-4)}(\Delta_i-\Delta_j),
\end{gathered}
}
for $[\boldsymbol{e}_n,\boldsymbol{e}_{n+2},\ell\boldsymbol{e}_1]$ and $[\boldsymbol{0},n\boldsymbol{e}_2,\ell\boldsymbol{e}_1]$, respectively.  Hence, these three-point correlation functions vanish unless the conformal dimensions $\Delta_i=\Delta_j$ even though the two quasi-primary operators are in different irreducible representations, as expected.  We note that for $\ell=1$ and $\ell-n$, respectively, there are no constraints since $N_{ij\bar{m}}=0$ for both cases.


\subsection{Conserved \texorpdfstring{$\boldsymbol{e}_1+\boldsymbol{e}_r$}{e1+er}}

For completeness, we analyse conserved currents in fermionic irreducible representations, namely $\boldsymbol{e}_1+\boldsymbol{e}_r$ with projection operator
\eqn{(\hat{\mathcal{P}}_{31}^{\boldsymbol{e}_1+\boldsymbol{e}_r})_{e'E'}^{\phantom{e'E'}Ee}=\A_{13E'}^{\phantom{13E'}E}\delta_{e'}^{\phantom{e'}e}-\frac{1}{d}(\Gamma_{13E'}\Gamma_{13}^E)_{e'}^{\phantom{e'}e}.}
We focus on $[\boldsymbol{0},\boldsymbol{e}_r,\boldsymbol{e}_1+\boldsymbol{e}_r]$ for which there are two tensor structures ($N_{ijm}=2$) given by
\eqn{(\FCF{1}{i}{j}{m}{1}{2})_{beE}=(\A_{12}\cdot\bar{\eta}_3)_E(C_\Gamma^{-1})_{be},\qquad(\FCF{2}{i}{j}{m}{1}{2})_{beE}=(\A_{12}\cdot\bar{\eta}_3)_E(\bar{\eta}_3\cdot\Gamma_{12}C_\Gamma^{-1})_{be}.}
Since $N_{ij\bar{m}}=2$, there are two conservation conditions for two tensor structures.  We thus expect that the original three-point correlation will vanish unless the conformal dimensions of the scalar and the spinor are related.

Using \eqref{Eq3pt3} with the help of \eqref{Eq3pt3subs}, we reach after several simplifications with the half-projectors and setting $\Delta_m=\Delta_m^*=d-1/2$ [see \eqref{EqU}]
\eqn{0=\aCF{1}{i}{j}{m}\frac{\Delta_i-\Delta_j-1/2}{d},\qquad0=\aCF{2}{i}{j}{m}\frac{\Delta_i-\Delta_j+1/2}{d},}
which imply that the three-point correlation function vanishes unless the conformal dimensions $\Delta_i$ and $\Delta_j$ are related, as expected.  Indeed, the first (second) tensor structure appears only when $\Delta_i=\Delta_j+1/2$ ($\Delta_i=\Delta_j-1/2$), reminiscent of the cases $[\boldsymbol{e}_n,\boldsymbol{e}_{n+2},\ell\boldsymbol{e}_1]$ and $[\boldsymbol{0},n\boldsymbol{e}_2,\ell\boldsymbol{e}_1]$ studied above.

As a final remark, we note that the conservation conditions never mix tensor structures with even and odd numbers of $\Gamma$-matrices, as can be seen directly from \eqref{Eq3pt1}, \eqref{Eq3pt2} and \eqref{Eq3pt3}.


\subsection{Conserved \texorpdfstring{$\boldsymbol{e}_3$}{e3}}

As a final explicit example, we study $[\boldsymbol{e}_1,\boldsymbol{e}_1,\boldsymbol{e}_3]$ with all quasi-primary operators being conserved.  First, we note that $N_{ijm}=1$ with the tensor structure given by
\eqn{(\FCF{1}{i}{j}{m}{1}{2})_{ABE_3E_2E_1}=\A_{12AE_1}\A_{12BE_2}(\A_{12}\cdot\bar{\eta}_3)_{E_3},}
while $N_{\bar{\imath}jm}=N_{i\bar{\jmath}m}=0$ and $N_{ij\bar{m}}=3$.  Therefore, there are no constraints originating from the conservation conditions for the two $\boldsymbol{e}_1$ apart from fixing $\Delta_i=\Delta_j=d-1$.  Demanding that the $\boldsymbol{e}_3$ is conserved however leads to the three constraints
\eqn{0=(\Delta_m+1)\aCF{1}{i}{j}{m},\qquad0=(\Delta_m+1)\aCF{1}{i}{j}{m},\qquad0=(\Delta_m+3-d)\aCF{1}{i}{j}{m},}
where the last one is satified for any three-point coefficient since $\Delta_m=d-3$ from \eqref{EqU}.  The first two constraints on the other hand force the three-point coefficient $\aCF{1}{i}{j}{m}$ to vanish, implying that there are no three-point correlation functions $[\boldsymbol{e}_1,\boldsymbol{e}_1,\boldsymbol{e}_3]$ when all three quasi-primary operators are conserved.

Moreover, this observation shows that conservation of the quasi-primary operator $\Op{m}{3}$ in the three-point correlation functions $\Vev{J_i(\eta_1)J_j(\eta_2)\Op{m}{3}}$ for two conserved vector currents and $\Vev{T(\eta_1)T(\eta_2)\Op{m}{3}}$ for two energy-momentum tensors are not necessarily trivial.


\subsection{Conserved Currents and Energy-Momentum Tensors}

\begin{table}[t]
\centering
\resizebox{11cm}{!}{%
\begin{tabular}{|c|c|c|}\hline
$\mathcal{O}_m$ & Parity $(-1)^\ell$ & Parity $-(-1)^\ell$\\\hline\hline
$\ell\boldsymbol{e}_1$ & \begin{tabular}{ccc}$\ell=0$&:&$2\to1$\\$\ell=1$ unconserved&:&$3\to1$\\$\ell=1$ conserved&:&$3\to2$\\$\ell\geq2$&:&$4\to2$\end{tabular} & \begin{tabular}{ccc}$\ell=0$&:&$0$\\$\ell\geq1$&:&$1\to0$\end{tabular}\\\hline
$\boldsymbol{e}_2+\ell\boldsymbol{e}_1$ & \begin{tabular}{ccc}$\ell=0$&:&$1\to0$\\$\ell\geq1$&:&$2\to1$\end{tabular} & \begin{tabular}{ccc}$\ell\geq0$&:&$2\to1$\end{tabular}\\\hline
$\boldsymbol{e}_3+\ell\boldsymbol{e}_1$ & \begin{tabular}{ccc}$\ell\geq0$&:&$1\to1$\end{tabular} & \begin{tabular}{ccc}$\ell\geq0$&:&$0$\end{tabular}\\\hline
$2\boldsymbol{e}_2+\ell\boldsymbol{e}_1$ & \begin{tabular}{ccc}$\ell\geq0$&:&$1\to1$\end{tabular} & \begin{tabular}{ccc}$\ell\geq0$&:&$0$\end{tabular}\\\hline
\end{tabular}
}
\caption{Number of independent tensor structures in three-point correlation functions $\Vev{J_i(\eta_1)J_j(\eta_2)\mathcal{O}_m(\eta_3)}$ of two conserved vector currents and one extra quasi-primary operator.  The notation $n\to m$ states that the initial $n$ tensor structures with a given parity are reduced to $m$ independent tensor structures with the same parity when the conservation conditions on the vector currents are imposed.}
\label{TabJJO}
\end{table}
With an eye on the four-point conformal bootstrap involving four conserved vector currents or four energy-momentum tensors, we compute the conservation conditions imposed on the three-point correlation functions $\Vev{J_i(\eta_1)J_j(\eta_2)\mathcal{O}_m(\eta_3)}$ and $\Vev{T(\eta_1)T(\eta_2)\mathcal{O}_m(\eta_3)}$.  The actual computations are done in Appendix \ref{SAppJEM}, here we only present in Tables \ref{TabJJO} and \ref{TabTTO} the implications of conservation on the number of independent tensor structures.

\begin{table}[t]
\centering
\resizebox{15cm}{!}{%
\begin{tabular}{|c|c||c|c|}\hline
$\mathcal{O}_m$ & Symmetric & $\mathcal{O}_m$ & Symmetric\\\hline\hline
$\ell\boldsymbol{e}_1$ & \begin{tabular}{ccc}$\ell=0$&:&$3\to1$\\$\ell\geq1$ odd&:&$4\to0$\\$\ell=2$ unconserved&:&$8\to2$\\$\ell=2$ conserved&:&$8\to3$\\$\ell\geq4$ even&:&$10\to3$\end{tabular} & $\boldsymbol{e}_2+\ell\boldsymbol{e}_1$ & \begin{tabular}{ccc}$\ell=0$&:&$2\to0$\\$\ell=1$&:&$6\to1$\\$\ell=2$&:&$7\to1$\\$\ell\geq3$ odd&:&$8\to2$\\$\ell\geq4$ even&:&$8\to2$\end{tabular}\\\hline
$\boldsymbol{e}_3+\ell\boldsymbol{e}_1$ & \begin{tabular}{ccc}$\ell=0$&:&$2\to1$\\$\ell\geq1$ odd&:&$1\to0$\\$\ell\geq2$ even&:&$4\to2$\end{tabular} & $2\boldsymbol{e}_2+\ell\boldsymbol{e}_1$ & \begin{tabular}{ccc}$\ell=0$&:&$5\to2$\\$\ell\geq1$ odd&:&$4\to1$\\$\ell\geq2$ even&:&$7\to3$\end{tabular}\\\hline
$\boldsymbol{e}_2+\boldsymbol{e}_3+\ell\boldsymbol{e}_1$ & \begin{tabular}{ccc}$\ell=0$&:&$1\to0$\\$\ell\geq1$&:&$2\to1$\end{tabular} & $2\boldsymbol{e}_3+\ell\boldsymbol{e}_1$ & \begin{tabular}{ccc}$\ell\geq0$ even&:&$1\to1$\\$\ell\geq1$ odd&:&$0$\end{tabular}\\\hline
$3\boldsymbol{e}_2+\ell\boldsymbol{e}_1$ & \begin{tabular}{ccc}$\ell=0$&:&$1\to0$\\$\ell\geq1$&:&$2\to1$\end{tabular} & $2\boldsymbol{e}_2+\boldsymbol{e}_3+\ell\boldsymbol{e}_1$ & \begin{tabular}{ccc}$\ell\geq0$ even&:&$1\to1$\\$\ell\geq1$ odd&:&$0$\end{tabular}\\\hline
$4\boldsymbol{e}_2+\ell\boldsymbol{e}_1$ & \begin{tabular}{ccc}$\ell\geq0$ even&:&$1\to1$\\$\ell\geq1$ odd&:&$0$\end{tabular} & &\\\hline
\end{tabular}
}
\caption{Number of independent tensor structures in three-point correlation functions $\Vev{T(\eta_1)T(\eta_2)\mathcal{O}_m(\eta_3)}$ of two energy-momentum tensors and one extra quasi-primary operator.  The notation $n\to m$ states that the initial $n$ symmetric tensor structures are reduced to $m$ independent symmetric tensor structures when the conservation conditions on the energy-momentum tensors are imposed.}
\label{TabTTO}
\end{table}
For $\Vev{J_i(\eta_1)J_j(\eta_2)\mathcal{O}_m(\eta_3)}$, the tensor structures are divided into two groups depending on their parity under the exchanged symmetry group $\mathbb{Z}_2$.  Since the energy-momentum tensors are identical, only the symmetric tensor structures are included for $\Vev{T(\eta_1)T(\eta_2)\mathcal{O}_m(\eta_3)}$.

Some of these results can be found elsewhere, see for example \cite{Schreier:1971um,Osborn:1993cr,Costa:2011mg}.


\section{Discussion and Conclusion}\label{SecConc}

In this paper we investigated conservation conditions in CFTs from the point of view of the embedding space formalism developed in \cite{Fortin:2019fvx,Fortin:2019dnq}.  We found that the differential operators implementing conservation in position space, which are not explicitly conformally covariant, can nevertheless be uplifted to the embedding space covariantly.  More precisely, using several identities, we showed that the position space differential operators relevant for conservation can be uplifted into associated embedding space differential operators expressed in terms of three contributions.  One of these contributions can be discarded since it vanishes when acting on quasi-primary operators.  Another contribution is not conformally covariant but vanishes when the unitarity bound is saturated.  Finally, the last contribution can be made explicitly conformally covariant and annihilates two-point correlation functions.

Hence, demanding that two-point correlation functions are conserved forces the conformal dimensions of the relevant quasi-primary operators to saturate the unitarity bound, leading to conformally-covariant differential operators implementing conservation in embedding space.  Our results, shown in \eqref{EqUAll}, exhibit a pleasing relationship: the link between the OPE differential operators and the conserved current differential operators is the same in position space and in embedding space.  In passing, we also used the OPE to demonstrate how the conservation conditions on higher-point correlation functions can be straightforwardly implemented from three-point, but not from two-point, correlation functions.

With the help of the Fock conditions and some projection operators, we then gave the complete action of the conserved current differential operators to compute conservation conditions of three-point correlation functions with quasi-primary operators in arbitrary irreducible representations of the Lorentz group.  Our technique does not necessitate the knowledge of all projection operators---only the projection operators of the conserved quasi-primary operators are necessary.

We then provided several examples, focusing mostly on $\Vev{JJ\mathcal{O}}$ and $\Vev{TT\mathcal{O}}$ where $J$ and $T$ are conserved vector currents and the energy-momentum tensor, respectively.  In doing so, we completed the analysis necessary to implement the four-point conformal bootstrap of four conserved vector currents and four energy-momentum tensors.  We also discussed the implications of Ward identities at coincident points for $\Vev{\mathcal{O}\mathcal{O}J}$ and $\Vev{\mathcal{O}\mathcal{O}T}$, providing a set of simple rules to determine their relations to two-point correlation functions.

Conserved quasi-primary operators in arbitrary irreducible representations of the Lorentz group other than $\boldsymbol{e}_1$ and $2\boldsymbol{e}_1$, commonly called higher-spin currents, can be studied straightforwardly from our approach, although theories with higher-spin currents are believed to be free.  It would be interesting to analyse conformal higher-spin currents from the point of view of the embedding space formalism used here, perhaps to find an alternative argument for the triviality of the associated theories.


\ack{
The work of JFF is supported by NSERC.  VP is supported in part by NSERC.  WJM is supported by the China Scholarship Council and in part by NSERC.  The work of WS is supported in part by DOE HEP grant DE-SC00-17660.
}


\setcounter{section}{0}
\renewcommand{\thesection}{\Alph{section}}

\section{Notation}\label{SAppNot}

In this appendix we briefly review the half-projectors appearing in the OPE \eqref{EqOPE} as well as in correlation functions.  They were first introduced in \cite{Fortin:2019dnq} where a more detailed discussion can be found.

Half-projectors $(\mathcal{T}_{ij}^{\boldsymbol{N}}\Gamma)$ appear in the OPE or in correlation functions to ensure that the latter transform properly under Lorentz transformations.  For a quasi-primary operator in embedding space living in the irreducible representation $\boldsymbol{N}$ of the Lorentz group, with all its Lorentz group indices chosen to be embedding space spinor indices, denoted here by $\{a\}$, an associated half-projector $(\mathcal{T}_{ij}^{\boldsymbol{N}}\Gamma)_{\{a\}}^{\{Aa\}}$ is included with the same set of embedding space spinor indices.  As such, the half-projectors are built from $\Gamma$-matrices in embedding space.

The associated set of embedding space vector (and spinor) indices, $\{Aa\}$, of the half-projectors are then contracted properly together through the tensor structures to generate singlets.  With this prescription, the OPE and correlation functions transform as expected through the transformation properties of the embedding space spinor indices $\{a\}$.  Obviously, the half-projectors are also built from the proper embedding space metrics to ensure transversality.

Once they have been built, their particular definition is no longer important since they can be shown to verify several powerful identities, as for example \eqref{EqIdTP}.  Hence, it is possible to rely on these identities and other important properties of the half-projectors, like their behavior under permutations of embedding space vector indices, dictated by the Fock conditions, to obtain final results.

This observation is exemplified for example in Section \ref{SecEx} and Appendix \ref{SAppJEM} where explicit formulas for half-projectors are not needed.


\section{Conserved Vector Currents and Energy-Momentum Tensors}\label{SAppJEM}

Motivated by the conformal bootstrap of four conserved vector currents and four energy-momentum tensors, in this appendix we complete the analysis of the conservation conditions for conserved vector currents $[\boldsymbol{e}_1,\boldsymbol{e}_1,\boldsymbol{N}]$ and the energy-momentum tensor $[2\boldsymbol{e}_1,2\boldsymbol{e}_1,\boldsymbol{N}]$.  The results are obtained from a straightforward application of the equations found in Section \ref{SecCF}, \textit{i.e.}\ the conservation conditions and the Fock conditions.  We stress again that only the explicit projection operators for $\boldsymbol{e}_1$ and $2\boldsymbol{e}_1$ are necessary, which are known.

For unconserved quasi-primary operators in irreducible representations $\boldsymbol{N}$, the projection operators for $\boldsymbol{N}$ are not necessary, only the Fock conditions are, which greatly simplifies the analysis.  Since conservation of the third quasi-primary operator is not necessarily trivial, as demonstrated in the $[\boldsymbol{e}_1,\boldsymbol{e}_1,\boldsymbol{e}_3]$ example above, we must know the projection operators for $\boldsymbol{N}$ when the third quasi-primary operator is conserved.  As not all the necessary projection operators are known, in the following we will assume that the third quasi-primary operator is not conserved, and only comment on the implications of setting $\Delta_m=\Delta_m^*$ in the conservation conditions coming from the first two quasi-primary operators.\footnote{Since it is believed that higher-spin currents lead to free theories (see \textit{e.g.}\ \cite{Maldacena:2011jn,Alba:2015upa,Alba:2013yda}), this simplification should not be too generous.  The cases where the third quasi-primary operators are standard conserved vector currents and energy-momentum tensors are described in detail in Appendix \ref{SAppWard}.}

Throughout this appendix, the tensor structures are written in terms of $\A_{12}\cdot\bar{\eta}_3$ and
\eqn{\A_{12;3}^{AB}=\A_{12}^{AB}+\frac{1}{2}(\A_{12}\cdot\bar{\eta}_3)^A(\A_{12}\cdot\bar{\eta}_3)^B,}[EqAp]
since these quantities transform simply under the exchange symmetry of the two conserved currents $J_i(\eta_1)\leftrightarrow J_i(\eta_2)$ and the two energy-momentum tensors $T(\eta_1)\leftrightarrow T(\eta_2)$.  Indeed, denoting the embedding space vector indices of the first (second) [third] quasi-primary operators by $\{A\}$ ($\{B\}$) [$\{E\}$], respectively, the non-trivial behaviors are
\eqn{
\begin{gathered}
\A_{12;3}^{AE}\to\A_{12;3}^{BE},\qquad\A_{12;3}^{BE}\to\A_{12;3}^{AE},\qquad(\A_{12}\cdot\bar{\eta}_3)_E\to-(\A_{12}\cdot\bar{\eta}_3)_E,\\
(\A_{12}\cdot\bar{\eta}_3)_A\to(\A_{12}\cdot\bar{\eta}_3)_B,\qquad(\A_{12}\cdot\bar{\eta}_3)_B\to(\A_{12}\cdot\bar{\eta}_3)_A,
\end{gathered}
}[EqES12]
under the exchange symmetry of the first two quasi-primary operators.  It is straightforward to rotate from the conserved basis of tensor structures above to either the OPE basis, the three-point basis, or any other convenient basis.


\subsection{Conserved Vector Currents}

We focus here on usual conserved vector currents $J_i(\eta)$ and $J_j(\eta)$ in the vector irreducible representation of the Lorentz group.  Considering three-point correlation functions $\Vev{J_i(\eta_1)J_j(\eta_2)\Op{m}{3}}$, there are four different types of possible exchanged quasi-primary operators.  They correspond to exchanged quasi-primary operators in irreducible representations $\ell\boldsymbol{e}_1$ for which there are five tensor structures, $\boldsymbol{e}_2+\ell\boldsymbol{e}_1$ with four tensor structures, $\boldsymbol{e}_3+\ell\boldsymbol{e}_1$ with one tensor structure, and $2\boldsymbol{e}_2+\ell\boldsymbol{e}_1$ with one tensor structure, respectively.

For exchanged quasi-primary operators in irreducible representations $\boldsymbol{e}_3+\ell\boldsymbol{e}_1$ and $2\boldsymbol{e}_2+\ell\boldsymbol{e}_1$, for which there are only one tensor structure, conservation conditions do not lead to any constraints since there are no tensor structures for the associated three-point correlation functions with one scalar, one vector, and the exchanged quasi-primary operators ($N_{\bar{\imath}jm}=N_{i\bar{\jmath}m}=0$).\footnote{Again, this statement may be corrected for conserved $\boldsymbol{e}_3+\ell\boldsymbol{e}_1$ and $2\boldsymbol{e}_2+\ell\boldsymbol{e}_1$.}  Therefore, conservation conditions only necessitate to set $\Delta_i=\Delta_j=d-1$ for these exchanged quasi-primary operators.

We can now focus on the first two groups of exchanged quasi-primary operators.

\subsubsection{\texorpdfstring{$\ell\boldsymbol{e}_1$}{le1}}

In the case of $\ell\boldsymbol{e}_1$ exchanged quasi-primary operators, the five tensor structures can be written as
\eqn{
\begin{gathered}
(\FCF{1}{i}{j}{m}{1}{2})=\A_{12;3AB}\left[\prod_{1\leq k\leq\ell}(\A_{12}\cdot\bar{\eta}_3)_{E_k}\right],\\
(\FCF{2}{i}{j}{m}{1}{2})=(\A_{12}\cdot\bar{\eta}_3)_A(\A_{12}\cdot\bar{\eta}_3)_B\left[\prod_{1\leq k\leq\ell}(\A_{12}\cdot\bar{\eta}_3)_{E_k}\right],\\
(\FCF{3}{i}{j}{m}{1}{2})=[\A_{12;3AE_1}(\A_{12}\cdot\bar{\eta}_3)_B-\A_{12;3BE_1}(\A_{12}\cdot\bar{\eta}_3)_A]\left[\prod_{2\leq k\leq\ell}(\A_{12}\cdot\bar{\eta}_3)_{E_k}\right],\\
(\FCF{4}{i}{j}{m}{1}{2})=[\A_{12;3AE_1}(\A_{12}\cdot\bar{\eta}_3)_B+\A_{12;3BE_1}(\A_{12}\cdot\bar{\eta}_3)_A]\left[\prod_{2\leq k\leq\ell}(\A_{12}\cdot\bar{\eta}_3)_{E_k}\right],\\
(\FCF{5}{i}{j}{m}{1}{2})=\A_{12;3AE_1}\A_{12;3BE_2}\left[\prod_{3\leq k\leq\ell}(\A_{12}\cdot\bar{\eta}_3)_{E_k}\right],
\end{gathered}
}[EqTSJJle1]
where the first two tensor structures exist for $\ell\geq0$, the third and the fourth tensor structures exist for $\ell\geq1$, while the last tensor structure exists only for $\ell\geq2$.  For convenience, tensor structures \eqref{EqTSJJle1} have been chosen to have definite parity under the possible exchange symmetry of the two conserved currents, with all the tensor structures having $(-1)^\ell$ parity except for $\FCF{4}{i}{j}{m}{1}{2}$ which has $-(-1)^\ell$ parity.

From the tensor structures \eqref{EqTSJJle1} as well as \eqref{Eq3pt2} and \eqref{Eq3pt2subs}, the conservation conditions for the second conserved current lead to the following $N_{i\bar{\jmath}m}=2$ constraints
\eqn{
\begin{gathered}
0=\frac{d+\ell-1}{4}(2\aCF{1}{i}{j}{m}+\aCF{5}{i}{j}{m})+(d-1-\Delta_m)\aCF{2}{i}{j}{m}-\frac{\Delta_m+\ell}{2}\aCF{3}{i}{j}{m}+\left(d-1-\frac{\Delta_m-\ell}{2}\right)\aCF{4}{i}{j}{m},\\
0=\ell\aCF{1}{i}{j}{m}+(d-2-\Delta_m)\aCF{3}{i}{j}{m}+(d-\Delta_m)\aCF{4}{i}{j}{m}+\frac{d+\ell-2}{2}\aCF{5}{i}{j}{m},
\end{gathered}
}[EqJcJle1]
while the conservation conditions for the first conserved current are obtained from \eqref{EqTSJJle1}, \eqref{Eq3pt1} and \eqref{Eq3pt1subs}, or more simply from \eqref{EqJcJle1} by substituting $\aCF{4}{i}{j}{m}\to-\aCF{4}{i}{j}{m}$, \textit{i.e.}\
\eqn{
\begin{gathered}
0=\frac{d+\ell-1}{4}(2\aCF{1}{i}{j}{m}+\aCF{5}{i}{j}{m})+(d-1-\Delta_m)\aCF{2}{i}{j}{m}-\frac{\Delta_m+\ell}{2}\aCF{3}{i}{j}{m}-\left(d-1-\frac{\Delta_m-\ell}{2}\right)\aCF{4}{i}{j}{m},\\
0=\ell\aCF{1}{i}{j}{m}+(d-2-\Delta_m)\aCF{3}{i}{j}{m}-(d-\Delta_m)\aCF{4}{i}{j}{m}+\frac{d+\ell-2}{2}\aCF{5}{i}{j}{m}.
\end{gathered}
}[EqcJJle1]
Since one constraint is linearly dependent, combining \eqref{EqJcJle1} and \eqref{EqcJJle1} leads to only three constraints given by
\eqn{
\begin{gathered}
\frac{d+\ell-1}{4}(2\aCF{1}{i}{j}{m}+\aCF{5}{i}{j}{m})+(d-1-\Delta_m)\aCF{2}{i}{j}{m}-\frac{\Delta_m+\ell}{2}\aCF{3}{i}{j}{m}=0,\qquad\aCF{4}{i}{j}{m}=0,\\
\ell\aCF{1}{i}{j}{m}+(d-2-\Delta_m)\aCF{3}{i}{j}{m}+\frac{d+\ell-2}{2}\aCF{5}{i}{j}{m}=0,
\end{gathered}
}[EqcJcJle1]
which shows that the sole tensor structure with parity $-(-1)^\ell$, namely $\FCF{4}{i}{j}{m}{1}{2}$, never appears.

For $\ell=0$, \eqref{EqcJcJle1} implies that of the two initial tensor structures, only one survives.  For $\ell=1$, \eqref{EqcJcJle1} forces the four initial tensor structures to combine such that there is only one, unless the $\boldsymbol{e}_1$ exchanged quasi-primary operator is also conserved (with $\Delta_m=d-1$) in which case there are two tensor structures remaining due to the fact that two constraints become linearly dependent (see Appendix \ref{SAppWard}).  Finally, for $\ell\geq2$, only two of the five initial tensor structures survive.

\subsubsection{\texorpdfstring{$\boldsymbol{e}_2+\ell\boldsymbol{e}_1$}{e2+le1}}

In the case of $\boldsymbol{e}_2+\ell\boldsymbol{e}_1$ exchanged quasi-primary operators, the four tensor structures can be written as
\eqn{
\begin{gathered}
(\FCF{1}{i}{j}{m}{1}{2})=\A_{12;3AF_1}\A_{12;3BF_2}\left[\prod_{1\leq k\leq\ell}(\A_{12}\cdot\bar{\eta}_3)_{E_k}\right],\\
(\FCF{2}{i}{j}{m}{1}{2})=[\A_{12;3AF_1}(\A_{12}\cdot\bar{\eta}_3)_B+\A_{12;3BF_1}(\A_{12}\cdot\bar{\eta}_3)_A](\A_{12}\cdot\bar{\eta}_3)_{F_2}\left[\prod_{1\leq k\leq\ell}(\A_{12}\cdot\bar{\eta}_3)_{E_k}\right],\\
(\FCF{3}{i}{j}{m}{1}{2})=[\A_{12;3AF_1}(\A_{12}\cdot\bar{\eta}_3)_B-\A_{12;3BF_1}(\A_{12}\cdot\bar{\eta}_3)_A](\A_{12}\cdot\bar{\eta}_3)_{F_2}\left[\prod_{1\leq k\leq\ell}(\A_{12}\cdot\bar{\eta}_3)_{E_k}\right],\\
(\FCF{4}{i}{j}{m}{1}{2})=(\A_{12;3AF_1}\A_{12;3BE_1}+\A_{12;3BF_1}\A_{12;3AE_1})(\A_{12}\cdot\bar{\eta}_3)_{F_2}\left[\prod_{2\leq k\leq\ell}(\A_{12}\cdot\bar{\eta}_3)_{E_k}\right],
\end{gathered}
}[EqTSJJe2le1]
where the indices $F_1$ and $F_2$ are the $\boldsymbol{e}_2$ indices and the indices $E_{1\leq k\leq\ell}$ are the $\ell\boldsymbol{e}_1$ indices.  The first three tensor structures in \eqref{EqTSJJe2le1} exist when $\ell\geq0$ while the last tensor structures exists only when $\ell\geq1$.  Again, the tensor structures \eqref{EqTSJJe2le1} are chosen such that their parity eigenvalues are well defined, with the first two tensor structures having parity $-(-1)^\ell$ while the parity of the last two tensor structures is $(-1)^\ell$.

Using \eqref{EqTSJJe2le1} as well as \eqref{Eq3pt2} and \eqref{Eq3pt2subs} leads to the conservation conditions (only one constraint since $N_{i\bar{\jmath}m}=1$)
\eqn{0=\frac{d+\ell-2}{2}\aCF{1}{i}{j}{m}+(d-\Delta_m)\aCF{2}{i}{j}{m}+(d-2-\Delta_m)\aCF{3}{i}{j}{m}+\frac{d+\ell}{2}\aCF{4}{i}{j}{m},}[EqJcJe2le1]
when demanding conservation of the second conserved current.  Changing the signs of $\aCF{3}{i}{j}{m}$ and $\aCF{4}{i}{j}{m}$ leads to
\eqn{0=\frac{d+\ell-2}{2}\aCF{1}{i}{j}{m}+(d-\Delta_m)\aCF{2}{i}{j}{m}-(d-2-\Delta_m)\aCF{3}{i}{j}{m}-\frac{d+\ell}{2}\aCF{4}{i}{j}{m},}[EqcJJe2le1]
which corresponds to the conservation conditions for the first conserved current.  Combining the constraints \eqref{EqJcJe2le1} and \eqref{EqcJJe2le1} implies the conservation conditions
\eqn{\frac{d+\ell-2}{2}\aCF{1}{i}{j}{m}+(d-\Delta_m)\aCF{2}{i}{j}{m}=0,\qquad(d-2-\Delta_m)\aCF{3}{i}{j}{m}+\frac{d+\ell}{2}\aCF{4}{i}{j}{m}=0.}[EqcJcJe2le1]

Consequently, \eqref{EqcJcJe2le1} states that there is only one (two) tensor structure(s) instead of the expected three (four) when $\ell=0$ ($\ell\geq1$).  For $\ell=0$, two tensor structures survive if the $\boldsymbol{e}_2$ is conserved (not taking into account its conservation conditions).


\subsection{Energy-Momentum Tensors}

We now turn our attention to the conservation conditions for the energy-momentum tensor $T(\eta)$ by considering three-point correlation functions $\Vev{T(\eta_1)T(\eta_2)\Op{m}{3}}$, where there are nine different types of exchanged quasi-primary operators denoted by their respective irreducible representations.  They are: $\ell\boldsymbol{e}_1$ with fourteen tensor structures, $\boldsymbol{e}_2+\ell\boldsymbol{e}_1$ with sixteen tensor structures, $\boldsymbol{e}_3+\ell\boldsymbol{e}_1$ with five tensor structures, $2\boldsymbol{e}_2+\ell\boldsymbol{e}_1$ with eleven tensor structures, $\boldsymbol{e}_2+\boldsymbol{e}_3+\ell\boldsymbol{e}_1$ with four tensor structures, $2\boldsymbol{e}_3+\ell\boldsymbol{e}_1$ with one tensor structure, $3\boldsymbol{e}_2+\ell\boldsymbol{e}_1$ with four tensor structures, $2\boldsymbol{e}_2+\boldsymbol{e}_3+\ell\boldsymbol{e}_1$ with one tensor structure, and $4\boldsymbol{e}_2+\ell\boldsymbol{e}_1$ with one tensor structure.

Therefore, for quasi-primary operators in the infinite towers of irreducible representations $2\boldsymbol{e}_3+\ell\boldsymbol{e}_1$, $2\boldsymbol{e}_2+\boldsymbol{e}_3+\ell\boldsymbol{e}_1$, and $4\boldsymbol{e}_2+\ell\boldsymbol{e}_1$, there are no constraints originating from the conservation conditions since their associated three-point correlation functions with one vector, one energy-momentum tensor, and the exchanged quasi-primary operators are zero identically ($N_{\bar{\imath}jm}=N_{i\bar{\jmath}m}=0$).\footnote{We stress that this statement may be modified for conserved exchanged quasi-primary operators.}  For them, it is only necessary to fix the conformal dimensions to $\Delta_i=\Delta_j=\Delta_T=d$.

It is important to note that, contrary to the previous case of two conserved vector currents which can be different (for example two components of the multiplet of conserved vector currents for a non-abelian global symmetry), the two energy-momentum tensors are obviously the same.  Hence, only tensor structures with parity $1$ appear, \textit{i.e.}\ tensor structures with parity $(-1)^\ell$ exist only for $\ell$ even while tensor structures with parity $-(-1)^\ell$ exist only for $\ell$ odd.

We are ready to study the conservation conditions for the remaining infinite towers of irreducible representations.

\subsubsection{\texorpdfstring{$\ell\boldsymbol{e}_1$}{le1}}

Considering exchanged quasi-primary operators in the irreducible representations $\ell\boldsymbol{e}_1$, there are fourteen tensor structures which we choose to be
\eqna{
(\FCF{1}{i}{j}{m}{1}{2})&=\A_{12;3A_1B_1}\A_{12;3A_2B_2}\left[\prod_{1\leq k\leq\ell}(\A_{12}\cdot\bar{\eta}_3)_{E_k}\right],\\
(\FCF{2}{i}{j}{m}{1}{2})&=\A_{12;3A_1B_1}(\A_{12}\cdot\bar{\eta}_3)_{A_2}(\A_{12}\cdot\bar{\eta}_3)_{B_2}\left[\prod_{1\leq k\leq\ell}(\A_{12}\cdot\bar{\eta}_3)_{E_k}\right],\\
(\FCF{3}{i}{j}{m}{1}{2})&=(\A_{12}\cdot\bar{\eta}_3)_{A_1}(\A_{12}\cdot\bar{\eta}_3)_{A_2}(\A_{12}\cdot\bar{\eta}_3)_{B_1}(\A_{12}\cdot\bar{\eta}_3)_{B_2}\left[\prod_{1\leq k\leq\ell}(\A_{12}\cdot\bar{\eta}_3)_{E_k}\right],\\
(\FCF{4}{i}{j}{m}{1}{2})&=[\A_{12;3A_1E_1}(\A_{12}\cdot\bar{\eta}_3)_{B_1}-\A_{12;3B_1E_1}(\A_{12}\cdot\bar{\eta}_3)_{A_1}]\A_{12;3A_2B_2}\left[\prod_{2\leq k\leq\ell}(\A_{12}\cdot\bar{\eta}_3)_{E_k}\right],\\
(\FCF{6}{i}{j}{m}{1}{2})&=[\A_{12;3A_1E_1}(\A_{12}\cdot\bar{\eta}_3)_{B_1}-\A_{12;3B_1E_1}(\A_{12}\cdot\bar{\eta}_3)_{A_1}]\\
&\phantom{=}\qquad\times(\A_{12}\cdot\bar{\eta}_3)_{A_2}(\A_{12}\cdot\bar{\eta}_3)_{B_2}\left[\prod_{2\leq k\leq\ell}(\A_{12}\cdot\bar{\eta}_3)_{E_k}\right],\\
(\FCF{8}{i}{j}{m}{1}{2})&=\A_{12;3A_1E_1}\A_{12;3B_1E_2}\A_{12;3A_2B_2}\left[\prod_{3\leq k\leq\ell}(\A_{12}\cdot\bar{\eta}_3)_{E_k}\right],\\
(\FCF{9}{i}{j}{m}{1}{2})&=\A_{12;3A_1E_1}\A_{12;3B_1E_2}(\A_{12}\cdot\bar{\eta}_3)_{A_2}(\A_{12}\cdot\bar{\eta}_3)_{B_2}\left[\prod_{3\leq k\leq\ell}(\A_{12}\cdot\bar{\eta}_3)_{E_k}\right],\\
(\FCF{10}{i}{j}{m}{1}{2})&=[\A_{12;3A_1E_1}\A_{12;3A_2E_2}(\A_{12}\cdot\bar{\eta}_3)_{B_1}(\A_{12}\cdot\bar{\eta}_3)_{B_2}\\
&\phantom{=}\qquad+\A_{12;3B_1E_1}\A_{12;3B_2E_2}(\A_{12}\cdot\bar{\eta}_3)_{A_1}(\A_{12}\cdot\bar{\eta}_3)_{A_2}]\left[\prod_{3\leq k\leq\ell}(\A_{12}\cdot\bar{\eta}_3)_{E_k}\right],\\
(\FCF{12}{i}{j}{m}{1}{2})&=[\A_{12;3A_1E_1}\A_{12;3A_2E_2}\A_{12;3B_1E_3}(\A_{12}\cdot\bar{\eta}_3)_{B_2}\\
&\phantom{=}\qquad-\A_{12;3B_1E_1}\A_{12;3B_2E_2}\A_{12;3A_1E_3}(\A_{12}\cdot\bar{\eta}_3)_{A_2}]\left[\prod_{4\leq k\leq\ell}(\A_{12}\cdot\bar{\eta}_3)_{E_k}\right],
}
and
\eqn{(\FCF{14}{i}{j}{m}{1}{2})=\A_{12;3A_1E_1}\A_{12;3A_2E_2}\A_{12;3B_1E_3}\A_{12;3B_2E_4}\left[\prod_{5\leq k\leq\ell}(\A_{12}\cdot\bar{\eta}_3)_{E_k}\right],}[EqTSTTle1even]
with parity $(-1)^\ell$, as well as
\eqna{
(\FCF{5}{i}{j}{m}{1}{2})&=[\A_{12;3A_1E_1}(\A_{12}\cdot\bar{\eta}_3)_{B_1}+\A_{12;3B_1E_1}(\A_{12}\cdot\bar{\eta}_3)_{A_1}]\A_{12;3A_2B_2}\left[\prod_{2\leq k\leq\ell}(\A_{12}\cdot\bar{\eta}_3)_{E_k}\right],\\
(\FCF{7}{i}{j}{m}{1}{2})&=[\A_{12;3A_1E_1}(\A_{12}\cdot\bar{\eta}_3)_{B_1}+\A_{12;3B_1E_1}(\A_{12}\cdot\bar{\eta}_3)_{A_1}]\\
&\phantom{=}\qquad\times(\A_{12}\cdot\bar{\eta}_3)_{A_2}(\A_{12}\cdot\bar{\eta}_3)_{B_2}\left[\prod_{2\leq k\leq\ell}(\A_{12}\cdot\bar{\eta}_3)_{E_k}\right],\\
(\FCF{11}{i}{j}{m}{1}{2})&=[\A_{12;3A_1E_1}\A_{12;3A_2E_2}(\A_{12}\cdot\bar{\eta}_3)_{B_1}(\A_{12}\cdot\bar{\eta}_3)_{B_2}\\
&\phantom{=}\qquad-\A_{12;3B_1E_1}\A_{12;3B_2E_2}(\A_{12}\cdot\bar{\eta}_3)_{A_1}(\A_{12}\cdot\bar{\eta}_3)_{A_2}]\left[\prod_{3\leq k\leq\ell}(\A_{12}\cdot\bar{\eta}_3)_{E_k}\right],\\
(\FCF{13}{i}{j}{m}{1}{2})&=[\A_{12;3A_1E_1}\A_{12;3A_2E_2}\A_{12;3B_1E_3}(\A_{12}\cdot\bar{\eta}_3)_{B_2}\\
&\phantom{=}\qquad+\A_{12;3B_1E_1}\A_{12;3B_2E_2}\A_{12;3A_1E_3}(\A_{12}\cdot\bar{\eta}_3)_{A_2}]\left[\prod_{4\leq k\leq\ell}(\A_{12}\cdot\bar{\eta}_3)_{E_k}\right],
}[EqTSTTle1odd]
with parity $-(-1)^\ell$.  Not taking into account parity, the first three tensor structures exist for $\ell\geq0$, the tensor structures numbered from $4$ to $7$ exist for $\ell\geq1$, the tensor structures numbered from $8$ to $11$ appear only when $\ell\geq2$, $(\FCF{12}{i}{j}{m}{1}{2})$ and $(\FCF{13}{i}{j}{m}{1}{2})$ necessitate $\ell\geq3$, and finally the last tensor structure exists only for $\ell\geq4$.

From the tensor structures \eqref{EqTSTTle1even} and \eqref{EqTSTTle1odd}, the conservation conditions \eqref{Eq3pt2} and \eqref{Eq3pt2subs}, and the separation of the constraints for even and odd tensor structures, we find the eight ($N_{\bar{\imath}jm}=N_{i\bar{\jmath}m}=8$) constraints for the parity $(-1)^\ell$ tensor structures
\eqna{
0&=[d(d+\ell+1)-\Delta_m-\ell-2](4\aCF{1}{i}{j}{m}+2\aCF{8}{i}{j}{m}+\aCF{14}{i}{j}{m})\\
&\phantom{=}\qquad+2d(2d+\ell-\Delta_m)(2\aCF{2}{i}{j}{m}+\aCF{9}{i}{j}{m}+2\aCF{10}{i}{j}{m})+16[d(d-1-\Delta_m)+\Delta_m+\ell+2]\aCF{3}{i}{j}{m}\\
&\phantom{=}\qquad-(d-2)(\Delta_m+\ell+2)(2\aCF{4}{i}{j}{m}+4\aCF{6}{i}{j}{m}+\aCF{12}{i}{j}{m}),\\
0&=4d\ell(\aCF{1}{i}{j}{m}+\aCF{2}{i}{j}{m})+2[d(2d+\ell-4-\Delta_m)+2(\Delta_m+\ell)]\aCF{4}{i}{j}{m}\\
&\phantom{=}\qquad+2[d(d-2-\Delta_m)+\Delta_m+\ell](4\aCF{6}{i}{j}{m}+\aCF{12}{i}{j}{m})+[d(2d+3\ell)-4(\Delta_m+\ell)]\aCF{8}{i}{j}{m}\\
&\phantom{=}\qquad+2d(2d+\ell-\Delta_m)\aCF{9}{i}{j}{m}+8[d(d-\Delta_m)+\Delta_m+\ell]\aCF{10}{i}{j}{m}+2[d(d+\ell)-\Delta_m-\ell]\aCF{14}{i}{j}{m},
}
\eqna{
0&=4\ell(\aCF{1}{i}{j}{m}-4\aCF{3}{i}{j}{m})+[d(d+\ell)-2\ell](2\aCF{4}{i}{j}{m}-\aCF{8}{i}{j}{m})+4[d(d-\Delta_m)-2\ell]\aCF{6}{i}{j}{m}\\
&\phantom{=}\qquad-2d(d-\Delta_m)\aCF{9}{i}{j}{m}-4d(d+\ell)\aCF{10}{i}{j}{m}+[d(d+2\ell+\Delta_m)-2\ell]\aCF{12}{i}{j}{m}\\
&\phantom{=}\qquad-[d(d+\ell)-\ell]\aCF{14}{i}{j}{m},\\
0&=4d\ell\aCF{1}{i}{j}{m}+2[d(d-1-\Delta_m)+2]\aCF{4}{i}{j}{m}+8\aCF{6}{i}{j}{m}\\
&\phantom{=}\qquad+[d(d+\ell-1)-4]\aCF{8}{i}{j}{m}+8\aCF{10}{i}{j}{m}+2\aCF{12}{i}{j}{m}-2\aCF{14}{i}{j}{m},\\
0&=4[d(d+\ell)-2]\aCF{1}{i}{j}{m}+4d(d-\Delta_m)\aCF{2}{i}{j}{m}+32\aCF{3}{i}{j}{m}-2[d(\Delta_m+\ell)-4]\aCF{4}{i}{j}{m}\\
&\phantom{=}\qquad+16\aCF{6}{i}{j}{m}+[d(d+\ell)-4]\aCF{8}{i}{j}{m}+4\aCF{12}{i}{j}{m}-2\aCF{14}{i}{j}{m},\\
0&=2(d-2)(\ell-1)\aCF{4}{i}{j}{m}-8(\ell-1)\aCF{6}{i}{j}{m}-[d(d+2\ell-3)-4(\ell-1)]\aCF{8}{i}{j}{m}-2d(d-\Delta_m)\aCF{9}{i}{j}{m}\\
&\phantom{=}\qquad-8(d+\ell-1)\aCF{10}{i}{j}{m}+2[d(\Delta_m+\ell)-\ell+1]\aCF{12}{i}{j}{m}-2[d(d+\ell-1)-\ell+1]\aCF{14}{i}{j}{m},\\
0&=(d-2)(\ell-2)\aCF{8}{i}{j}{m}+4(\ell-2)\aCF{10}{i}{j}{m}+d(d-2-\Delta_m)\aCF{12}{i}{j}{m}+[d(d+\ell-2)-\ell+2]\aCF{14}{i}{j}{m},\\
0&=2d(\ell-1)\aCF{4}{i}{j}{m}+[d\ell-2(\Delta_m+\ell-2)]\aCF{8}{i}{j}{m}+2d\aCF{9}{i}{j}{m}+4[d(d-1-\Delta_m)+\Delta_m+\ell-2]\aCF{10}{i}{j}{m}\\
&\phantom{=}\qquad+d(2d+\ell-4-\Delta_m)\aCF{12}{i}{j}{m}+[d(d+\ell-1)-\Delta_m-\ell+2]\aCF{14}{i}{j}{m},
}[EqcTcTle1even]
and the parity $-(-1)^\ell$ tensor structures
\eqna{
0&=[d(4d+3\ell+2-\Delta_m)-2(\Delta_m+\ell+2)](2\aCF{5}{i}{j}{m}+\aCF{13}{i}{j}{m})\\
&\phantom{=}\qquad+4[d(4d+\ell-2-3\Delta_m)+2(\Delta_m+\ell+2)]\aCF{7}{i}{j}{m}-4(d-2)(\Delta_m+\ell+2)\aCF{11}{i}{j}{m},\\
0&=[d(2d+3\ell-\Delta_m)-2(\Delta_m+\ell)]\aCF{5}{i}{j}{m}+4[d(d-\Delta_m)+\Delta_m+\ell]\aCF{7}{i}{j}{m}\\
&\phantom{=}\qquad+4[d(d-2-\Delta_m)+\Delta_m+\ell]\aCF{11}{i}{j}{m}+[d(3d+2\ell-\Delta_m)-\Delta_m-\ell]\aCF{13}{i}{j}{m},\\
0&=2[d(d+\ell)-2\ell](\aCF{5}{i}{j}{m}-2\aCF{11}{i}{j}{m})+4[d(d-\Delta_m)+2\ell]\aCF{7}{i}{j}{m}+[d(3d+2\ell-\Delta_m)-2\ell]\aCF{13}{i}{j}{m},\\
0&=[d(d+1-\Delta_m)-2]\aCF{5}{i}{j}{m}+4\aCF{7}{i}{j}{m}+4\aCF{11}{i}{j}{m}-\aCF{13}{i}{j}{m},\\
0&=[d(2d+\ell-\Delta_m)-4]\aCF{5}{i}{j}{m}+8\aCF{7}{i}{j}{m}+8\aCF{11}{i}{j}{m}-2\aCF{13}{i}{j}{m},\\
0&=(d-2)(\ell-1)\aCF{5}{i}{j}{m}+4(\ell-1)\aCF{7}{i}{j}{m}-4(d-\ell+1)\aCF{11}{i}{j}{m}+[d(2d+\ell-\Delta_m)-\ell+1]\aCF{13}{i}{j}{m},\\
0&=4(\ell-2)\aCF{11}{i}{j}{m}+d(d+2-\Delta_m)\aCF{13}{i}{j}{m},\\
0&=2d(\ell-1)\aCF{5}{i}{j}{m}+4[d(d-1-\Delta_m)+\Delta_m+\ell-2]\aCF{11}{i}{j}{m}+d(2d+\ell-\Delta_m)\aCF{13}{i}{j}{m}.
}[EqcTcTle1odd]
It is easy to see that one of the parity $(-1)^\ell$ constraints in \eqref{EqcTcTle1even} is linearly dependent while the constraints \eqref{EqcTcTle1odd} for the parity $-(-1)^\ell$ tensor structures force $\aCF{5}{i}{j}{m}=\aCF{7}{i}{j}{m}=\aCF{11}{i}{j}{m}=\aCF{13}{i}{j}{m}=0$.  Hence there are seven constraints for parity $(-1)^\ell$ tensor structures and the tensor structures with parity $-(-1)^\ell$ can be discarded.

We are thus left with one tensor structure for $\ell=0$, two tensor structures for $\ell=2$ (three, as can be seen from the rank, when the third quasi-primary operator is conserved, see also Appendix \ref{SAppWard}), three tensor structures for $\ell\geq4$ and no tensor structures for $\ell$ odd.

\subsubsection{\texorpdfstring{$\boldsymbol{e}_2+\ell\boldsymbol{e}_1$}{e2+le1}}

There are sixteen tensor structures for exchanged quasi-primary operators in the irreducible representations $\boldsymbol{e}_2+\ell\boldsymbol{e}_1$.  They can be chosen as
\eqna{
(\FCF{1}{i}{j}{m}{1}{2})&=\A_{12;3A_1F_1}\A_{12;3B_1F_2}\A_{12;3A_2B_2}\left[\prod_{1\leq k\leq\ell}(\A_{12}\cdot\bar{\eta}_3)_{E_k}\right],\\
(\FCF{2}{i}{j}{m}{1}{2})&=\A_{12;3A_1F_1}\A_{12;3B_1F_2}(\A_{12}\cdot\bar{\eta}_3)_{A_2}(\A_{12}\cdot\bar{\eta}_3)_{B_2}\left[\prod_{1\leq k\leq\ell}(\A_{12}\cdot\bar{\eta}_3)_{E_k}\right],\\
(\FCF{3}{i}{j}{m}{1}{2})&=[\A_{12;3A_1F_1}(\A_{12}\cdot\bar{\eta}_3)_{B_1}+\A_{12;3B_1F_1}(\A_{12}\cdot\bar{\eta}_3)_{A_1}]\\
&\phantom{=}\qquad\times\A_{12;3A_2B_2}(\A_{12}\cdot\bar{\eta}_3)_{F_2}\left[\prod_{1\leq k\leq\ell}(\A_{12}\cdot\bar{\eta}_3)_{E_k}\right],\\
(\FCF{5}{i}{j}{m}{1}{2})&=[\A_{12;3A_1F_1}(\A_{12}\cdot\bar{\eta}_3)_{B_1}+\A_{12;3B_1F_1}(\A_{12}\cdot\bar{\eta}_3)_{A_1}]\\
&\phantom{=}\qquad\times(\A_{12}\cdot\bar{\eta}_3)_{A_2}(\A_{12}\cdot\bar{\eta}_3)_{B_2}(\A_{12}\cdot\bar{\eta}_3)_{F_2}\left[\prod_{1\leq k\leq\ell}(\A_{12}\cdot\bar{\eta}_3)_{E_k}\right],\\
(\FCF{7}{i}{j}{m}{1}{2})&=\A_{12;3A_1F_1}\A_{12;3B_1F_2}[\A_{12;3A_2E_1}(\A_{12}\cdot\bar{\eta}_3)_{B_2}\\
&\phantom{=}\qquad-\A_{12;3B_2E_1}(\A_{12}\cdot\bar{\eta}_3)_{A_2}]\left[\prod_{2\leq k\leq\ell}(\A_{12}\cdot\bar{\eta}_3)_{E_k}\right],\\
(\FCF{9}{i}{j}{m}{1}{2})&=[\A_{12;3A_1F_1}\A_{12;3A_2E_1}(\A_{12}\cdot\bar{\eta}_3)_{B_1}(\A_{12}\cdot\bar{\eta}_3)_{B_2}\\
&\phantom{=}\qquad-\A_{12;3B_1F_1}\A_{12;3B_2E_1}(\A_{12}\cdot\bar{\eta}_3)_{A_1}(\A_{12}\cdot\bar{\eta}_3)_{A_2}](\A_{12}\cdot\bar{\eta}_3)_{F_2}\left[\prod_{2\leq k\leq\ell}(\A_{12}\cdot\bar{\eta}_3)_{E_k}\right],\\
(\FCF{13}{i}{j}{m}{1}{2})&=\A_{12;3A_1F_1}\A_{12;3B_1F_2}\A_{12;3A_2E_1}\A_{12;3B_2E_2}\left[\prod_{3\leq k\leq\ell}(\A_{12}\cdot\bar{\eta}_3)_{E_k}\right],\\
(\FCF{14}{i}{j}{m}{1}{2})&=[\A_{12;3A_1F_1}(\A_{12}\cdot\bar{\eta}_3)_{B_1}+\A_{12;3B_1F_1}(\A_{12}\cdot\bar{\eta}_3)_{A_1}]\\
&\phantom{=}\qquad\times\A_{12;3A_2E_1}\A_{12;3B_2E_2}(\A_{12}\cdot\bar{\eta}_3)_{F_2}\left[\prod_{3\leq k\leq\ell}(\A_{12}\cdot\bar{\eta}_3)_{E_k}\right],
}[EqTSTTe2le1even]
for $-(-1)^\ell$ parity as well as
\eqna{
(\FCF{4}{i}{j}{m}{1}{2})&=[\A_{12;3A_1F_1}(\A_{12}\cdot\bar{\eta}_3)_{B_1}-\A_{12;3B_1F_1}(\A_{12}\cdot\bar{\eta}_3)_{A_1}]\\
&\phantom{=}\qquad\times\A_{12;3A_2B_2}(\A_{12}\cdot\bar{\eta}_3)_{F_2}\left[\prod_{1\leq k\leq\ell}(\A_{12}\cdot\bar{\eta}_3)_{E_k}\right],\\
(\FCF{6}{i}{j}{m}{1}{2})&=[\A_{12;3A_1F_1}(\A_{12}\cdot\bar{\eta}_3)_{B_1}-\A_{12;3B_1F_1}(\A_{12}\cdot\bar{\eta}_3)_{A_1}]\\
&\phantom{=}\qquad\times(\A_{12}\cdot\bar{\eta}_3)_{A_2}(\A_{12}\cdot\bar{\eta}_3)_{B_2}(\A_{12}\cdot\bar{\eta}_3)_{F_2}\left[\prod_{1\leq k\leq\ell}(\A_{12}\cdot\bar{\eta}_3)_{E_k}\right],\\
(\FCF{8}{i}{j}{m}{1}{2})&=\A_{12;3A_1F_1}\A_{12;3B_1F_2}[\A_{12;3A_2E_1}(\A_{12}\cdot\bar{\eta}_3)_{B_2}\\
&\phantom{=}\qquad+\A_{12;3B_2E_1}(\A_{12}\cdot\bar{\eta}_3)_{A_2}]\left[\prod_{2\leq k\leq\ell}(\A_{12}\cdot\bar{\eta}_3)_{E_k}\right],\\
(\FCF{10}{i}{j}{m}{1}{2})&=[\A_{12;3A_1F_1}\A_{12;3A_2E_1}(\A_{12}\cdot\bar{\eta}_3)_{B_1}(\A_{12}\cdot\bar{\eta}_3)_{B_2}\\
&\phantom{=}\qquad+\A_{12;3B_1F_1}\A_{12;3B_2E_1}(\A_{12}\cdot\bar{\eta}_3)_{A_1}(\A_{12}\cdot\bar{\eta}_3)_{A_2}](\A_{12}\cdot\bar{\eta}_3)_{F_2}\left[\prod_{2\leq k\leq\ell}(\A_{12}\cdot\bar{\eta}_3)_{E_k}\right],\\
(\FCF{11}{i}{j}{m}{1}{2})&=[\A_{12;3A_1F_1}\A_{12;3B_1E_1}+\A_{12;3B_1F_1}\A_{12;3A_1E_1}]\\
&\phantom{=}\qquad\times\A_{12;3A_2B_2}(\A_{12}\cdot\bar{\eta}_3)_{F_2}\left[\prod_{2\leq k\leq\ell}(\A_{12}\cdot\bar{\eta}_3)_{E_k}\right],\\
(\FCF{12}{i}{j}{m}{1}{2})&=[\A_{12;3A_1F_1}\A_{12;3B_1E_1}+\A_{12;3B_1F_1}\A_{12;3A_1E_1}]\\
&\phantom{=}\qquad\times(\A_{12}\cdot\bar{\eta}_3)_{A_2}(\A_{12}\cdot\bar{\eta}_3)_{B_2}(\A_{12}\cdot\bar{\eta}_3)_{F_2}\left[\prod_{2\leq k\leq\ell}(\A_{12}\cdot\bar{\eta}_3)_{E_k}\right],\\
(\FCF{15}{i}{j}{m}{1}{2})&=[\A_{12;3A_1F_1}(\A_{12}\cdot\bar{\eta}_3)_{B_1}-\A_{12;3B_1F_1}(\A_{12}\cdot\bar{\eta}_3)_{A_1}]\\
&\phantom{=}\qquad\times\A_{12;3A_2E_1}\A_{12;3B_2E_2}(\A_{12}\cdot\bar{\eta}_3)_{F_2}\left[\prod_{3\leq k\leq\ell}(\A_{12}\cdot\bar{\eta}_3)_{E_k}\right],\\
(\FCF{16}{i}{j}{m}{1}{2})&=[\A_{12;3A_1F_1}\A_{12;3B_1E_1}+\A_{12;3B_1F_1}\A_{12;3A_1E_1}]\\
&\phantom{=}\qquad\times\A_{12;3A_2E_2}\A_{12;3B_2E_3}(\A_{12}\cdot\bar{\eta}_3)_{F_2}\left[\prod_{4\leq k\leq\ell}(\A_{12}\cdot\bar{\eta}_3)_{E_k}\right],
}[EqTSTTe2le1odd]
for parity $(-1)^\ell$.  Here the indices $F_1$ and $F_2$ are the $\boldsymbol{e}_2$ indices.  Disregarding parity, the first six tensor structures are there for all $\ell\geq0$, the tensor structures numbered from $7$ to $12$ exist for $\ell\geq1$, the tensor structures numbered from $13$ to $15$ necessitate $\ell\geq2$, while $\ell\geq3$ for the last tensor structure $(\FCF{16}{i}{j}{m}{1}{2})$ to appear.

The conservation conditions obtained from the tensor structures \eqref{EqTSTTe2le1even} and \eqref{EqTSTTe2le1odd}, using \eqref{Eq3pt2}, \eqref{Eq3pt2subs}, and the independence of the even and odd tensor structures, are given by ($N_{\bar{\imath}jm}=N_{i\bar{\jmath}m}=8$)
\eqna{
0&=2d(d+\ell)\aCF{1}{i}{j}{m}+2d(2d+\ell-2-\Delta_m)\aCF{2}{i}{j}{m}+2[d(2d+\ell+2-\Delta_m)-2(\Delta_m+\ell+2)]\aCF{3}{i}{j}{m}\\
&\phantom{=}\qquad+8[d(d-\Delta_m)+\Delta_m+\ell+2]\aCF{5}{i}{j}{m}-(d-2)(\Delta_m+\ell+2)\aCF{7}{i}{j}{m}\\
&\phantom{=}\qquad+4[d(d-2-\Delta_m)+\Delta_m+\ell+2]\aCF{9}{i}{j}{m}+[d(d+\ell)-\Delta_m-\ell-2]\aCF{13}{i}{j}{m}\\
&\phantom{=}\qquad+d(2d+\ell+2-\Delta_m)\aCF{14}{i}{j}{m},\\
0&=d(d+\ell+1)\aCF{1}{i}{j}{m}+2d(d-\Delta_m)\aCF{2}{i}{j}{m}-2(d+2\ell+2)\aCF{3}{i}{j}{m}+8(\ell+1)\aCF{5}{i}{j}{m}\\
&\phantom{=}\qquad-[d(d+2\ell-2+\Delta_m)-2(\ell+1)]\aCF{7}{i}{j}{m}-4(d-\ell-1)\aCF{9}{i}{j}{m}\\
&\phantom{=}\qquad+[d(d+\ell-1)-\ell-1]\aCF{13}{i}{j}{m}+d(d+2-\Delta_m)\aCF{14}{i}{j}{m},\\
0&=(d+\ell+2)(2\aCF{1}{i}{j}{m}-4\aCF{3}{i}{j}{m}+4\aCF{9}{i}{j}{m}+\aCF{13}{i}{j}{m}-2\aCF{14}{i}{j}{m})\\
&\phantom{=}\qquad+4(d-\Delta_m)(\aCF{2}{i}{j}{m}-2\aCF{5}{i}{j}{m})-2(\Delta_m+\ell+2)\aCF{7}{i}{j}{m},\\
0&=d\ell\aCF{1}{i}{j}{m}-2(d-2)\ell\aCF{3}{i}{j}{m}-8\ell\aCF{5}{i}{j}{m}+[d(2d+\ell-4-\Delta_m)-2\ell]\aCF{7}{i}{j}{m}\\
&\phantom{=}\qquad+4(2d-\ell)\aCF{9}{i}{j}{m}+(2d+\ell)\aCF{13}{i}{j}{m}-d(2d+\ell+4-\Delta_m)\aCF{14}{i}{j}{m},\\
0&=d\ell(\aCF{1}{i}{j}{m}+2\aCF{3}{i}{j}{m})+d(2d+\ell-4-\Delta_m)\aCF{7}{i}{j}{m}+4[d(d-1-\Delta_m)+\Delta_m+\ell]\aCF{9}{i}{j}{m}\\
&\phantom{=}\qquad+[d(d+\ell-1)-\Delta_m-\ell]\aCF{13}{i}{j}{m}+d(2d+\ell-\Delta_m)\aCF{14}{i}{j}{m},\\
0&=d(d+\ell-1)\aCF{1}{i}{j}{m}+2[d(d+1-\Delta_m)-2]\aCF{3}{i}{j}{m}+8\aCF{5}{i}{j}{m}+2\aCF{7}{i}{j}{m}+4\aCF{9}{i}{j}{m}-\aCF{13}{i}{j}{m},\\
0&=2d\ell\aCF{1}{i}{j}{m}+2d(d-1-\Delta_m)\aCF{7}{i}{j}{m}+8\aCF{9}{i}{j}{m}+[d(d+\ell)-2]\aCF{13}{i}{j}{m}-2d\aCF{14}{i}{j}{m},\\
0&=8(\ell-1)\aCF{9}{i}{j}{m}+[d(d+\ell-4)-2(\ell-1)]\aCF{13}{i}{j}{m}+2d(d+2-\Delta_m)\aCF{14}{i}{j}{m},
}[EqcTcTe2le1even]
for the parity $-(-1)^\ell$ tensor structures while they are
\eqna{
0&=2[d(2d+\ell-2-\Delta_m)+2(\Delta_m+\ell+2)]\aCF{4}{i}{j}{m}+8[d(d-2-\Delta_m)+\Delta_m+\ell+2]\aCF{6}{i}{j}{m}\\
&\phantom{=}\qquad+[d(4d+3\ell-2-\Delta_m)-2(\Delta_m+\ell+2)]\aCF{8}{i}{j}{m}+4[d(d-\Delta_m)+\Delta_m+\ell+2]\aCF{10}{i}{j}{m}\\
&\phantom{=}\qquad+2[d(d+\ell+2)-2(\Delta_m+\ell+2)]\aCF{11}{i}{j}{m}+2d(2d+\ell+2-\Delta_m)\aCF{12}{i}{j}{m}\\
&\phantom{=}\qquad+d(2d+\ell-2-\Delta_m)\aCF{15}{i}{j}{m}+[d(d+\ell+2)-\Delta_m-\ell-2]\aCF{16}{i}{j}{m},\\
0&=2(d+2\ell+2)\aCF{4}{i}{j}{m}+8(\ell+1)\aCF{6}{i}{j}{m}+[d(3d+2\ell-2-\Delta_m)-2(\ell+1)]\aCF{8}{i}{j}{m}\\
&\phantom{=}\qquad+4(d+\ell+1)\aCF{10}{i}{j}{m}+[d(d+\ell-1)-4(\ell+1)]\aCF{11}{i}{j}{m}+2d(d-\Delta_m)\aCF{12}{i}{j}{m}\\
&\phantom{=}\qquad+d(d-2-\Delta_m)\aCF{15}{i}{j}{m}+[d(d+\ell+1)-\ell-1]\aCF{16}{i}{j}{m},\\
0&=(d+\ell+2)(4\aCF{4}{i}{j}{m}-4\aCF{10}{i}{j}{m}-2\aCF{11}{i}{j}{m}+2\aCF{15}{i}{j}{m}-\aCF{16}{i}{j}{m})\\
&\phantom{=}\qquad+4(d-\Delta_m)(2\aCF{6}{i}{j}{m}-\aCF{12}{i}{j}{m})+2(2d+\ell+2-\Delta_m)\aCF{8}{i}{j}{m},
}
\eqna{
0&=2(d-2)\ell\aCF{4}{i}{j}{m}-8\ell\aCF{6}{i}{j}{m}-[d(\Delta_m+\ell-4)-2\ell]\aCF{8}{i}{j}{m}-4(2d+\ell)\aCF{10}{i}{j}{m}\\
&\phantom{=}\qquad-[d(2d+3\ell)-4\ell]\aCF{11}{i}{j}{m}-4d(d-\Delta_m)\aCF{12}{i}{j}{m}\\
&\phantom{=}\qquad+d(\Delta_m+\ell+4)\aCF{15}{i}{j}{m}-[2d(d+\ell+1)-\ell]\aCF{16}{i}{j}{m},\\
0&=2d\ell\aCF{4}{i}{j}{m}+d(2d+\ell-2-\Delta_m)(\aCF{8}{i}{j}{m}+\aCF{15}{i}{j}{m})+4[d(d-1-\Delta_m)+\Delta_m+\ell]\aCF{10}{i}{j}{m}\\
&\phantom{=}\qquad+[d(\ell+2)-4(\Delta_m+\ell)]\aCF{11}{i}{j}{m}+4d\aCF{12}{i}{j}{m}+[d(d+\ell+1)-\Delta_m-\ell]\aCF{16}{i}{j}{m},\\
0&=2[d(d-1-\Delta_m)+2]\aCF{4}{i}{j}{m}+8\aCF{6}{i}{j}{m}-2\aCF{8}{i}{j}{m}+4\aCF{10}{i}{j}{m}+[d(d+\ell+1)-4]\aCF{11}{i}{j}{m}-\aCF{16}{i}{j}{m},\\
0&=2d(d+1-\Delta_m)\aCF{8}{i}{j}{m}+8\aCF{10}{i}{j}{m}-2(d\ell+4)\aCF{11}{i}{j}{m}+2d\aCF{15}{i}{j}{m}-[d(d+\ell)+2]\aCF{16}{i}{j}{m},\\
0&=8(\ell-1)\aCF{10}{i}{j}{m}+4(d-2)(\ell-1)\aCF{11}{i}{j}{m}+2d(d-2-\Delta_m)\aCF{15}{i}{j}{m}+[3d(d+\ell)-2(\ell-1)]\aCF{16}{i}{j}{m},
}[EqcTcTe2le1odd]
for the parity $(-1)^\ell$ tensor structures.  For both \eqref{EqcTcTe2le1even} and \eqref{EqcTcTe2le1odd}, there are two constraints that are linearly dependent, leaving six constraints for both parity $-(-1)^\ell$ and parity $(-1)^\ell$ tensor structures.

Focusing on fixed $\ell$, we conclude that there is zero (or possibly one if the $\boldsymbol{e}_2$ is conserved although its conservation conditions have not been checked) tensor structure for $\ell=0$, one tensor structure for $\ell=1$, one tensor structure for $\ell=2$, and two tensor structures for $\ell\geq3$.

\subsubsection{\texorpdfstring{$\boldsymbol{e}_3+\ell\boldsymbol{e}_1$}{e3+le1}}

The number of independent tensor structures for exchanged quasi-primary operators in the irreducible representations $\boldsymbol{e}_3+\ell\boldsymbol{e}_1$ is only five.  Taking $F_{1\leq k\leq3}$ as the $\boldsymbol{e}_3$ indices, we select 
\eqna{
(\FCF{1}{i}{j}{m}{1}{2})&=\A_{12;3A_1F_1}\A_{12;3B_1F_2}\A_{12;3A_2B_2}(\A_{12}\cdot\bar{\eta}_3)_{F_3}\left[\prod_{1\leq k\leq\ell}(\A_{12}\cdot\bar{\eta}_3)_{E_k}\right],\\
(\FCF{2}{i}{j}{m}{1}{2})&=\A_{12;3A_1F_1}\A_{12;3B_1F_2}(\A_{12}\cdot\bar{\eta}_3)_{A_2}(\A_{12}\cdot\bar{\eta}_3)_{B_2}(\A_{12}\cdot\bar{\eta}_3)_{F_3}\left[\prod_{1\leq k\leq\ell}(\A_{12}\cdot\bar{\eta}_3)_{E_k}\right],\\
(\FCF{3}{i}{j}{m}{1}{2})&=\A_{12;3A_1F_1}\A_{12;3B_1F_2}[\A_{12;3A_2E_1}(\A_{12}\cdot\bar{\eta}_3)_{B_2}-\A_{12;3B_2E_1}(\A_{12}\cdot\bar{\eta}_3)_{A_2}]\\
&\phantom{=}\qquad\times(\A_{12}\cdot\bar{\eta}_3)_{F_3}\left[\prod_{2\leq k\leq\ell}(\A_{12}\cdot\bar{\eta}_3)_{E_k}\right],\\
(\FCF{4}{i}{j}{m}{1}{2})&=\A_{12;3A_1F_1}\A_{12;3B_1F_2}[\A_{12;3A_2E_1}(\A_{12}\cdot\bar{\eta}_3)_{B_2}+\A_{12;3B_2E_1}(\A_{12}\cdot\bar{\eta}_3)_{A_2}]\\
&\phantom{=}\qquad\times(\A_{12}\cdot\bar{\eta}_3)_{F_3}\left[\prod_{2\leq k\leq\ell}(\A_{12}\cdot\bar{\eta}_3)_{E_k}\right],\\
(\FCF{5}{i}{j}{m}{1}{2})&=\A_{12;3A_1F_1}\A_{12;3B_1F_2}\A_{12;3A_2E_1}\A_{12;3B_2E_2}(\A_{12}\cdot\bar{\eta}_3)_{F_3}\left[\prod_{3\leq k\leq\ell}(\A_{12}\cdot\bar{\eta}_3)_{E_k}\right],
}[EqTSTTe3le1]
with all tensor structures having parity $(-1)^\ell$ expect the fourth that has parity $-(-1)^\ell$.  Moreover, forgetting about parity, \eqref{EqTSTTe3le1} shows that the first two tensor structures exist for any $\ell\geq0$, the third and the fourth appear when $\ell\geq1$, while the fifth necessitates $\ell\geq2$.

Since $N_{\bar{\imath}jm}=N_{i\bar{\jmath}m}=2$, we have two conservation constraints for each parity.  Thus we already expect that the fourth tensor structure, the only one with parity $-(-1)^\ell$, will not appear.  Putting the tensor structures \eqref{EqTSTTe3le1} of the same parity in \eqref{Eq3pt2} using \eqref{Eq3pt2subs} leads to
\eqn{
\begin{gathered}
0=2(d+\ell+1)\aCF{1}{i}{j}{m}+4(d-\Delta_m)\aCF{2}{i}{j}{m}-2(\Delta_m+\ell+1)\aCF{3}{i}{j}{m}+(d+\ell+1)\aCF{5}{i}{j}{m},\\
0=2\ell\aCF{1}{i}{j}{m}+2(d-1-\Delta_m)\aCF{3}{i}{j}{m}+(d+\ell)\aCF{5}{i}{j}{m},
\end{gathered}
}[EqcTcTe3le1even]
for the tensor structures with parity $(-1)^\ell$ as well as
\eqn{0=(2d+\ell+1-\Delta_m)\aCF{4}{i}{j}{m},\qquad0=(d+1-\Delta_m)\aCF{4}{i}{j}{m},}[EqcTcTe3le1odd]
for the tensor structures with parity $-(-1)^\ell$.  As expected, the fourth tensor structure never appears, therefore there are no tensor structures for odd $\ell$.  Moreover, for general $\ell$, the two conservation constraints \eqref{EqcTcTe3le1even} are linearly independent, which implies that there are two tensor structures, both with parity $(-1)^\ell$.

It is easy to see that the second constraint in \eqref{EqcTcTe3le1even} disappears when $\ell=0$, hence there is only one tensor structure for $\ell=0$.  For $\ell\geq1$ there are no tensor structures if $\ell$ is odd and two tensor structures if $\ell$ is even.

\subsubsection{\texorpdfstring{$2\boldsymbol{e}_2+\ell\boldsymbol{e}_1$}{2e2+le1}}

Considering exchanged quasi-primary operators in the irreducible representations $2\boldsymbol{e}_2+\ell\boldsymbol{e}_1$, there are eleven tensor structures.  Using $F_1$ and $F_2$ for the indices of the first $\boldsymbol{e}_2$ and $G_1$ and $G_2$ for the indices of the second $\boldsymbol{e}_2$, one choice of tensor structures is
\eqna{
(\FCF{1}{i}{j}{m}{1}{2})&=\A_{12;3A_1F_1}\A_{12;3B_1F_2}\A_{12;3A_2G_1}\A_{12;3B_2G_2}\left[\prod_{1\leq k\leq\ell}(\A_{12}\cdot\bar{\eta}_3)_{E_k}\right],\\
(\FCF{2}{i}{j}{m}{1}{2})&=\A_{12;3A_1F_1}\A_{12;3B_1F_2}[\A_{12;3A_2G_1}(\A_{12}\cdot\bar{\eta}_3)_{B_2}+\A_{12;3B_2G_1}(\A_{12}\cdot\bar{\eta}_3)_{A_2}]\\
&\phantom{=}\qquad\times(\A_{12}\cdot\bar{\eta}_3)_{G_2}\left[\prod_{1\leq k\leq\ell}(\A_{12}\cdot\bar{\eta}_3)_{E_k}\right],\\
(\FCF{4}{i}{j}{m}{1}{2})&=\A_{12;3A_1F_1}\A_{12;3B_1G_1}\A_{12;3A_2B_2}(\A_{12}\cdot\bar{\eta}_3)_{F_2}(\A_{12}\cdot\bar{\eta}_3)_{G_2}\left[\prod_{1\leq k\leq\ell}(\A_{12}\cdot\bar{\eta}_3)_{E_k}\right],
}
\eqna{
(\FCF{5}{i}{j}{m}{1}{2})&=\A_{12;3A_1F_1}\A_{12;3B_1G_1}(\A_{12}\cdot\bar{\eta}_3)_{A_2}(\A_{12}\cdot\bar{\eta}_3)_{B_2}(\A_{12}\cdot\bar{\eta}_3)_{F_2}(\A_{12}\cdot\bar{\eta}_3)_{G_2}\left[\prod_{1\leq k\leq\ell}(\A_{12}\cdot\bar{\eta}_3)_{E_k}\right],\\
(\FCF{6}{i}{j}{m}{1}{2})&=[\A_{12;3A_1F_1}\A_{12;3A_2G_1}(\A_{12}\cdot\bar{\eta}_3)_{B_1}(\A_{12}\cdot\bar{\eta}_3)_{B_2}+\A_{12;3B_1F_1}\A_{12;3B_2G_1}(\A_{12}\cdot\bar{\eta}_3)_{A_1}(\A_{12}\cdot\bar{\eta}_3)_{A_2}]\\
&\phantom{=}\qquad\times(\A_{12}\cdot\bar{\eta}_3)_{F_2}(\A_{12}\cdot\bar{\eta}_3)_{G_2}\left[\prod_{1\leq k\leq\ell}(\A_{12}\cdot\bar{\eta}_3)_{E_k}\right],\\
(\FCF{8}{i}{j}{m}{1}{2})&=[\A_{12;3A_1F_1}\A_{12;3A_2G_1}\A_{12;3B_1E_1}(\A_{12}\cdot\bar{\eta}_3)_{B_2}-\A_{12;3B_1F_1}\A_{12;3B_2G_1}\A_{12;3A_1E_1}(\A_{12}\cdot\bar{\eta}_3)_{A_2}]\\
&\phantom{=}\qquad\times(\A_{12}\cdot\bar{\eta}_3)_{F_2}(\A_{12}\cdot\bar{\eta}_3)_{G_2}\left[\prod_{2\leq k\leq\ell}(\A_{12}\cdot\bar{\eta}_3)_{E_k}\right],\\
(\FCF{11}{i}{j}{m}{1}{2})&=\A_{12;3A_1F_1}\A_{12;3B_1G_1}\A_{12;3A_2E_1}\A_{12;3B_2E_2}(\A_{12}\cdot\bar{\eta}_3)_{F_2}(\A_{12}\cdot\bar{\eta}_3)_{G_2}\left[\prod_{3\leq k\leq\ell}(\A_{12}\cdot\bar{\eta}_3)_{E_k}\right],
}[EqTSTT2e2le1even]
for parity $(-1)^\ell$ and
\eqna{
(\FCF{3}{i}{j}{m}{1}{2})&=\A_{12;3A_1F_1}\A_{12;3B_1F_2}[\A_{12;3A_2G_1}(\A_{12}\cdot\bar{\eta}_3)_{B_2}-\A_{12;3B_2G_1}(\A_{12}\cdot\bar{\eta}_3)_{A_2}]\\
&\phantom{=}\qquad\times(\A_{12}\cdot\bar{\eta}_3)_{G_2}\left[\prod_{1\leq k\leq\ell}(\A_{12}\cdot\bar{\eta}_3)_{E_k}\right],\\
(\FCF{7}{i}{j}{m}{1}{2})&=[\A_{12;3A_1F_1}\A_{12;3A_2G_1}(\A_{12}\cdot\bar{\eta}_3)_{B_1}(\A_{12}\cdot\bar{\eta}_3)_{B_2}-\A_{12;3B_1F_1}\A_{12;3B_2G_1}(\A_{12}\cdot\bar{\eta}_3)_{A_1}(\A_{12}\cdot\bar{\eta}_3)_{A_2}]\\
&\phantom{=}\qquad\times(\A_{12}\cdot\bar{\eta}_3)_{F_2}(\A_{12}\cdot\bar{\eta}_3)_{G_2}\left[\prod_{1\leq k\leq\ell}(\A_{12}\cdot\bar{\eta}_3)_{E_k}\right],\\
(\FCF{9}{i}{j}{m}{1}{2})&=[\A_{12;3A_1F_1}\A_{12;3A_2G_1}\A_{12;3B_1E_1}(\A_{12}\cdot\bar{\eta}_3)_{B_2}+\A_{12;3B_1F_1}\A_{12;3B_2G_1}\A_{12;3A_1E_1}(\A_{12}\cdot\bar{\eta}_3)_{A_2}]\\
&\phantom{=}\qquad\times(\A_{12}\cdot\bar{\eta}_3)_{F_2}(\A_{12}\cdot\bar{\eta}_3)_{G_2}\left[\prod_{2\leq k\leq\ell}(\A_{12}\cdot\bar{\eta}_3)_{E_k}\right],\\
(\FCF{10}{i}{j}{m}{1}{2})&=\A_{12;3A_1F_1}\A_{12;3B_1F_2}[\A_{12;3A_2G_1}\A_{12;3B_2E_1}+\A_{12;3B_2G_1}\A_{12;3A_2E_1}]\\
&\phantom{=}\qquad\times(\A_{12}\cdot\bar{\eta}_3)_{G_2}\left[\prod_{2\leq k\leq\ell}(\A_{12}\cdot\bar{\eta}_3)_{E_k}\right],
}[EqTSTT2e2le1odd]
for parity $-(-1)^\ell$.  From \eqref{EqTSTT2e2le1even} and \eqref{EqTSTT2e2le1odd}, there are six tensor structures for $\ell=0$, ten tensor structures for $\ell=1$, and eleven tensor structures for $\ell\geq2$ when parity is not taken into account.

With $N_{\bar{\imath}jm}=N_{i\bar{\jmath}m}=4$, there are four conservation conditions for each parity type.  Once again, we compute the conservation conditions starting from the tensor structures \eqref{EqTSTT2e2le1even} and \eqref{EqTSTT2e2le1odd} with the help of \eqref{Eq3pt2} and \eqref{Eq3pt2subs}, for both parities, and obtain
\eqna{
0&=2[d(d+\ell)-\Delta_m-\ell-2]\aCF{1}{i}{j}{m}+2d(2d+\ell-\Delta_m)(\aCF{2}{i}{j}{m}+\aCF{8}{i}{j}{m})+2[d-2(\Delta_m+\ell+2)]\aCF{4}{i}{j}{m}\\
&\phantom{=}\qquad+4d\aCF{5}{i}{j}{m}+8[d(d-1-\Delta_m)+\Delta_m+\ell+2]\aCF{6}{i}{j}{m}+d\aCF{11}{i}{j}{m},\\
0&=2(d+\ell)\aCF{1}{i}{j}{m}-2(\Delta_m+\ell-2)\aCF{2}{i}{j}{m}-2(d+\ell+2)\aCF{4}{i}{j}{m}\\
&\phantom{=}\qquad-4(d-\Delta_m)\aCF{5}{i}{j}{m}-16\aCF{6}{i}{j}{m}+8\aCF{8}{i}{j}{m}-(d+\ell+2)\aCF{11}{i}{j}{m},\\
0&=2[d(d+\ell-1)-2]\aCF{1}{i}{j}{m}+2d(d+1-\Delta_m)\aCF{2}{i}{j}{m}-2[d(\ell+1)+4]\aCF{4}{i}{j}{m}\\
&\phantom{=}\qquad+16\aCF{6}{i}{j}{m}+4d\aCF{8}{i}{j}{m}-d(d+\ell+1)\aCF{11}{i}{j}{m},\\
0&=2\ell(\aCF{1}{i}{j}{m}-4\aCF{6}{i}{j}{m})-2(d-2)\ell\aCF{4}{i}{j}{m}-2d(d-2-\Delta_m)\aCF{8}{i}{j}{m}-d(d+\ell+2)\aCF{11}{i}{j}{m},
}[EqcTcT2e2le1even]
for the tensor structures with parity $(-1)^\ell$ as well as
\eqna{
0&=d(2d+\ell-2-\Delta_m)\aCF{3}{i}{j}{m}+4[d(d-1-\Delta_m)+\Delta_m+\ell+2]\aCF{7}{i}{j}{m}\\
&\phantom{=}\qquad+d(2d+\ell-\Delta_m)\aCF{9}{i}{j}{m}+[d(d+\ell+1)-\Delta_m-\ell-2]\aCF{10}{i}{j}{m},\\
0&=(2d+\ell-2-\Delta_m)\aCF{3}{i}{j}{m}+8\aCF{7}{i}{j}{m}-4\aCF{9}{i}{j}{m}+2\aCF{10}{i}{j}{m},\\
0&=d(d-1-\Delta_m)\aCF{3}{i}{j}{m}+8\aCF{7}{i}{j}{m}-2d\aCF{9}{i}{j}{m}+2(d-1)\aCF{10}{i}{j}{m},\\
0&=4\ell\aCF{7}{i}{j}{m}+d(d+2-\Delta_m)\aCF{9}{i}{j}{m}+[d(d+\ell-2)-\ell]\aCF{10}{i}{j}{m},
}[EqcTcT2e2le1odd]
for the tensor structures with parity $-(-1)^\ell$.  With large enough $\ell$, none of the four constraints in \eqref{EqcTcT2e2le1even} are linearly dependent while one constraint in \eqref{EqcTcT2e2le1odd} is linearly dependent.  Thus we are left with three parity $(-1)^\ell$ tensor structures and one parity $-(-1)^\ell$ tensor structure.

In general, for even $\ell$ there are two tensor structures when $\ell=0$ and three tensor structures when $\ell\geq2$, while for odd $\ell$ there is only one tensor structure.

\subsubsection{\texorpdfstring{$\boldsymbol{e}_2+\boldsymbol{e}_3+\ell\boldsymbol{e}_1$}{e2+e3+le1}}

Considering exchanged quasi-primary operators in the irreducible representations $\boldsymbol{e}_2+\boldsymbol{e}_3+\ell\boldsymbol{e}_1$, there are four independent tensor structures.  With $F_{1\leq k\leq2}$ the $\boldsymbol{e}_2$ indices and $G_{1\leq k\leq3}$ the $\boldsymbol{e}_3$ indices, they can be chosen to be
\eqna{
(\FCF{1}{i}{j}{m}{1}{2})&=\A_{12;3A_1G_1}\A_{12;3B_1G_2}\A_{12;3A_2F_1}\A_{12;3B_2F_2}(\A_{12}\cdot\bar{\eta}_3)_{G_3}\left[\prod_{1\leq k\leq\ell}(\A_{12}\cdot\bar{\eta}_3)_{E_k}\right],\\
(\FCF{2}{i}{j}{m}{1}{2})&=\A_{12;3A_1G_1}\A_{12;3B_1G_2}[\A_{12;3A_2F_1}(\A_{12}\cdot\bar{\eta}_3)_{B_2}+\A_{12;3B_2F_1}(\A_{12}\cdot\bar{\eta}_3)_{A_2}]\\
&\phantom{=}\qquad\times(\A_{12}\cdot\bar{\eta}_3)_{F_2}(\A_{12}\cdot\bar{\eta}_3)_{G_3}\left[\prod_{1\leq k\leq\ell}(\A_{12}\cdot\bar{\eta}_3)_{E_k}\right],\\
(\FCF{3}{i}{j}{m}{1}{2})&=\A_{12;3A_1G_1}\A_{12;3B_1G_2}[\A_{12;3A_2F_1}(\A_{12}\cdot\bar{\eta}_3)_{B_2}-\A_{12;3B_2F_1}(\A_{12}\cdot\bar{\eta}_3)_{A_2}]\\
&\phantom{=}\qquad\times(\A_{12}\cdot\bar{\eta}_3)_{F_2}(\A_{12}\cdot\bar{\eta}_3)_{G_3}\left[\prod_{1\leq k\leq\ell}(\A_{12}\cdot\bar{\eta}_3)_{E_k}\right],\\
(\FCF{4}{i}{j}{m}{1}{2})&=\A_{12;3A_1G_1}\A_{12;3B_1G_2}[\A_{12;3A_2F_1}\A_{12;3B_2E_1}+\A_{12;3B_2F_1}\A_{12;3A_2E_1}]\\
&\phantom{=}\qquad\times(\A_{12}\cdot\bar{\eta}_3)_{F_2}(\A_{12}\cdot\bar{\eta}_3)_{G_3}\left[\prod_{2\leq k\leq\ell}(\A_{12}\cdot\bar{\eta}_3)_{E_k}\right],
}[EqTSTTe2e3le1]
where the first two tensor structures have parity $-(-1)^\ell$ while the last two have parity $(-1)^\ell$.  Clearly, from \eqref{EqTSTTe2e3le1} there are three tensor structures when $\ell=0$ and four tensor structures when $\ell\geq1$ if parity is not considered.

Since $N_{\bar{\imath}jm}=N_{i\bar{\jmath}m}=1$, there is only one conservation condition by parity.  Employing \eqref{Eq3pt2} with \eqref{Eq3pt2subs}, the conservation conditions for the tensor structures \eqref{EqTSTTe2e3le1} are
\eqn{
\begin{gathered}
0=(d+\ell)\aCF{1}{i}{j}{m}+2(d+1-\Delta_m)\aCF{2}{i}{j}{m},\\
0=2(d-1-\Delta_m)\aCF{3}{i}{j}{m}+(d+\ell+2)\aCF{4}{i}{j}{m}.
\end{gathered}
}[EqcTcTe2e3le1]
Hence, generically there is only one tensor structure per parity.  We note that for $\ell=0$, the second conservation constraint becomes $(d-1-\Delta_m)\aCF{3}{i}{j}{m}=0$ which is satisfied trivially for a conserved $\boldsymbol{e}_2+\boldsymbol{e}_3$.

As a consequence of \eqref{EqcTcTe2e3le1}, there are no tensor structures when $\ell=0$ unless the $\boldsymbol{e}_2+\boldsymbol{e}_3$ is conserved in which case the number of tensor structures may be one (not taking into account its own conservation conditions), and one tensor structure when $\ell\geq1$.

\subsubsection{\texorpdfstring{$3\boldsymbol{e}_2+\ell\boldsymbol{e}_1$}{3e2+le1}}

The final non-trivial case consists of exchanged quasi-primary operators in the irreducible representations $3\boldsymbol{e}_2+\ell\boldsymbol{e}_1$, for which there are four independent tensor structures.  Taking $F_{1\leq k\leq2}$, $G_{1\leq k\leq2}$ and $H_{1\leq k\leq2}$ as the three groups of $\boldsymbol{e}_2$ indices, we choose the tensor structures to be
\eqna{
(\FCF{1}{i}{j}{m}{1}{2})&=\A_{12;3A_1F_1}\A_{12;3B_1F_2}\A_{12;3A_2G_1}\A_{12;3B_2H_1}(\A_{12}\cdot\bar{\eta}_3)_{G_2}(\A_{12}\cdot\bar{\eta}_3)_{H_2}\left[\prod_{1\leq k\leq\ell}(\A_{12}\cdot\bar{\eta}_3)_{E_k}\right],\\
(\FCF{2}{i}{j}{m}{1}{2})&=[\A_{12;3A_1F_1}(\A_{12}\cdot\bar{\eta}_3)_{B_1}+\A_{12;3B_1F_1}(\A_{12}\cdot\bar{\eta}_3)_{A_1}]\A_{12;3A_2G_1}\A_{12;3B_2H_1}\\
&\phantom{=}\qquad\times(\A_{12}\cdot\bar{\eta}_3)_{F_2}(\A_{12}\cdot\bar{\eta}_3)_{G_2}(\A_{12}\cdot\bar{\eta}_3)_{H_2}\left[\prod_{1\leq k\leq\ell}(\A_{12}\cdot\bar{\eta}_3)_{E_k}\right],\\
(\FCF{3}{i}{j}{m}{1}{2})&=[\A_{12;3A_1F_1}(\A_{12}\cdot\bar{\eta}_3)_{B_1}-\A_{12;3B_1F_1}(\A_{12}\cdot\bar{\eta}_3)_{A_1}]\A_{12;3A_2G_1}\A_{12;3B_2H_1}\\
&\phantom{=}\qquad\times(\A_{12}\cdot\bar{\eta}_3)_{F_2}(\A_{12}\cdot\bar{\eta}_3)_{G_2}(\A_{12}\cdot\bar{\eta}_3)_{H_2}\left[\prod_{1\leq k\leq\ell}(\A_{12}\cdot\bar{\eta}_3)_{E_k}\right],\\
(\FCF{4}{i}{j}{m}{1}{2})&=\A_{12;3A_1F_1}\A_{12;3B_1G_1}[\A_{12;3A_2H_1}\A_{12;3B_2E_1}+\A_{12;3B_2H_1}\A_{12;3A_2E_1}]\\
&\phantom{=}\qquad\times(\A_{12}\cdot\bar{\eta}_3)_{F_2}(\A_{12}\cdot\bar{\eta}_3)_{G_2}(\A_{12}\cdot\bar{\eta}_3)_{H_2}\left[\prod_{2\leq k\leq\ell}(\A_{12}\cdot\bar{\eta}_3)_{E_k}\right].
}[EqTSTT3e2le1]
Here the parity of the first two tensor structures is $-(-1)^\ell$ and the parity of the last two tensor structures is $(-1)^\ell$.  Moreover, the fourth tensor structure does not exist when $\ell=0$.

For each parity, there is only one conservation condition since $N_{\bar{\imath}jm}=N_{i\bar{\jmath}m}=1$.  Thus, plugging \eqref{EqTSTT3e2le1} in \eqref{Eq3pt2} with the help of \eqref{Eq3pt2subs} leads to the conservation conditions
\eqn{
\begin{gathered}
0=(d+\ell)\aCF{1}{i}{j}{m}+2(d+2-\Delta_m)\aCF{2}{i}{j}{m},\\
0=2(d-2-\Delta_m)\aCF{3}{i}{j}{m}+(d+\ell+4)\aCF{4}{i}{j}{m},
\end{gathered}
}[EqcTcT3e2le1]
which implies that there is in general only one tensor structure per parity.

In conclusion, there would be only one tensor structure when $\ell=0$ but its parity is $-(-1)^\ell$, hence the number of tensor structures for $\ell=0$ is zero while there is one tensor structure when $\ell\geq1$.


\section{Conserved Vector Currents and Energy-Momentum Tensors: Coincident Points}\label{SAppWard}

For completeness, in this appendix we discuss the conformal Ward identities at coincident points for conserved vector currents and the energy-momentum tensor.  Thus we study three-point correlation functions of the form $\Vev{\mathcal{O}\mathcal{O}J}$ and $\Vev{\mathcal{O}\mathcal{O}T}$.  We consider several examples, including $\mathcal{O}$ in some defining irreducible representations as well as $\mathcal{O}=J$ and $\mathcal{O}=T$ (see \cite{Osborn:1993cr}).

In general, for a conserved vector current $J$ it is possible to use invariance under special conformal transformations to show that conformal Ward identities at coincident points relate $\Vev{\mathcal{O}\mathcal{O}J}$ to $\Vev{\delta_J\mathcal{O}\mathcal{O}}$ where $\delta_J\mathcal{O}$ is the variation of $\mathcal{O}$ under a $J$-symmetry transformation.  In other words, the three-point coefficients of $\Vev{\mathcal{O}\mathcal{O}J}$ can be related to the charge under the symmetry generated by $J$ of the quasi-primary operator $\mathcal{O}$ and the two-point coefficient of $\Vev{\mathcal{O}\mathcal{O}}$ \cite{Cardy:1987dg,Osborn:1993cr}, as well as some pre-factors including the area $S_d=2\pi^{d/2}/\Gamma(d/2)$ of the $(d-1)$-dimensional sphere.\footnote{This statement is also true for the dilatation current built from the contraction of the energy-momentum tensor and its position space coordinate $x\cdot T$, with the charge of $\mathcal{O}$ under the dilatation symmetry being the conformal dimension $\Delta_\mathcal{O}$.}  Formally, in position space this statement corresponds to
\eqn{\int_{B_1}d^dx_3\,\partial_{3\mu}\Vev{\mathcal{O}^{(x)}(x_1)\mathcal{O}^{(x)}(x_2)J^{(x)\mu}(x_3)}=\Vev{\delta_J\mathcal{O}^{(x)}(x_1)\mathcal{O}^{(x)}(x_2)},}[EqCP]
where $B_1$ is a ball centered at $x_1$ that does not encompass $x_2$.  We stress that this occurs only when the two remaining quasi-primary operators have the same conformal dimension and are in the same irreducible representation.  In other cases the conformal Ward identities at coincident points for $J$ and $x\cdot T$ are verified by the conformal Ward identities at non-coincident points.

Following the usual procedure in position space, stated in \eqref{EqCP}, but for conserved vector currents and the energy-momentum tensor in the embedding space formalism, we find that
\eqna{
\Vev{\delta_J\Op{i}{1}\Op{j}{2}}&=\frac{(\mathcal{T}_{12}^{\boldsymbol{N}_i}\Gamma)^{\{Aa\}}(\mathcal{T}_{21}^{\boldsymbol{N}_j}\Gamma)^{\{Bb\}}}{\ee{1}{2}{\tau_\mathcal{O}}}(-2)^{(d-2)/2}\\
&\phantom{=}\qquad\times\left[\left.\sum_{r=1}^{N_{ijm}}\aCF{r}{i}{j}{m}(\FCF{r}{i}{j}{m}{1}{2})_{\{aA\}\{bB\}E}\right|_{\substack{\A_{12;3FE}\to0\\(\A_{12}\cdot\bar{\eta}_3)_E\to-1}}\right]_S,
}[EqOOJ]
and
\eqna{
\Vev{\delta_{x\cdot T}\Op{i}{1}\Op{j}{2}}&=\frac{(\mathcal{T}_{12}^{\boldsymbol{N}_i}\Gamma)^{\{Aa\}}(\mathcal{T}_{21}^{\boldsymbol{N}_j}\Gamma)^{\{Bb\}}}{\ee{1}{2}{\tau_\mathcal{O}}}(-2)^{(d-2)/2}\\
&\qquad\times\left[\left.\sum_{r=1}^{N_{ijm}}\aCF{r}{i}{j}{m}(\FCF{r}{i}{j}{m}{1}{2})_{\{aA\}\{bB\}E_2E_1}\right|_{\substack{\A_{12;3F_1E_1}\A_{12;3F_2E_2}\to\frac{1}{2d}\A_{12;3F_1F_2}\\\A_{12;3FE_1}(\A_{12}\cdot\bar{\eta}_3)_{E_2}\to0\\(\A_{12}\cdot\bar{\eta}_3)_{E_1}\A_{12;3FE_2}\to0\\(\A_{12}\cdot\bar{\eta}_3)_{E_1}(\A_{12}\cdot\bar{\eta}_3)_{E_2}\to\frac{d-1}{d}}}\right]_S,
}[EqOOT]
with $\Op{i}{1}$ and $\Op{j}{2}$ sharing the same conformal dimension $\Delta_i=\Delta_j=\Delta_\mathcal{O}$ and the same irreducible representation $\boldsymbol{N}_i=\boldsymbol{N}_j=\boldsymbol{N}^\mathcal{O}$.\footnote{More precisely, in irreducible representations that are contragredient reflections of each other.}  When comparing to \cite{Osborn:1993cr}, we note that the peculiar factors $(-2)^{(d-2)/2}$ in \eqref{EqOOJ} and \eqref{EqOOT} appear from the embedding space formalism, where $-2\ee{1}{2}{}\propto(x_1-x_2)^2$.  These factors disappear when using the standard normalization suited for position space, as in \cite{Osborn:1993cr}.

The last substitution in \eqref{EqOOJ} and \eqref{EqOOT}, denoted $S$, implements integration along the $(d-1)$-dimensional sphere that bounds the ball $B_1$ and corresponds to
\eqn{
\begin{gathered}
\prod_{1\leq k\leq2n-1}(\A_{12}\cdot\bar{\eta}_3)_{F_k}\to0,\\
\prod_{1\leq k\leq2n}(\A_{12}\cdot\bar{\eta}_3)_{F_k}\to\frac{(-2)^n(1/2)_n}{(d/2)_n}S_d\A_{12(F_1F_2}\cdots\A_{12F_{2n-1}F_{2n})}.
\end{gathered}
}[EqS]
We note that the tensor structures must first be re-expressed in terms of $\A_{12}$ using the definition \eqref{EqAp} for $\A_{12;3}$ before carrying out the $S$-substitution \eqref{EqS}.

Finally, in the following, we normalize the necessary two-point correlation functions as
\eqn{\Vev{\Op{i}{1}\Op{j}{2}}=(\mathcal{T}_{12}^{\boldsymbol{N}_i}\Gamma)\cdot(\mathcal{T}_{21}^{\boldsymbol{N}_j}\Gamma)\frac{c_\mathcal{O}}{\ee{1}{2}{\tau_\mathcal{O}}},\qquad\cOPE{}{i}{j}{\1}=c_\mathcal{O}\delta_{ij},}[EqOO]
for unconserved quasi-primary operators,
\eqn{\Vev{J_i(\eta_1)J_j(\eta_2)}=(\mathcal{T}_{12}^{\boldsymbol{e}_1}\Gamma)\cdot(\mathcal{T}_{21}^{\boldsymbol{e}_1}\Gamma)\frac{c_J\delta_{ij}}{\ee{1}{2}{d-1}},\qquad\cOPE{}{i}{j}{\1}=c_J\delta_{ij},}[EqJJ]
for conserved vector currents, as well as
\eqn{\Vev{T(\eta_1)T(\eta_2)}=(\mathcal{T}_{12}^{2\boldsymbol{e}_1}\Gamma)\cdot(\mathcal{T}_{21}^{2\boldsymbol{e}_1}\Gamma)\frac{c_T}{\ee{1}{2}{d}},\qquad\cOPE{}{T}{T}{\1}=c_T,}[EqTT]
for the energy-momentum tensor.  Clearly, $c_\mathcal{O}$ can be set to $1$ by rescaling the unconserved quasi-primary operators $\mathcal{O}$.  This is not the case for the conserved vector currents and the energy-momentum tensor, hence $c_J$ and $c_T$ are genuine quantities of CFTs.

Before proceeding, we also determine the behavior of the tensor structure building blocks $\A_{12;3}$ [see \eqref{EqAp}] and $\A_{12}\cdot\bar{\eta}_3$ under the exchange symmetry of the last two quasi-primary operators which is relevant when the three quasi-primary operators are in the same irreducible representation.  Under the exchange symmetry of the first two quasi-primary operators, the tensor structure building blocks transform as in \eqref{EqES12} while under the exchange symmetry of the last two quasi-primary operators we have
\eqn{
\begin{gathered}
\A_{12;3}^{AB}\to\A_{12;3}^{AE},\qquad\A_{12;3}^{AE}\to\A_{12;3}^{AB},\qquad(\A_{12}\cdot\bar{\eta}_3)_A\to-(\A_{12}\cdot\bar{\eta}_3)_A,\\
(\A_{12}\cdot\bar{\eta}_3)_B\to-(\A_{12}\cdot\bar{\eta}_3)_E,\qquad(\A_{12}\cdot\bar{\eta}_3)_E\to-(\A_{12}\cdot\bar{\eta}_3)_B,
\end{gathered}
}[EqES23]
with the unspecified transformations being trivial.  Since \eqref{EqES12} and \eqref{EqES23} generate the exchange symmetry group $S_3$, the other transformations are not necessary.


\subsection{\texorpdfstring{$\Vev{\mathcal{O}\mathcal{O}J}$ with $\mathcal{O}$ in the $n$-index antisymmetric representation}{OOJ}}

Considering $\Vev{\Op{i}{1}\Op{j}{2}J_m(\eta_3)}$ with $\mathcal{O}_i$ and $\mathcal{O}_j$ in the $n$-index antisymmetric representation, there are four (only one for scalars where $n=0$) tensor structures that we choose to be
\eqn{
\begin{gathered}
(\FCF{1}{i}{j}{m}{1}{2})=(\A_{12}\cdot\bar{\eta}_3)_E\left[\prod_{1\leq k\leq n}\A_{12A_kB_k}\right],\\
(\FCF{2}{i}{j}{m}{1}{2})=(\A_{12}\cdot\bar{\eta}_3)_{A_1}(\A_{12}\cdot\bar{\eta}_3)_{B_1}(\A_{12}\cdot\bar{\eta}_3)_E\left[\prod_{2\leq k\leq n}\A_{12A_kB_k}\right],\\
(\FCF{3}{i}{j}{m}{1}{2})=[\A_{12;3A_1E}(\A_{12}\cdot\bar{\eta}_3)_{B_1}-\A_{12;3B_1E}(\A_{12}\cdot\bar{\eta}_3)_{A_1}]\left[\prod_{2\leq k\leq n}\A_{12A_kB_k}\right],\\
(\FCF{4}{i}{j}{m}{1}{2})=[\A_{12;3A_1E}(\A_{12}\cdot\bar{\eta}_3)_{B_1}+\A_{12;3B_1E}(\A_{12}\cdot\bar{\eta}_3)_{A_1}]\left[\prod_{2\leq k\leq n}\A_{12A_kB_k}\right].
\end{gathered}
}[EqTSOOJ]
From \eqref{EqES12}, the first three tensor structures are odd while the fourth has even parity under the exchange symmetry group.  Moreover, to simplify the analysis, we took $\A_{12AB}$ instead of $\A_{12;3AB}$.

The conservation conditions at non-coincident points are obtained directly from \eqref{Eq3pt3} and \eqref{Eq3pt3subs}.  With the tensor structures \eqref{EqTSOOJ}, they are ($N_{ij\bar{m}}=2$ for $n\geq1$)
\eqn{d\aCF{4}{i}{j}{m}=0,\qquad\aCF{4}{i}{j}{m}=0,}
which imply that the fourth tensor structure does not occur.  Looking for the conservation conditions at coincident points, the rules of \eqref{EqOOJ} and \eqref{EqS} lead to
\eqn{-\frac{(-2)^{(d-2)/2}}{d}[d\aCF{1}{i}{j}{m}-2\aCF{2}{i}{j}{m}]S_d=c_\mathcal{O}(T_m)_{ij},}[EqOOJcp]
taking into account that $\delta_{J_m}\mathcal{O}_i=(T_m)_{ij}\mathcal{O}_j$ and the two-point correlation functions \eqref{EqOO}.  For $n=0$, \eqref{EqOOJcp} says that the sole three-point coefficient (which is $\aCF{1}{i}{j}{m}$) is proportional to its charge $(T_m)_{ij}$ under the $J$-symmetry.


\subsection{\texorpdfstring{$\Vev{\mathcal{O}\mathcal{O}T}$ with $\mathcal{O}$ in the $n$-index antisymmetric representation}{OOT}}

The same exercise can be performed for $\mathcal{O}_i$ and $\mathcal{O}_j$ in the $n$-index antisymmetric representation and the energy-momentum tensor, \textit{i.e.} $\Vev{\Op{i}{1}\Op{j}{2}T(\eta_3)}$.  In this case, there are six (only one for scalars where $n=0$ and five for vectors where $n=1$) tensor structures that we select as
\eqn{
\begin{gathered}
(\FCF{1}{i}{j}{m}{1}{2})=(\A_{12}\cdot\bar{\eta}_3)_{E_1}(\A_{12}\cdot\bar{\eta}_3)_{E_2}\left[\prod_{1\leq k\leq n}\A_{12A_kB_k}\right],\\
(\FCF{2}{i}{j}{m}{1}{2})=(\A_{12}\cdot\bar{\eta}_3)_{A_1}(\A_{12}\cdot\bar{\eta}_3)_{B_1}(\A_{12}\cdot\bar{\eta}_3)_{E_1}(\A_{12}\cdot\bar{\eta}_3)_{E_2}\left[\prod_{2\leq k\leq n}\A_{12A_kB_k}\right],\\
(\FCF{3}{i}{j}{m}{1}{2})=[\A_{12;3A_1E_1}(\A_{12}\cdot\bar{\eta}_3)_{B_1}-\A_{12;3B_1E_1}(\A_{12}\cdot\bar{\eta}_3)_{A_1}](\A_{12}\cdot\bar{\eta}_3)_{E_2}\left[\prod_{2\leq k\leq n}\A_{12A_kB_k}\right],\\
(\FCF{4}{i}{j}{m}{1}{2})=[\A_{12;3A_1E_1}(\A_{12}\cdot\bar{\eta}_3)_{B_1}+\A_{12;3B_1E_1}(\A_{12}\cdot\bar{\eta}_3)_{A_1}](\A_{12}\cdot\bar{\eta}_3)_{E_2}\left[\prod_{2\leq k\leq n}\A_{12A_kB_k}\right],\\
(\FCF{5}{i}{j}{m}{1}{2})=\A_{12;3A_1E_1}\A_{12;3B_1E_2}\left[\prod_{2\leq k\leq n}\A_{12A_kB_k}\right],\\
(\FCF{6}{i}{j}{m}{1}{2})=\A_{12;3A_1E_1}\A_{12;3B_1E_2}(\A_{12}\cdot\bar{\eta}_3)_{A_2}(\A_{12}\cdot\bar{\eta}_3)_{B_2}\left[\prod_{3\leq k\leq n}\A_{12A_kB_k}\right].
\end{gathered}
}[EqTSOOT]
From the behavior under the exchange symmetry group \eqref{EqES12}, all tensor structures are even apart from the fourth tensor structure that is odd.  Again we chose $\A_{12AB}$ instead of $\A_{12;3AB}$ to simplify the subsequent analysis.

Looking at the conservation conditions at non-coincident points with the help of \eqref{Eq3pt3}, \eqref{Eq3pt3subs}, and the tensor structures \eqref{EqTSOOT}, we find the following constraints ($N_{ij\bar{m}}=4$ for $n\geq1$)
\eqn{8\aCF{2}{i}{j}{m}+(d^2-2)\aCF{5}{i}{j}{m}-2(d+2)\aCF{6}{i}{j}{m}=0,\qquad\aCF{4}{i}{j}{m}=0,}
where two constraints are automatically satisfied since $\Delta_i=\Delta_j=\Delta_\mathcal{O}$.  Therefore, the fourth tensor structure cannot appear in the three-point correlation functions $\Vev{\Op{i}{1}\Op{j}{2}T(\eta_3)}$.  Considering conservation conditions at coincident points instead, the general method of \eqref{EqOOT} and \eqref{EqS} implies that
\eqn{\frac{(-2)^{(d-2)/2}}{2d^2}\{(d-1)[2d\aCF{1}{i}{j}{m}-4\aCF{2}{i}{j}{m}+\aCF{5}{i}{j}{m}]-2\aCF{6}{i}{j}{m}\}S_d=\Delta_\mathcal{O}c_\mathcal{O},}[EqOOTcp]
where $\delta_{x\cdot T}\mathcal{O}_i=\Delta_\mathcal{O}$.  For $n=0$, there is only one three-point coefficient (namely $\aCF{1}{i}{j}{m}$) and \eqref{EqOOTcp} gives the correct relation with the conformal dimension \cite{Cardy:1987dg}.


\subsection{\texorpdfstring{$\Vev{JJJ}$}{JJJ}}

With three conserved vector currents $\Vev{J_i(\eta_1)J_j(\eta_2)J_m(\eta_3)}$, the tensor structures can be chosen as in \eqref{EqTSJJle1} with $\ell=1$, \textit{i.e.}\
\eqn{
\begin{gathered}
(\FCF{1}{i}{j}{m}{1}{2})=\A_{12;3AB}(\A_{12}\cdot\bar{\eta}_3)_E,\\
(\FCF{2}{i}{j}{m}{1}{2})=(\A_{12}\cdot\bar{\eta}_3)_A(\A_{12}\cdot\bar{\eta}_3)_B(\A_{12}\cdot\bar{\eta}_3)_E,\\
(\FCF{3}{i}{j}{m}{1}{2})=\A_{12;3AE}(\A_{12}\cdot\bar{\eta}_3)_B-\A_{12;3BE}(\A_{12}\cdot\bar{\eta}_3)_A,\\
(\FCF{4}{i}{j}{m}{1}{2})=\A_{12;3AE}(\A_{12}\cdot\bar{\eta}_3)_B+\A_{12;3BE}(\A_{12}\cdot\bar{\eta}_3)_A,
\end{gathered}
}
which transform well under $J_i(\eta_1)\leftrightarrow J_j(\eta_2)$ but not under $J_j(\eta_2)\leftrightarrow J_m(\eta_3)$.  A better tensor structure basis which divides into irreducible representations of the exchange symmetry group $S_3$ is
\eqn{
\begin{gathered}
(\tFCF{1}{i}{j}{m}{1}{2})=\A_{12;3AB}(\A_{12}\cdot\bar{\eta}_3)_E+\A_{12;3AE}(\A_{12}\cdot\bar{\eta}_3)_B-\A_{12;3BE}(\A_{12}\cdot\bar{\eta}_3)_A,\\
(\tFCF{2}{i}{j}{m}{1}{2})=(\A_{12}\cdot\bar{\eta}_3)_A(\A_{12}\cdot\bar{\eta}_3)_B(\A_{12}\cdot\bar{\eta}_3)_E,\\
(\tFCF{3}{i}{j}{m}{1}{2})=\A_{12;3AB}(\A_{12}\cdot\bar{\eta}_3)_E+\A_{12;3AE}(\A_{12}\cdot\bar{\eta}_3)_B+2\A_{12;3BE}(\A_{12}\cdot\bar{\eta}_3)_A,\\
(\tFCF{4}{i}{j}{m}{1}{2})=\A_{12;3AE}(\A_{12}\cdot\bar{\eta}_3)_B+\A_{12;3BE}(\A_{12}\cdot\bar{\eta}_3)_A,
\end{gathered}
}[EqTSJJJ]
with the relations
\eqn{
\begin{gathered}
(\tFCF{1}{i}{j}{m}{1}{2})=(\FCF{1}{i}{j}{m}{1}{2})+(\FCF{3}{i}{j}{m}{1}{2}),\qquad(\tFCF{2}{i}{j}{m}{1}{2})=(\FCF{2}{i}{j}{m}{1}{2}),\\
(\tFCF{3}{i}{j}{m}{1}{2})=(\FCF{1}{i}{j}{m}{1}{2})-\frac{1}{2}(\FCF{3}{i}{j}{m}{1}{2})+\frac{3}{2}(\FCF{4}{i}{j}{m}{1}{2}),\qquad(\tFCF{4}{i}{j}{m}{1}{2})=(\FCF{4}{i}{j}{m}{1}{2}).
\end{gathered}
}
From \eqref{EqTSJJJ} it is easy to see that the first two tensor structures are each in the sign representation of $S_3$ while the last two tensor structures together form the standard representation of $S_3$.  Moreover, the link between the three-point coefficients in the two bases is simply
\eqn{
\begin{gathered}
\taCF{1}{i}{j}{m}=\frac{1}{3}\aCF{1}{i}{j}{m}+\frac{2}{3}\aCF{3}{i}{j}{m},\qquad\taCF{2}{i}{j}{m}=\aCF{2}{i}{j}{m},\\
\taCF{3}{i}{j}{m}=\frac{2}{3}\aCF{1}{i}{j}{m}-\frac{2}{3}\aCF{3}{i}{j}{m},\qquad\taCF{4}{i}{j}{m}=-\aCF{1}{i}{j}{m}+\aCF{3}{i}{j}{m}+\aCF{4}{i}{j}{m}.
\end{gathered}
}[EqaJJJ]

We can now make general statements from the tensor structure basis \eqref{EqTSJJJ} which are correct whether the $\boldsymbol{e}_1$ are conserved or not.  For example, since there are no tensor structures in the trivial representation of $S_3$, we conclude that the three-point correlation functions $[\boldsymbol{e}_1,\boldsymbol{e}_1,\boldsymbol{e}_1]$ of three identical $\boldsymbol{e}_1$ vanish.  The same is true when only two $\boldsymbol{e}_1$ are identical.\footnote{A conformal version of the Landau-Yang theorem.}  For three different $\boldsymbol{e}_1$, the three-point coefficients $\taCF{1}{i}{j}{m}$ and $\taCF{2}{i}{j}{m}$ must be in the sign representation of $S_3$ while $\taCF{3}{i}{j}{m}$ and $\taCF{4}{i}{j}{m}$ must lie in the standard representation of $S_3$, where the $S_3$ acts on the indices $i$, $j$, and $m$ in the expected manner.

In terms of the three-point coefficients \eqref{EqaJJJ}, the conservation conditions \eqref{EqcJcJle1} for $\ell=1$ and $\Delta_m=d-1$ lead to $\taCF{3}{i}{j}{m}=\taCF{4}{i}{j}{m}=0$ which imply that the last two tensor structures in \eqref{EqTSJJJ}, which are in the standard representation of $S_3$, vanish.  This was expected since the conservation conditions \eqref{EqcJcJle1} already forced the vanishing of one of the two tensor structures appearing in the standard representation, and since it mixes with the other tensor structure under the $S_3$ action, they both had to vanish.

At this point, it is straightforward to find the conservation conditions for the last conserved vector current using \eqref{Eq3pt3} and \eqref{Eq3pt3subs}.  A direct computation shows that these conservation conditions are automatically satisfied.  This observation was however guaranteed since the two remaining tensor structures are in the sign representation of the exchange symmetry group.

Therefore, three-point correlation functions $\Vev{J_i(\eta_1)J_j(\eta_2)J_m(\eta_3)}$ have two tensor structures, namely $(\tFCF{1}{i}{j}{m}{1}{2})$ and $(\tFCF{2}{i}{j}{m}{1}{2})$, with three-point coefficients $\taCF{1}{i}{j}{m}$ and $\taCF{2}{i}{j}{m}$ that are in the sign representation of the exchange symmetry group $S_3$.  For usual conserved vector currents of global symmetries, the three-point coefficients $\taCF{a}{i}{j}{m}={}_a\beta f_{ijm}$ are thus proportional to the structure constants $f_{ijm}$ of the associated Lie algebras, since they are fully antisymmetric.  Hence the three-point correlation functions are
\eqn{\Vev{J_i(\eta_1)J_j(\eta_2)J_m(\eta_3)}=f_{ijm}\frac{(\mathcal{T}_{12}^{\boldsymbol{e}_1}\Gamma)(\mathcal{T}_{21}^{\boldsymbol{e}_1}\Gamma)(\mathcal{T}_{31}^{\boldsymbol{e}_1}\Gamma)\cdot[{}_1\beta(\tFCF{1}{i}{j}{m}{1}{2})+{}_2\beta(\tFCF{2}{i}{j}{m}{1}{2})]}{\ee{1}{2}{(d-3)/2}\ee{1}{3}{(d-2)/2}\ee{2}{3}{(d-1)/2}},}
but the two three-point coefficients ${}_a\beta$ are partially fixed through the Ward identity at coincident points.  Indeed, since $\delta_{J_m}J_i=f_{mij}J_j$, following \eqref{EqOOJ} with the substitution \eqref{EqS} we find that \cite{Osborn:1993cr,Li:2015itl}
\eqn{-\frac{(-2)^{(d-2)/2}}{d}[(d-1){}_1\beta-2{}_2\beta]S_d=c_J,}[EqJJJcp]
where the two-point correlation functions of conserved vector currents are given in \eqref{EqJJ}.  As a consequence of \eqref{EqJJJcp}, there is only one independent three-point coefficient in the three-point correlation functions $\Vev{J_i(\eta_1)J_j(\eta_2)J_m(\eta_3)}$.


\subsection{\texorpdfstring{$\Vev{JJT}$}{JJT}}

Since the energy-momentum tensor is in a different irreducible representation than the vector currents, the tensor structures \eqref{EqTSJJle1} with $\ell=2$ for the three-point correlation functions of two conserved vector currents and one energy-momentum tensor $\Vev{J_i(\eta_1)J_j(\eta_2)T(\eta_3)}$ are already convenient with respect to the exchange $\mathbb{Z}_2$ symmetry.

The conservation conditions \eqref{EqcJcJle1} with $\ell=2$ and $\Delta_m=d$ force the sole parity odd tensor structure to vanish while the four parity even tensor structures are restricted to only three independent tensor structures.  Since the remaining tensor structures are in the trivial representation, three-point correlation functions of two identical conserved currents and the energy-momentum tensors exist.

Combining the conservation conditions originating from \eqref{Eq3pt3} and \eqref{Eq3pt3subs} with \eqref{EqcJcJle1} leads to
\eqn{
\begin{gathered}
(d+1)(2\aCF{1}{i}{j}{m}+\aCF{5}{i}{j}{m})-4\aCF{2}{i}{j}{m}-2(d+2)\aCF{3}{i}{j}{m}=0,\qquad\aCF{4}{i}{j}{m}=0,\\
4\aCF{1}{i}{j}{m}-4\aCF{3}{i}{j}{m}+d\aCF{5}{i}{j}{m}=0,\qquad4\aCF{1}{i}{j}{m}+8\aCF{2}{i}{j}{m}+(d^2-2)\aCF{5}{i}{j}{m}=0,
\end{gathered}
}
but the unique new conservation condition is linearly dependent.  Hence demanding conservation of the energy-momentum tensor in $\Vev{J_i(\eta_1)J_j(\eta_2)T(\eta_3)}$ does not lead to any extra constraints apart from setting $\Delta_m=d$.

As for the Ward identity at coincident points for the three conserved vector currents, a simple computation following \eqref{EqOOT} and \eqref{EqS} shows that
\eqn{\frac{(-2)^{(d-2)/2}}{2d^2}[2(d-1)\aCF{1}{i}{j}{m}-4\aCF{2}{i}{j}{m}+\aCF{5}{i}{j}{m}]S_d=c_J,}[EqJJTcp]
where we used \eqref{EqJJ} and $\Delta_J=d-1$ \cite{Osborn:1993cr,Li:2015itl}.  Once again, \eqref{EqJJTcp} leads to a relation between the three-point coefficients, the charge $\Delta_J=d-1$ and the two-point coefficient $c_J$.


\subsection{\texorpdfstring{$\Vev{TTJ}$}{TTJ}}

The case of two energy-momentum tensors and one conserved vector current $\Vev{T(\eta_1)T(\eta_2)J_m(\eta_3)}$ is trivial.  Indeed, because the energy-momentum tensors are identical, only tensor structures that are even under the exchange $\mathbb{Z}_2$ symmetry survive.  Thus the tensor structures of interest are \eqref{EqTSTTle1odd} with $\ell=1$, and the conservation conditions \eqref{EqcTcTle1odd} imply that the three-point correlation functions vanish.  Imposing conservation for the vector current cannot change this result.

Moreover, with respect to Ward identities at coincident points, it is clear that no new information can be obtained since $\delta_JT=0$ by consistency.


\subsection{\texorpdfstring{$\Vev{TTT}$}{TTT}}

Considering the three-point correlation function $\Vev{T(\eta_1)T(\eta_2)T(\eta_3)}$ of three (identical) energy-momentum tensors, the tensor structures must be in the trivial representation of the exchange symmetry group $S_3$.  The proper tensor structures can be built from \eqref{EqTSTTle1even} with $\ell=2$ as
\eqna{
(\tFCF{1}{i}{j}{m}{1}{2})&=\A_{12;3A_1B_1}\A_{12;3A_2B_2}(\A_{12}\cdot\bar{\eta}_3)_{E_1}(\A_{12}\cdot\bar{\eta}_3)_{E_2}+\A_{12;3A_1E_1}\A_{12;3A_2E_2}(\A_{12}\cdot\bar{\eta}_3)_{B_1}(\A_{12}\cdot\bar{\eta}_3)_{B_2}\\
&\phantom{=}\qquad+\A_{12;3B_1E_1}\A_{12;3B_2E_2}(\A_{12}\cdot\bar{\eta}_3)_{A_1}(\A_{12}\cdot\bar{\eta}_3)_{A_2},\\
(\tFCF{2}{i}{j}{m}{1}{2})&=[\A_{12;3A_1B_1}(\A_{12}\cdot\bar{\eta}_3)_{E_1}+\A_{12;3A_1E_1}(\A_{12}\cdot\bar{\eta}_3)_{B_1}-\A_{12;3B_1E_1}(\A_{12}\cdot\bar{\eta}_3)_{A_1}]\\
&\phantom{=}\qquad\times(\A_{12}\cdot\bar{\eta}_3)_{A_2}(\A_{12}\cdot\bar{\eta}_3)_{B_2}(\A_{12}\cdot\bar{\eta}_3)_{E_2},\\
(\tFCF{3}{i}{j}{m}{1}{2})&=(\A_{12}\cdot\bar{\eta}_3)_{A_1}(\A_{12}\cdot\bar{\eta}_3)_{A_2}(\A_{12}\cdot\bar{\eta}_3)_{B_1}(\A_{12}\cdot\bar{\eta}_3)_{B_2}(\A_{12}\cdot\bar{\eta}_3)_{E_1}(\A_{12}\cdot\bar{\eta}_3)_{E_2},
}
\eqna{
(\tFCF{4}{i}{j}{m}{1}{2})&=\A_{12;3A_1E_1}\A_{12;3A_2B_2}(\A_{12}\cdot\bar{\eta}_3)_{B_1}(\A_{12}\cdot\bar{\eta}_3)_{E_2}-\A_{12;3B_1E_1}\A_{12;3A_2B_2}(\A_{12}\cdot\bar{\eta}_3)_{A_1}(\A_{12}\cdot\bar{\eta}_3)_{E_2}\\
&\phantom{=}\qquad-\A_{12;3A_1E_1}\A_{12;3B_1E_2}(\A_{12}\cdot\bar{\eta}_3)_{A_2}(\A_{12}\cdot\bar{\eta}_3)_{B_2},\\
(\tFCF{5}{i}{j}{m}{1}{2})&=\A_{12;3A_1E_1}\A_{12;3B_1E_2}\A_{12;3A_2B_2},
}[EqTSTTT]
where
\eqn{
\begin{gathered}
(\tFCF{1}{i}{j}{m}{1}{2})=(\FCF{1}{i}{j}{m}{1}{2})+(\FCF{10}{i}{j}{m}{1}{2}),\qquad(\tFCF{2}{i}{j}{m}{1}{2})=(\FCF{2}{i}{j}{m}{1}{2})+(\FCF{6}{i}{j}{m}{1}{2}),\\
(\tFCF{3}{i}{j}{m}{1}{2})=(\FCF{3}{i}{j}{m}{1}{2}),\qquad(\tFCF{4}{i}{j}{m}{1}{2})=(\FCF{4}{i}{j}{m}{1}{2})-(\FCF{9}{i}{j}{m}{1}{2}),\qquad(\tFCF{5}{i}{j}{m}{1}{2})=(\FCF{8}{i}{j}{m}{1}{2}).
\end{gathered}
}
From \eqref{EqTSTTT}, we conclude that there are five different tensor structures for three-point correlation functions $[2\boldsymbol{e}_1,2\boldsymbol{e}_1,2\boldsymbol{e}_1]$ with identical quasi-primary operators.

Imposing conservation reduces the set of independent tensor structures to three, with the conservation conditions given by
\eqn{
\begin{gathered}
8(d+1)\taCF{1}{i}{j}{m}+8\taCF{2}{i}{j}{m}-2(d-2)\taCF{4}{i}{j}{m}+(d^2+d-4)\taCF{5}{i}{j}{m}=0,\\
4(d^2+2d-2)\taCF{1}{i}{j}{m}+16\taCF{2}{i}{j}{m}+32\taCF{3}{i}{j}{m}-2(d^2+2d-4)\taCF{4}{i}{j}{m}+(d^2+2d-4)\taCF{5}{i}{j}{m}=0.
\end{gathered}
}
This statement implies that setting $\Delta_m=d$ in \eqref{EqcTcTle1even} is sufficient to impose conservation of the last $2\boldsymbol{e}_1$ quasi-primary operator.

Considering the Ward identity at coincident points, \eqref{EqOOT} and \eqref{EqS} imply that
\eqn{\frac{(-2)^{(d-2)/2}}{2d(d+2)}[2(d^3-d^2-2d+6)\taCF{1}{i}{j}{m}-4d(d-1)\taCF{2}{i}{j}{m}+16(d-1)\taCF{3}{i}{j}{m}+2d\taCF{4}{i}{j}{m}+(d^2-2)\taCF{5}{i}{j}{m}]S_d=c_T,}[EqTTTcp]
taking into account \eqref{EqTT} and $\Delta_T=d$ \cite{Osborn:1993cr,Li:2015itl}.  Equation \eqref{EqTTTcp}, which relates the three-point coefficients to $c_T$, can be written in terms of three independent three-point coefficients with the help of the conservation conditions at non-coincident points \eqref{EqcTcTle1even} with $\Delta_m=d$.


\bibliography{ConservedCurrents}

\end{document}